\documentclass[numberedappendix,appendixfloats]{emulateapj}
\usepackage{epstopdf}
\usepackage{graphicx}
\usepackage{epsfig}
\usepackage{amsmath}
\usepackage{wasysym}
\usepackage{txfonts}
\pdfoutput=1
\usepackage[pdftex]{color,xcolor}
\usepackage[backref]{hyperref}
\usepackage{hyperref}
\hypersetup{pdfauthor=Christoph Mordasini}
\hypersetup{backref=true, pagebackref=true, hyperindex=true, breaklinks=true,colorlinks=true,urlcolor=blue, linkcolor=blue,  citecolor=blue,pagecolor=red, bookmarks=true, bookmarksopen=true}
\usepackage{epstopdf}
\usepackage{lipsum}
\usepackage{natbib}
\usepackage{mathtools}
\usepackage{fancyvrb}
\usepackage{cprotect}
\usepackage{verbatimbox}

\bibpunct{(}{)}{;}{a}{}{,}
\usepackage{setspace}

\def\mearth{M_\oplus}
\def\Mearth{M$_\oplus$}
\def\AU{AU}

\def\msun{M_\odot}
\def\Msun{M$_\odot$}
\def\rsun{{\rm R}_\odot}

\def\f1{f_{\rm I}}
\def\mj{{\rm M}_{\textrm{\tiny \jupiter }}}

\newcommand{\rj}{{\rm R}_{\textrm{\tiny \jupiter}}}

\def\Rjup{R$_{\rm Jup}$}

\def\beq{\begin{equation}}
\def\eeq{\end{equation}}

\def\t2{\tau_{\rm II}}

\def\sigmas0{\Sigma_{\rm s,0}}

\newcommand{\ColorJournal}{See the electronic edition of the Journal for a color version of this figure.}

\newcommand{\Tice}{180}
\newcommand{\water}{H$_2$O}
\newcommand{\Cmonoxide}{CO}
\newcommand{\Cdioxide}{CO$_2$}
\newcommand{\methane}{CH$_4$}
\newcommand{\ammonia}{NH$_3$}
\newcommand{\HHe}{H$+$He}

\newcommand{\cs}{~}  
\newcommand{\cm}{\,}  
\newcommand{\ca}{\,}  
\newcommand{\cu}{\,}  
\newcommand{\cf}{~}  
\newcommand{\ct}{~}  

\def\apjl{ApJ}                
\def\apjs{ApJS}               
\def\ao{Appl.Optics}          
\def\aap{A\&A}                

\def\({\left(}
\def\){\right)}
\def\<{\left<}
\def\>{\right>}

\newcommand{\rch}[1]{{{#1}}}
\newcommand{\rt}[1]{{{#1}}}
\newcommand{\rcp}[1]{{{#1}}}

\VerbatimFootnotes

\shorttitle{Exoplanet formation and observable spectra}
\shortauthors{C. Mordasini et al.}

\begin{document}

\title{The imprint of exoplanet formation history on \\ observable present-day spectra \rt{of hot Jupiters}}


\author{C. Mordasini\altaffilmark{1,2},  
        R. van Boekel\altaffilmark{1}, P. Molli\`{e}re\altaffilmark{1},  Th. Henning\altaffilmark{1},
        Bj\"orn Benneke\altaffilmark{3}}
\altaffiltext{1}{Max-Planck-Institut f\"ur Astronomie, K\"onigstuhl 17, D-69117 Heidelberg, Germany; boekel@mpia.de, molliere@mpia.de, henning@mpia.de}
\altaffiltext{2}{Physikalisches Institut, University of Bern, Sidlerstrasse 5, 3012 Bern, Switzerland; christoph.mordasini@space.unibe.ch}
\altaffiltext{3}{Division of Geological and Planetary Sciences, California Institute of Technology, Pasadena, CA 91125, USA; bbenneke@caltech.edu}

\begin{abstract}
{The composition of a planet's atmosphere is determined by its formation, evolution, and present-day insolation. A planet's spectrum therefore \rt{may} hold clues on its origins. We present a ``chain'' of models, linking the formation of a planet to its \rt{observable} present-day spectrum. The chain links include (1) the planet's formation and migration, (2) its long-term thermodynamic evolution, (3) a variety of disk chemistry models, (4) a non-gray atmospheric model, and (5) a radiometric model to obtain simulated spectroscopic observations with JWST and ARIEL. In our standard chemistry model the inner disk is depleted in refractory carbon as in the Solar System and in white dwarfs polluted by extrasolar planetesimals. Our main findings are: (1) Envelope enrichment by planetesimal impacts during formation dominates the final planetary atmospheric composition of hot Jupiters. We investigate two, under this finding, prototypical formation pathways: a formation inside or outside the water iceline, called ``dry'' and ``wet'' planets, respectively. (2) Both the ``dry'' and ``wet'' planets are oxygen-rich (C/O$<$1) due to the oxygen-rich nature of the solid building blocks. The ``dry'' planet's C/O ratio is $<$0.2 for standard carbon depletion, while the ``wet'' planet has \rt{typical} C/O values between 0.1 and 0.5 \rt{depending mainly on the clathrate formation efficiency}. Only non-standard disk chemistries without carbon depletion lead to carbon-rich C/O ratios $>$1 for the ``dry'' planet. (3) \rt{While we consistently find C/O ratios $<$1, they still vary significantly. To link a formation history to a specific C/O,} a better understanding of the disk chemistry is \rt{thus needed}.}

\end{abstract}

\section{Introduction}
\label{sec:introduction}
One of the most fascinating aspects of the recent observational progress in exoplanet science are the first spectra of  planets around other stars. Spectra probe the atmosphere which is a window into the composition of a planet. This composition, in turn, may give critical insights into the formation and migration history of the planet.  
A planet's composition depends on the composition of the host star, the structure and chemistry of the protoplanetary disk, the locations where the planet accreted, the composition of the accreted gas and solids, the properties (size, strength) of the accreted bodies like planetesimals or pebbles, the mixing or separation of the different materials inside the planet, the interaction and exchange between the interior and the atmosphere, the stellar radiation field, etc. Therefore, each formation track of a planet will leave - potentially in a convoluted way -  an imprint in the atmospheric composition. This means that atmospheric spectra might contain a multitude of clues to planetary formation that cannot be provided by other observational techniques.
 
For the Solar System planets atmospheric observations show that Jupiter is enriched in carbon by about a factor four relative to the sun, Saturn by a factor $\sim$10, while Uranus and Neptune are enriched by a factor $\sim$90 \citep{guillotgautier2014}. This trend of decreasing enrichment with increasing mass was recently found to apply also to WASP-43b \citep{kreidbergbean2014}.  For the core accretion formation model \citep[e.g.,][]{alibertmordasini2005} such a trend is a natural prediction \citep[for a quantification, see][]{mordasiniklahr2014}, but not necessarily for the competing direct collapse model. Thus, spectra can help to distinguish formation models of the Solar System.

 Regarding exoplanets, the atmospheric composition may in particular also give clues on the  formation of hot Jupiters, which are currently the best characterized class of exoplanets. The discovery of a Jovian planet at an orbital distance of only 0.05\cu\AU \ from its star by \cite{1995Natur.378..355M} was a surprise. Theoretical planet formation models had rather predicted \citep[e.g.,][]{1995Sci...267..360B} that giant planets should be found several AU away. The mechanism that was underestimated was orbital migration
  \citep{1980ApJ...241..425G}. As a reaction, orbital migration due to gravitational interaction with the protoplanetary gas disk was included in planet formation theory as a key mechanism \citep[e.g.,][]{linbodenheimer1996}. Disk migration predicts that planetary orbits are coplanar with the stellar equatorial plane (but see also \citealt{batygin2012}). The subsequent discovery of highly inclined or even retrograde hot Jupiters has therefore again challenged theory \citep[e.g.,][]{2010ApJ...718L.145W}. Alternative scenarios bringing giant planets close to the host star were developed. The most important scenarios are planet-planet scattering in unstable systems of planets and Kozai migration due to the presence of an outer perturber \citep[e.g.][]{2010A&A...524A..25T}. These mechanism take place after the dispersal of the protoplanetary disk and can lead to highly inclined planets. To date, it is debated if disk migration or scattering/Kozai is the dominant mechanism leading to close-in planets \citep[e.g.,][]{cridabatygin2014}.

An interesting novel approach of constraining which migration processes acted on the planet during or after its formation is to evaluate whether the formation process, chiefly the planet's location(s) in the  disk during its formation, leaves an observable spectral signature.
If one could, e.g., deduce from the spectrum that a hot Jupiter has accreted exclusively outside the water iceline, this would make disk migration through the inner part of the disk unlikely as the processes that brought this planet close to the star. 

The reason is that the planet would accrete matter while migrating through the inner disk \citep{foggnelson2007a}. Apart from being able to constrain possible migration scenarios a successful link between a planet's formation and its spectrum would be very interesting on its own, providing a historical record of the formation of individual planets.


\rcp{The first attempts to link the planetary formation process to \rch{exo}planetary compositions have in part been sparked by
a retrieval analysis which suggested that WASP-12b, a hot Jupiter around a G0 main-sequence star, is carbon-rich\footnote{In this work we define oxygen-rich and carbon-rich as C/O$<$1 and $>$1 by number, respectively. This is different from the absolute enrichment level in C and O \rt{and the sub/super-stellar C/O distinction}.} with C/O $\gtrsim$ 1 \citep{madhusudhanharrington2011}. Further claims of a C/O$>$1, and a corresponding carbon rich chemistry including absorbers such as HCN and C$_2$H$_2$, have been made by \citet{stevensonbean2014}. Both of these assessments rely on \emph{Spitzer} eclipse photometry, impeding the conclusive detection of a carbon-bearing molecule in the atmosphere of this planet thus far. Studies contesting the claim of a carbon-rich WASP-12b include \citet{crossfieldbarman2012,swainderoo2013,lineknutson2014,benneke2015,kreidbergline2015}. In these studies the retrieved C/O ratio may reach super-solar ($\gtrsim$0.56) values, but the (7 $\sigma$) detection of H$_2$O firmly rules out an atmosphere with a carbon-rich chemistry (C/O$>$1), if equilibrium chemistry is assumed \citep{kreidbergline2015,benneke2015}. As stated in \citet{stevensonbean2014}, an oxygen-rich atmosphere would require unrealistically large CO$_2$ abundances to fit the planet's photometric emission data. In this case higher SNR dayside emission spectroscopy  may resolve these inconsistencies.} \rcp{Even though the  data quality thus can currently still inhibits conclusive statements about atmospheric compositions, the question of how the formation process constrains the planetary composition is interesting and should be studied in any case for the reasons outlined above.}

\rch{It is important to note that the existing studies attempting to link planetary formation and composition can be divided into two classes: in the first class the planetary formation process itself is included in the analysis. In the second class the planet formation process is not modeled. Here the disk gas and solid composition as a function of time and location in the disk is investigated and the results are used to infer the composition of gaseous planets forming at this location and time. In this second class planets with C/O~$>$~1 may only be formed if the planet's metal enrichment is dominated by the accreted gas.}

\rch{In the study presented here we will show that planets formed under the core accretion paradigm, with masses  typical of hot Jupiters and below, have an enrichment dominated by planetesimal accretion. We show this by explicitly modeling the planetary formation process and the planetesimal accretion process. We further show that this trend predicted by core accretion agrees well with measurements of the bulk and atmospheric abundances of exoplanets and Solar System planets.}

The studies which exist to this day include \rch{\citet{MousisMarboeuf2009,MousisLunine2009}}; \citet{obergmurray-clay2011,ali-dipmousis2014,thiabaudmarboeuf2014,hellingwoitke2014,marboeufthiabaud2014a,marboeufthiabaud2014b,madhusudhanamin2014,thiabaudmarboeuf2015}; \rch{\citet{CridlandPudritz2016}}.
These studies vary widely in their scopes: 

\rch{In the context of Jupiter and Saturn, \citet{MousisMarboeuf2009} combine the planet formation model of \citet{alibertmordasini2005} with a model for the formation of  clathrates and pure condensates. They assume that the observed atmospheric enrichment in volatiles originates from the vaporization of icy planetesimals entering the envelopes of the growing planets. They show that  for Jupiter this leads to an enrichment of both the atmosphere and interior that is in agreement with observations. Their results indicate that large amounts of icy solids have been incorporated into Jupiter's and Saturn's envelope.} 

\rch{In the context of exoplanets,} \citet{obergmurray-clay2011} constrain possible planetary C/O ratios based on the disk volatile ice lines but do not model the planet formation process. \rch{In this work the possibility of planets with C/O~$\rightarrow$1 may only arise for planets which have their enrichment dominated by gas accretion and only if they form between the CO$_2$ and CO icelines.}

\citet{hellingwoitke2014} carry out a more detailed analysis of the volatile components within a pre-stellar core and protoplanetary disk, 
modeling the volatile gas and ice abundances as a function of time in static core and disk models. They also model how cloud formation
ensues in planetary atmospheres of various abundances and C/O ratios, but do not model the formation of the planets in the disk.
\rch{Similar to \citet{obergmurray-clay2011}} they find that super-\rch{stellar} C/O ratios (but $\lesssim$ 1) in the disk gas are possible, mainly between the CO$_2$ and CO icelines.

The first studies to more self-consistently link the planet formation process to the final elemental abundances within the planet \rch{in the context of exoplanets} were performed by \citet{thiabaudmarboeuf2014,marboeufthiabaud2014a,marboeufthiabaud2014b} and \citet{thiabaudmarboeuf2015}. They modeled planetesimal formation by assuming refractory and volatile condensation in an initial protoplanetary disk, and then let the gas disk evolve viscously while modeling the planet formation via the core accretion paradigm. \citet{thiabaudmarboeuf2015} find that the gas giants forming in their models have low C/O ratios\footnote{The solar C/O ratio is $\sim$ 0.56 \citep{asplundgrevesse2009}.} unless there is a lack of mixing between the envelope and the accreted solids. \rch{They also find that if icy planetesimals fully sublimate into the planets' gaseous envelope, then their effect on the planetary C/O ratio is dominant compared to the contribution of the accreted gas.} 

\citet{madhusudhanamin2014} use a simplified description of a planetary population synthesis forming 1 M$_{\rm Jup}$ planets by both core accretion and gravitational instability with final semi-major axes of 0.1 AU. They study whether migration mechanisms might be constrained by the resulting planetary compositions. Disk migration and disk-free migration processes for a planet forming within a viscously evolving disk are treated, keeping track of the matter accreted at various orbital distances in both planetesimal (rocky and/or icy) and gaseous form (including volatiles).
Only type II disk migration is modeled and the growth of the planet before opening the gap in type II migration is neglected. This could be a non-trivial assumption, as the planets are possibly strongly enriched before opening of the gap and prior to runaway gas accretion \citep{fortneymordasini2013}. \rch{Therefore, planets in this study which form via core accretion, but have a sub-\rch{stellar} enrichment, are likely caused by the lack of modeling the planet's formation before the type II migration sets in.} Furthermore, rapid type I migration could result in a strong enrichment which originates farther outside in the disk where different planetesimal and gas compositions are likely present.
\citet{madhusudhanamin2014} consider two different compositional models for the volatiles and refractory disk, one including carbon grains based on protoplanetary disk observations and one without carbon grains. They find that planets which formed in the outer regions of the disk \rch{may} have sub-\rch{stellar} C and O abundances \rch{if the planets are dominated by gas enrichment} and C/O ratios ranging from \rch{stellar} to super-\rch{stellar} values. This class of planets, if found close to its star at 0.1 AU must therefore have moved in after disk dispersal, suggesting a disk-free migration mechanism. Planets which formed in the inner regions of the disk are found to have super-\rch{stellar} C and O abundances and \rch{stellar} and sub-\rch{stellar} C/O ratios.

\rch{In the work presented here we take the previous approaches a step further and} directly investigate whether the formation process leaves visible imprints in the planetary spectra and whether these can be used to constrain planetary formation and migration theory. This is achieved by constructing a ``chain'' of models directly linking the formation, evolution, and present-day spectral appearance of the planets, where the output of one chain link serves self-consistently as input for the next one. With this ``chain'' we furthermore want to study the range of resulting planetary C/O ratios. 

There are five chain links in our model: In the \textit{first chain link} we fully model the planet's formation via core accretion in a gas and planetesimal disk, yielding the planetary core and envelope masses. The viscous evolution of the disk is modeled as well as type I and II disk migration. The fate of planetesimals during infall into the protoplanet is also directly treated, so that it is known which solids enrich the H/He envelope, and which ones reach the solid core. In the \textit{second chain link}, after the planet has formed, we evolve it to 5 Gyr using a planet evolution model that describes the thermodynamic evolution with the initial conditions given by the formation model. We solve the planetary structure equations including atmospheric escape and using a double-gray atmospheric model with appropriately scaled solar opacities given the envelope's bulk enrichment from formation. 

As the composition of volatiles and refractories in the disk in- and outside of the iceline is currently not well understood, the planetary formation model merely tracks the mass fractions of accreted volatiles and refractories. Then, at 5 Gyr, in the \textit{third chain link}, a chemistry model translates these bulk compositions yielded by the formation model into elemental abundances in the planet's atmosphere.
Combining various volatile and refractory compositional models and turning on or off effects such as clathrate formation or
volatile flushing inside of icelines we have 152 different compositional models outside of the iceline and 54 inside. In the \textit{fourth chain link} we use the planetary elemental abundances from the chemistry model and the radius and luminosity from the evolutionary model to calculate the planet's emission and transmission spectra with self-consistent non-gray atmospheric models.  Finally, in the \textit{fifth chain link} we use the spectra to simulate secondary eclipse observations with the JWST and ARIEL using the EclipseSim package to see whether the different spectral imprints can be distinguished. 

With this linked approach, we want to make the aforementioned earlier predictions regarding the imprint of formation on the planetary composition and its expression in spectra more comprehensive and coherent and take a step towards exoplanetology which will be at the focus of upcoming observational studies on extrasolar planets. To demonstrate this we apply our chain model to the example of two \rch{prototypical} planets that eventually become hot Jupiters. First, a ``dry Jupiter'', a Jovian-mass planet that forms exclusively inside of the water iceline and migrates close to its host star by disk migration. Second, a ``wet Saturn'', a Saturnian-mass planet that forms fully outside of the water iceline, and gets to its final positions close to the host star by a dynamical interaction like planet-planet scattering or Kozai mechanism. \rch{As we find that the enrichment of planets with masses typical for hot Jupiter and below is dominated by planetesimal accretion it is important to study the two under this result  fundamentally different, and prototypical, cases of how a planet's enrichment can vary as a function of the planet's formation location. This lead to the choice of looking at the ``dry'' and ``wet'' planet.} We find that for some assumptions for the disk chemistry, clear imprints \rch{on the spectra} exist, while for others, it is difficult to distinguish the formation histories.

We introduce our model in Section \ref{sec:methods} and show our calculations and results in Section \ref{sec:results}.
A discussion and summary can be found in Section \ref{sec:discussion}.

\section{Methods}\label{sec:methods}
In this section we describe the methods used to model the connection between the formation history of a planet and its present day atmospheric spectrum with a chain of linked models as outlined briefly in Section \ref{sec:introduction}. The five chain links (formation, evolution, abundances, spectra, observations), 
their sequential relation and their various submodules and processes are summarized in Figure \ref{fig:chain_fig}.
Further details on the submodules and processes are given in the appendices and will be referenced throughout this section.

\begin{figure*}[htb!]
\includegraphics[angle=0,width=\textwidth]{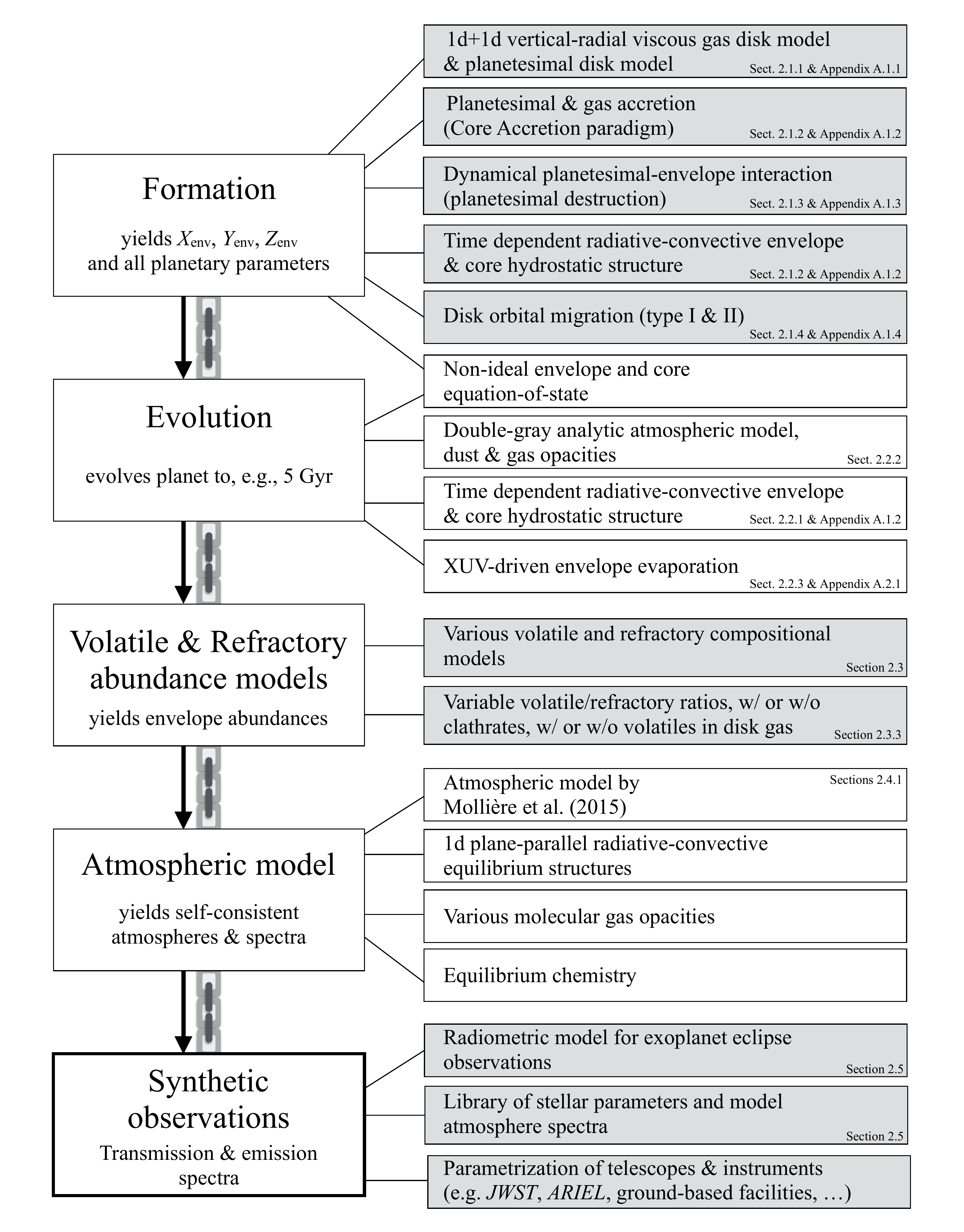}
\caption{Schematic overview showing the individual links in our chain of models ({\it left column}) and the various processes treated by them ({\it right column}).}
\label{fig:chain_fig}
\vspace*{10mm}
\end{figure*}

\subsection{Formation model}\label{sec:methods:formation_model}
Acting as the first chain link, our global planet formation model that describes the planet's accretion is based on the core accretion paradigm \citep{perricameron1974,mizunonakazawa1978} coupled self-consistently to disk evolution \citep{lynden-bellpringle1974} and orbital migration \citep{linpapaloizou1986}. The model has been described in detail in several previous works (\citealt{alibertmordasini2004,alibertmordasini2005}, \citealt{mordasinialibert2012b}) with a recent overview in \citet{mordasinimolliere2014}. Therefore, we only give a short summary here. The following four elements of the formation model are particularly important for the current work:

\subsubsection{Disk model \& initial conditions}
\label{sec:methods:disk_model}
The temporal evolution of the protoplanetary gas disk is described with a 1+1D (vertical and radial) viscous disk model. Our model includes the effects of turbulent viscosity in the $\alpha$-approximation, photoevaporation by the star and
from external sources, and mass accretion onto the planet. Further information on the gas disk evolution can be found in Appendix
\ref{appendix:chain_form_gas_disk}.
The model for the disk of planetesimals is very simple. It is assumed that very early in the evolution of the disk, planetesimals form quickly and rapidly grow to a size of 100 \,km, for example due to gravoturbulent formation \citep{johansenoishi2007,cuzzihogan2008,2011A&A...529A..62J}.  At this size, they interact only weakly with the disk gas, and do not significantly migrate radially. The surface density of planetesimals evolves only due to accretion and ejection of planetesimals by the protoplanet. The initial radial profile of the solids is given by the initial surface density profile of the gas multiplied by a factor 0.04\cm$\times$\cm$10^{\rm [Fe/H]}$ where [Fe/H] is the stellar metallicity, an initial condition of the model. We put 0.04 here instead of Z$_\odot$ $\sim$ 0.015 \citep{2003ApJ...591.1220L} as the inner parts
of a disk where planets form are typically enriched in solids when compared to the outer regions of the disk due to
dust drift \citep{rozyczkakornet2004}.
The change of the surface density at the water iceline is also taken into account (see \citealt{mordasinialibert2009a}).
The position of the water iceline is calculated as the distance where the temperature drops below \Tice\cu K in the initial disk structure.

It is clear that the concept of an ``initial'' disk profile is ill defined, since the disk gradually forms as a byproduct of the formation of the star, and this can lead to a significant radial movement of the iceline \citep[e.g.][]{mindullemond2011}. However, for the results of this study, the precise location of the iceline is not critical: We will concentrate on two planets having formed either significantly in- or outside the iceline
for our results in Section \ref{sec:results}, so the most important assumption simply is that the planets do not cross the iceline during their formation.

\subsubsection{Accretion of the solid core and of the gaseous envelope}\label{sec:methods:accretion_model}
The protoplanet's solid accretion rate is obtained  with a Safronov-type rate equation, considering its gravitationally enhanced cross-section, as described in \citet{pollackhubickyj1996}. As mentioned in Section \ref{sec:methods:disk_model} above, the surface density of planetesimals is coupled to the accretion of planetesimals by the planet. The decreasing amount of planetesimals increases the growth time scale of the protoplanetary core over time and eventually leads to an end of planetesimal accretion. We give more information on the modeling of the planetesimal accretion in Appendix \ref{appendix:chain_form_accretion}.

The accretion rate of gas is found by solving a slightly simplified set of internal structure equations of the planet's 1D radial structure in the quasi-hydrostatic approximation \citep{bodenheimerpollack1986}. In the first accretion phase the planet is still attached to the disk nebula, and the planet's accretion rate is regulated by the envelope's Kelvin-Helmholtz contraction.
Once the gas accretion rate given by the planet's contraction increases over the disk's ability to deliver gas to the planet (runaway gas accretion), the planet detaches from the disk and contracts quickly. In this phase the planet's
accretion rate is solely determined by the disk's ability to deliver gas to the planet \citep{tanigawatanaka2015}.
We assume that the accretion shock radiates away all gravitational potential energy freed during the gas infall onto the planet,
producing low entropy, so-called ``cold start'' planets \citep{marleyfortney2007}. The post-formation Kelvin-Helmholtz timescales of the rather low-mass planets we consider are still relatively short, therefore this assumption is irrelevant for the planet's structure at a Gyr-age, which is the age for which we  calculate the planetary spectra later. More details on how we model the planet's gas accretion and on the equations we solve in this process can be found in Appendix \ref{appendix:chain_form_accretion}.

\subsubsection{Planetesimal-protoplanet interaction: envelope enrichment}\label{sec:methods:planetesimalimpacts}
An element of the formation model that is particularly important in this work is the model for the interaction of the planetesimals with the protoplanet's gaseous  envelope when a planetesimal is accreted. It yields the fraction of the mass of an impacting planetesimal that is deposited in the envelope, enriching the gas in heavy elements \citep{podolakpollack1988,mordasinialibert2006}. Any planetesimal material surviving the flight through the envelope is added to the central solid core. The deposition of planetesimal material in the envelope is key for the resulting chemical composition of planets \rch{(\citealt{MousisMarboeuf2009}}; \citealt{fortneymordasini2013}), including the Jovian- and Saturnian-mass planets studied below.  For very massive giant planets, the composition of the gas becomes important, too \citep{helledbodenheimer2010,mordasiniklahr2014}. \rch{This is  further discussed in Sect. \ref{sect:importanceplanetesimals}}. 

The impact model determines the radial mass deposition profile by numerically integrating the trajectory of a planetesimal of initial mass $M_{\rm pl}$ during its flight through the protoplanetary envelope under the actions of gravity, gas drag, thermal ablation, and aerodynamical disruption. The equations we solve for this \rch{and the parameters we use} are described in Appendix \ref{appendix:chain_form_planetesimal_enve_interact}. \rch{An important parameter for the outcome of a planetesimal's infall into the protoplanetary envelope is the initial size of the planetesimal. For 100 km sized planetesimals as assumed in this work, we found in population syntheses \citep[see][]{fortneymordasini2013} that giant planets contain solid cores with masses between 6 and 12 $\mearth$ at the end of the formation phase. The two planets studied in this work have for comparison core masses of 7.3 and 7.9 $\mearth$ (Sect. \ref{sec:results:accretion_history}).  For 1 km planetesimals, the core masses of the giant planets are reduced to about 1.2 to 3.5 $\mearth$. Additional planetesimals that are accreted are instead enriching the H/He envelope. These masses are a consequence of a self-shielding of growing cores against planetesimal impacts \citep{mordasinialibert2006}: in general, the more massive a core becomes, the higher also the mass of its surrounding H/He envelope. This envelope increasingly shields the core against further direct impacts of larger and larger planetesimals. This causes an auto-regulation of the maximal core mass of gaseous planets originating from  direct impacts of planetesimals.  It is however possible that the core mass gets altered during the formation and evolution phase because of sedimentation of ablated material or core dissolution \citep[e.g.,][]{stevenson1982}. These  processes  are currently neglected in our model.}

Note the following inconsistency in our current model: while we keep track of the radial mass deposition and resulting envelope enrichment, we computationally nevertheless add all accreted planetesimal mass to the (computational) core, as our model is currently not able to handle a compositionally varying equation of state (EOS). Computationally, the planet therefore consists of a solid core surrounded by a pure \HHe \ envelope that is described with the EOS of \citet{saumonchabrier1995}. This should be critically kept in mind, since the very strong enrichment that occurs during the formation of the planets as found below is known to lower the critical core mass \citep{horiikoma2011}. We currently work on including the actual envelope composition in the formation and evolution model \citep{venturinialibert2014}.

\subsubsection{Orbital migration}\label{sec:methods:migration_model}
Several processes can lead to a radial displacement of a planet, like the interaction with the disk of planetesimals \citep[e.g.][]{levisonthommes2010} or with other protoplanets \citep[e.g.][]{fordrasio2006}, Kozai migration \citep[e.g.][]{fabryckytremaine2007}, or classical disk migration due to the exchange of angular momentum with the gaseous disk \citep[e.g.][]{linpapaloizou1986}. The only process that is explicitly modeled in this work is classical disk migration. Disk migration occurs in two regimes. At low masses, planets undergo type I migration \citep{tanakatakeuchi2002a}. In contrast to the original work of \citet{tanakatakeuchi2002a} for isothermal disks we use a significantly revised version of type I disk migration \citep{baruteaucrida2013}.
In this description the actual direction of migration for realistic disk thermodynamics can  also be directed outwards.
A detailed description of our non-isothermal migration model  is given in  \citet{dittkristmordasini2014}.

Once a planet becomes sufficiently massive to open up a gap in the gaseous disk (of order 100\cu\Mearth), it passes into type II migration. We use the transition criterion of \citet{cridamorbidelli2006} to determine a planet's migration regime.
The equations we solve for modeling type I and type II  migration are described in Appendix \ref{appendix:chain_form_migration}.

We note that the concurrent formation of several protoplanets can modify the migration behavior of individual planets. In this work, we study the formation of only one planet per disk since we focus on the link of formation and resulting spectra for two prototypical cases. The consequences of the formation of several planets are discussed in \citet{alibertcarron2013}, while the impact on the composition is discussed in \citet{thiabaudmarboeuf2014,thiabaudmarboeuf2015} and \citet{marboeufthiabaud2014a,marboeufthiabaud2014b}.

For planets forming completely outside the iceline, but which still end up close to their stars as hot Jupiters,
we assume that the planet is brought close to the star by a few-body interaction (planet-planet scattering, Kozai migration)
and tidally circularized at 0.04 AU. This process is not actually modeled. The details of this process are not important for our conclusions as long as the interaction occurs at a sufficiently early time (which is likely, \citealt{malmbergdavies2010}), and without accretion of significant amounts of material (planetesimals, other protoplanets) that has formed inside of the water ice line. The results of \citet{matsumuraida2013} and \citet{mustildavies2015} indicate that this is a good approximation.

\subsection{Planet evolution model}
\label{sec:methods:planet structure}
Once the protoplanetary gas disk has dispersed or the planet has migrated to the inner border of the (computational) disk,
the evolutionary phase starts, modeled by our second chain link. In this phase, no accretion of gas and planetesimals occurs anymore.
Nonetheless, the mass of the planet can still change due to atmospheric escape.
For the calculation of the final observable we are interested in in this paper (the planetary spectra),
the calculations in this chain link yield the planet's radius, mass, and internal temperature (or luminosity) at an age of 5 Gyr.
In this phase, a temporal evolution of the composition of the atmosphere could also occur, but this effect is currently neglected.
Our evolutionary model has been described in details in \citet{mordasinialibert2012b} and \citet{jinmordasini2014},
therefore we here only give a short overview of the physical processes that are included.

\subsubsection{Interior}\label{sec:methods:interior}
To calculate the temporal evolution of the interior (its cooling and contraction) the same basic 1D internal structure equations are solved 
as in the formation phase (see Section\cs\ref{sec:methods:accretion_model} and Appendix \ref{appendix:chain_form_accretion}) using, 
however, different outer  boundary conditions. These boundary conditions are provided by an atmospheric model. 

The evolutionary model is self-consistently linked to the outcome of the formation phase in the sense that not only the planetary bulk composition resulting from formation is taken as initial condition, but also the entropy in the deep convective zone. The impact of the formation  on the luminosity and radius at young ages  is therefore automatically included. Both during the formation and evolution phase, we use the Schwarzschild criterion to determine if a layer is convective or radiative, and assume that in the convective parts, the radial entropy gradient vanishes. Both during formation and evolution, the planet consists of a deep convective interior  that contains almost all the mass and a surrounding radiative zone. The size of the radiative zone increases as the planet cools during evolution \citep{guillotshowman2002} and contains of order 1 \% of the mass at late times. 

Following the usual paradigm of fully convective interiors in giant planets, semi-convection that could occur due to compositional gradients is  neglected \citep{stevenson1985,lecontechabrier2012}. We note that in view of the findings in Section\cs\ref{sec:results:accretion_history}, where during formation weakly enriched \HHe \ is accreted on top of strongly enriched gas, semiconvection could occur \citep{stevenson1985,vazanhelled2015}. On the other hand, luminosities are high during formation, which favors vigorous convection, meaning that detailed calculations are necessary to clarify this point in future work. 

We find below that the enrichment of the atmosphere is mainly due to planetesimal accretion relatively early during formation. In light of this the assumption of large scale convection is crucial, because for a fully convective interior, one can assume a homogenous chemical composition, as convective eddies are very efficient in homogenizing it \citep{vazanhelled2015}. Convection is thus the justification for a second assumption, namely that for the chemical calculations, the planet's envelope (and atmosphere, see below) are uniformly mixed, so that all planetesimal material that has been dissolved in the envelope during formation contributes to the finally measured enrichment.

If semi-convection does in reality occur, it would mean that some of the highly enriched material accreted in the early phase of the formation of the planet does not contribute to the final atmospheric composition, since it would be buried below the purer gas accreted during gas runaway accretion. In this case, the composition of this gas and the planetesimals accreted only in the final stages would determine the final observable composition \citep{thiabaudmarboeuf2015}.

Other important assumptions in the evolutionary model are that the core does not dissolve (which could further enrich the envelope, \citealt{guillotstevenson2004}) and that no special bloating mechanisms occur \citep[for a recent overview, see][]{baraffechabrier2014}. For planets forming outside the iceline but finally becoming hot Jupiters one requires few-body interactions followed by tidal circularization to bring the planets to their final close-in orbits. In this case bloating would occur during the circularization phase.  However, at the ages of several Gyrs at which hot Jupiters are typically observed this should no longer be important \citep{lecontechabrier2010}.

Regarding  additional physics included in the evolutionary model, the radius of the solid core is calculated taking into account its composition (ice mass fraction) and the compression by the surrounding gas using the  modified polytropic equation of state of \citet{seagerkuchner2007}. The heating due to radioactive decay in the planet's core is included in the planet's luminosity budget while the effect of the thermal cooling of the core is currently neglected \citep{lopezfortney2013b}. However, as a (hot) Jupiter's internal evolution is dominated by envelope cooling and contraction, this has a small effect on the planet's thermal evolution. 
  
Note that during the evolution, there is the same inconsistency between the chemical and the cooling model as during formation: In the cooling model, the planets actually consists of a (computational) core that contains all the planetesimals accreted during formation surrounded by a pure \HHe \ envelope. In the chemical model, it is in contrast assumed that the planetesimal material that was deposited during the impacts in the envelope stays there, homogeneously mixed with the \HHe. It is clear that due to this inconsistency, the radii and luminosities predicted with the evolutionary model must  be considered approximative. The effect of the distribution of the heavy elements in a giant planet on its evolution has been studied in detail by \citet{baraffechabrier2008}. For a Jovian planet with a metal mass fraction $Z$ from  0.20 to 0.50, mixing the metals homogeneously into the \HHe \ leads to radii that are typically 4 to 12\% smaller than putting all solids in the core. The difference in radii of up to 12 \% leads to differences in log$(g)$ of $\sim$ 0.1.
The resulting effect of such small differences in log$(g)$ on planetary emission spectra of hot Jupiters are very small \citep{sudarskyburrows2003,molliereboekel2015} and the internal thermodynamic evolution for hot Jupiters is much less important for
the spectral shape than the SED and strength of the insolation.
For transmission spectroscopy the associated change in the atmospheric scale height could lead to non-negligible   changes in
the absolute planetary transit radii, but not so much in the shape of the transmission spectrum, except for the amplitude of the
absorption features, which varies with varying $g$ \citep[see, e.g.,][]{fortneyshabram2010}.
For comparison, the envelope $Z$ resulting from planetesimal accretion of the planets considered below are 0.09 and 0.28. 

\subsubsection{Atmospheric model for the evolutionary phase}\label{sec:methods:atmosphere}
During the planet's evolution we use an improved version of the double-gray atmospheric model of  \citet{guillot2010} as outer boundary condition  (see also \citealt{henghayek2012,hengmendoca2014}). This model yields atmospheric pressure-temperature profiles which are in fair agreement with detailed radiative transfer calculations of atmospheres of strongly irradiated giant planets. As described in \citet{jinmordasini2014}, we use tabulated data giving the central model parameter  $\gamma$, which is the ratio of the optical to the infrared opacity. The data have been derived by comparison with the atmospheric models of \citet{fortneylodders2008}.

For the optical depth calculations we use Rosseland mean opacities from the \citet{freedmanmarley2008} tables for
solar and scaled solar abundances, depending on the actual enrichment $Z$ of the planetary envelope.
In this way, the effect of the chemical composition on the atmospheric structure and cooling of the planet is taken into account, even if only
in an approximate way,  as  the tables of \citet{freedmanmarley2008} only give the opacity for a scaled solar composition gas and not for the specific composition of the atmosphere as predicted by chemistry and atomic abundances resulting from given volatile and refractory composition models.

The impact of the high opacity on the radius evolution can be significant \citep{burrowshubeny2007}. In our simulations, we find differences of up to 0.16\cu\Rjup\ at 5\cu Gyrs (Section\cs\ref{sec:results:evolution}), therefore we work on replacing the current atmosphere model with one that takes the actual composition self-consistently into account, as demonstrated by \citet{fortneyikoma2011} for the Solar System. As discussed in Section \ref{sec:methods:interior} the effect of radius variations of the order of 10\% on  the shape of the emission spectra of hot Jupiters are, however, quite small but could show as  changes in the absolute values of planetary transit spectra.

We assume that the chemical composition of the atmosphere is identical to the interior composition and constant in time.  For cold giant planets like Jupiter, the convective zone reaches close to the photosphere leading to mixing, whereas for hot Jupiters, a deep radiative zone separates them \citep{guillotshowman2002}. While our assumption that this deep radiative zones does not lead to a separation of interior and atmospheric in terms of composition must be further tested with detailed models, we carried out the following back-of-the-envelope estimation:
For mixing, we ask for $v_{\rm D} > v_{\rm settle}$ where $v_{\rm settle}$ is the settling speed of a given particle and $v_{\rm D} = K_{zz}H_P$ is the diffusive mixing velocity over a pressure scale height $H_P$ with the vertical eddy diffusion coefficient $K_{zz}$. Using hydrostatic equilibrium one finds
\beq
K_{zz} > \frac{k_{\rm B}T}{g\mu m_{\rm H}}v_{\rm settle} \ ,
\eeq
with the Boltzmann constant $k_{\rm B}$, the temperature $T$, the gravitational acceleration $g$, the mean molecular weight $\mu$ and the hydrogen mass $m_{\rm H}$. One can then compare typical settling velocities \citep[see][their Fig. 1]{parmentiershowman2013} of particles of a size of up to 0.1-1 $\mu$m in H--He-dominated, Jupiter-like planets with vertical eddy diffusion values typically found in GCM simulations \citep[see, e.g.][]{mosesvisscher2011,agundezparmentier2014}. One finds that the radiative parts of the atmospheres can quite possibly mix small particles from the convective envelope into the radiative atmosphere, with $v_{\rm D} \gtrsim v_{\rm settle}$.  It is clear that future work should address this process -as also semiconvection in the interior-  in more detail building for example on \citet{chamberlainhunten1987,spiegelsilverio2009}.

\subsubsection{Envelope evaporation}
\label{sec:methods:evaporation}
Close-in planets are exposed to intense UV and X-ray irradiation from their host star, especially at young ages. This can drive atmospheric escape \citep[e.g.][]{lammerselsis2003,baraffeselsis2004,erkaevkulikov2007a,murray-claychiang2009,owenjackson2012a,lopezfortney2013}. For hot Jupiters, the escape is hydrodynamic and can be driven either by X-rays or EUV \citep{owenjackson2012a}. In our evolutionary model \citep[see][]{jinmordasini2014},  the envelope evaporation rate due to XUV irradiation is modeled at high EUV fluxes with a radiation-recombination limited rate \citep{murray-claychiang2009}, and with an energy limited rate at lower fluxes.
The equations we solve to model envelope evaporation can be found in Appendix \ref{appendix:chain_evo_evap}.

For the relatively massive planets studied in this work, envelope evaporation does occur, but is not a dominating process, reducing the envelope mass only on a 1-15 \% level over the star's lifetime (see Section\cs\ref{sec:results:evolution}). Due to this, a potential mass fractionation during the escape process \citep{huntenpepin1987a} which could modify the atmospheric composition should not be important, especially because the escape fluxes of hydrogen are expected to be sufficiently high in hot Jupiters in order to drag along heavy species \citep{koskinenharris2013}. 

\subsection{Elemental abundance models}\label{sec:methods:stoichiometry}
We now describe the third chain link which is the model of the elemental composition of the building blocks of the planets and the disk chemistry. We adopt a composition that is inherited from the ISM \citep{gaidos2015} but allow the disk chemistry to alter the ISM refractory composition in a single, but crucial, way: namely that carbon grains present in the ISM material can be destroyed in the inner parts of the disk by oxidizing reactions \citep{2001A&A...378..192G,leebergin2010}. 

The formation model yields the mass fractions of gaseous, refractory, and volatile material that make up the envelope of the planet, but does not yet specify what these materials are in terms of elemental composition. This is done with this chain link, assuming a high number of different possible elemental compositions of the building blocks.

All refractories in the envelope are stemming from planetesimal accretion, while the volatiles can stem from both icy planetesimals and volatile gas accretion. As mentioned earlier in Sections \ref{sec:methods:interior} and \ref{sec:methods:atmosphere} we assume that the envelope is well mixed throughout its formation and evolution phase and that the envelope and the atmosphere have the same 
elemental compositions, without any settling occurring in the atmospheres.

As stated in Section \ref{sec:methods:disk_model} we assume that the planetesimals form early in the evolution of the disk, and quickly reach their final size of 100\cu km. They will contain refractory material, and at those locations in the disk where the midplane temperature \emph{at the time of planetesimal formation} is sufficiently low for one or more volatile species to freeze out, these species will also be included in the planetesimals in the form of ice. Furthermore, we assume that \emph{all} refractory material and all (condensed) volatiles are incorporated in the planetesimals. Lastly, because of their large size, we assume that the planetesimals do not drift radially, and that they retain their original composition (i.e. there is no out-gassing). Hence, the composition of the planetesimals that are accreted at each location in the disk depends on the local temperature during planetesimal formation, and on the assumed overall composition of the refractory and volatile material in the disk.

After the planets have formed we can use the volatile and refractory mass fractions in the planetary envelope and combine them with different compositional models for the volatile and refractory material, specifying the disk chemistry. This then yields the atomic elemental abundances in the envelope and, therefore, the atmosphere. This post-processing allows us to test and study the effects of various different compositional models without having to couple them self-consistently to the formation and evolutionary model every time.

In the sections below we will discuss our various  models for the refractory and volatile elemental compositions. We have used a large number of different models in order to understand how the disk chemistry influences the final predicted spectrum of the planets.

\subsubsection{Refractory material}\label{sec:methods:refractories}
In our basic model the refractory material is assumed to be a mixture of iron (Fe), carbon (C), and silicates with Enstatite-type stoichiometry (MgSiO$_3$) \rch{\citep[see, e.g.,][]{2005Icar..179..158M,2007A&A...462..667M}}. We assume that iron always makes up 1/3 of the refractory mass, but note that the iron does not play an important role in the current considerations. The carbon and oxygen-bearing refractory species in contrast play an important role for the heavy element abundances in the atmosphere of the formed planets. In particular for the planets forming within the water iceline, where the atmospheric heavy element budget is dominated by the refractory material accreted in planetesimals, the assumed C/silicate mass ratio determines the final C/O ratio and whether a C-rich or O-rich chemistry will prevail. In the case where the planet forms outside
the water iceline, large amounts of oxygen and potentially also carbon are accreted in the form of icy volatiles, and the assumed refractory composition plays a smaller, but still significant, role.

In the local ISM the C/silicate mass ratio is approximately 0.5 \citep{1997ApJ...475..565D}. We adopt this value, \emph{for the outer parts of the disk} (outside 5 AU) such that the default refractory composition contains mass fractions of 2/9 in carbon and 4/9 in silicates (and 1/3 in iron). We also also investigate a carbon-poor composition (carbon mass fraction 1/9, silicates 5/9) and a carbon-rich composition (carbon mass fraction 1/3, silicates 1/3).
Furthermore we explore an alternative composition of the refractory material, after \cite{1994ApJ...421..615P}. This consists, in terms of mass fractions, of 8.3\% olivine silicates (Mg$_2$SiO$_4$), 25.0\% pyroxene silicates (MgSiO$_3$), 10.4\% iron (Fe), 10.1\% troilite (FeS), and 46.2\% organic ``CHON'' material. The CHON material has an elemental composition of  C:H:O:N $=$ 1:1:0.5:0.12 by mass.
In total this leaves us with 4 different refractory compositional models.

\subsubsection{Refractory material: Carbon depletion}
\emph{In the inner part of the disk} our default refractory model is one of the 4 described above, but with the carbon mass fraction
decreasing in a power law fashion from its nominal value within a given model to 2$\times$10$^{-4}$ times the nominal value when going from 5 to 1 AU. Inside 1 AU the carbon mass fraction is kept at the 2$\times$10$^{-4}$-depletion.

\rch{In our non-carbon-depleted compositional model the refractory material inherits its bulk composition from the ISM dust \citep{gaidos2015}.} The effects of evaporation and subsequent re-formation of solids along a condensation sequence that are important in the inner disk at early times \citep[e.g.][]{1974RvGSP..12...71G} are thus ignored, as well as any gas-solid reactions that may occur \citep[e.g.][]{2001A&A...378..192G,leebergin2010}.
Yet, considering measurements in the Solar System it seems to be the case that in the solar nebula there was a strong gradient in the carbon content of the refractory material: while comets have an approximately solar carbon abundance \citep[e.g.][]{1987A&A...187..859G,2005Icar..179..158M}, asteroids (as probed by meteorites) are deficient in carbon by a factor of $\gtrsim$\ca10 \citep[e.g.][]{1988RSPTA.325..535W}, and the Earth is estimated to be carbon deficient in its bulk composition by a factor of $\approx$\ca10$^4$ \citep[e.g.][]{2001E&PSL.185...49A}, despite the high surface abundance of carbon. \rch{This picture has recently been further refined by \citet{2015PNAS..112.8965B} who found that the carbon-to-silicon ratio in chondrites decreases from carbonaceous chondrites ($\sim$5\%~solar) to ordinary chondrites ($\sim$1\%~solar) to enstatite chondrites ($\sim$0.5\%~solar). This is interesting because the formation location of carbonaceous chondrites, ordinary chondrites and enstatite chondrites is stated to be at $\sim$2.75~AU, $\sim$2.15~AU and $\sim$1.75~AU, respectively \citep{2012AREPS..40..251M}. Therefore the amount of carbon found in chondrites reproduces the depletion of carbon in the inner Solar System, as well as the fact that the strength of this depletion increases with decreasing distance to the star.}

\rch{Outside the Solar System, a qualitatively similar picture arises from the abundance patterns of metal-enriched white dwarf photospheres that are thought to have recently accreted rocky bodies with low carbon content \citep[see, e.g.,][]{2013MNRAS.432.1955F,Wilson01072016}.}

\begin{figure}[t]
\includegraphics[angle=0,width=\columnwidth]{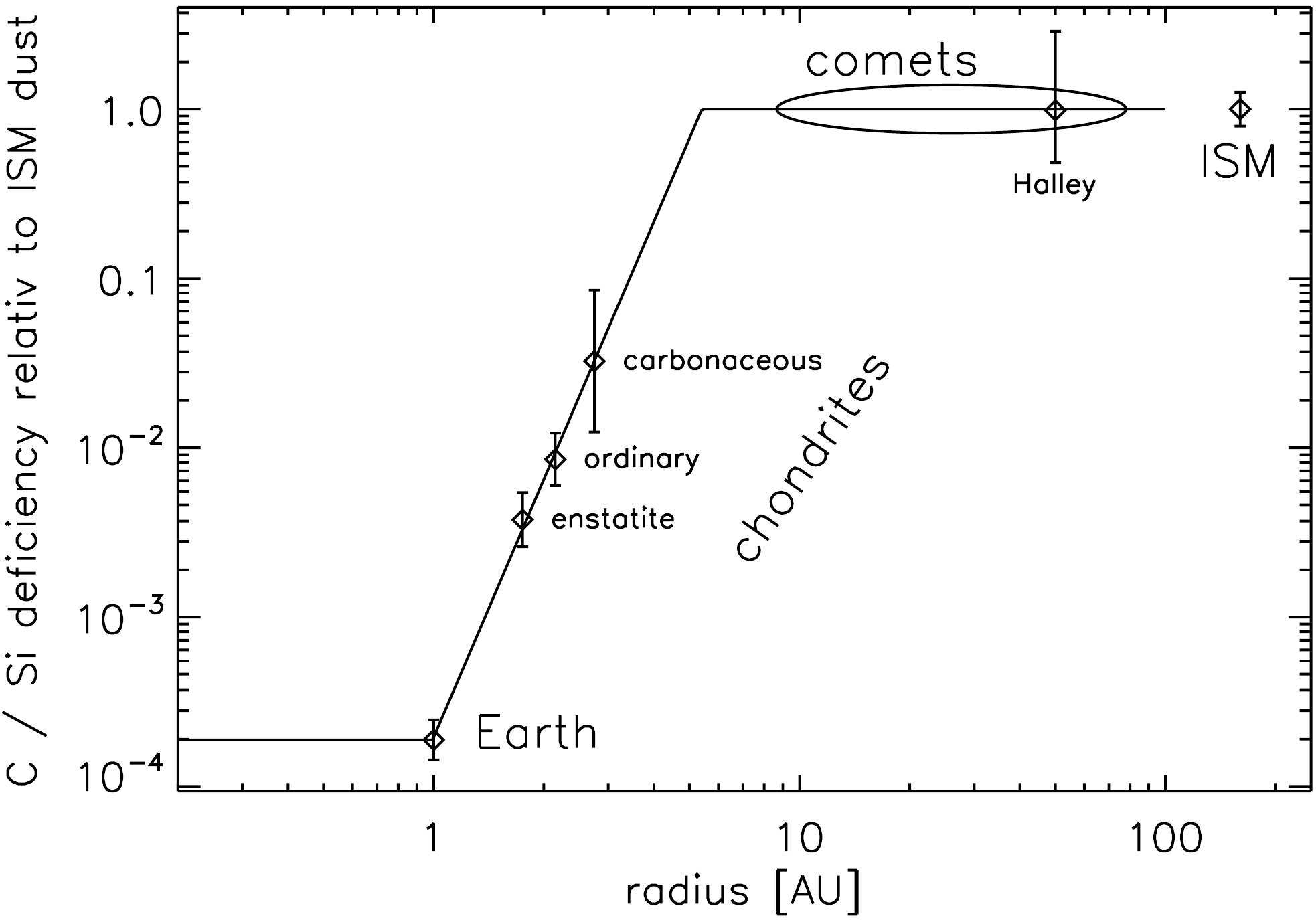}
\caption{\label{fig:Cdef} Parameterized description for carbon deficiency used to explore the effect of carbon-poor refractory material in the inner disk, as observed in the Solar System (see Section\cs\ref{sec:methods:refractories}).}
\end{figure}

Therefore, in order to investigate the effect of carbon deficiency in the solids in the inner disk we deplete the carbon in the 
planetesimals in our nominal model in a way that qualitatively mimics the composition of rocky bodies in the Solar System, leading to the aforementioned depletion by a factor of about 2$\times$10$^{-4}$ inside 1 AU (see Figure\cf\ref{fig:Cdef}).

This does not affect the carbon abundance of planets forming in the outer parts of the disk, but drastically reduces the carbon abundance of a planet forming in the inner parts of the disk. \rch{It is clear that the adopted value of carbon reduction factor is partially ad hoc as it is unclear if the specific value derived for the Solar System can be extrapolated to other systems.} Clearly this represents an area for improvement in future work. \rch{Changing the maximum depletion factor from about 10$^{-4}$ to only 10$^{-1}$ did not affect one of the main results obtained in the following sections, namely that with carbon depletion,  planets forming both in- and outside of the water iceline have a C/O$<$1.  }
For the time being we also investigate cases without carbon depletion in the inner regions of the disk, but these cases are not what we consider to be our nominal model.

In total, considering our 4 refractory abundance models and the carbon depletion switch, we end up with 8 different models
for the refractory composition.

\begin{table*}[t]
\centering
\begin{tabular}{l|c|ccccc|cl}
Source                 & ID & \water & \Cmonoxide & \Cdioxide & \methane & \ammonia & Reference \\
\hline \hline
Pure \water  & v0 &    100 &    0 &    0 &    0 &    0 & --- \\
Solar nebula model & v1 &  100 &    0 &    0 &   65 &   18 & \cite{2003ApJ...591.1220L}  \\
Comets                 & v2 &  100 &   10 &    5 &    1 &    1 & \cite{2004come.book..391B}  \\
Comets                 & v3 &  100 &    6 &   19 &    0 &    0 & \cite{leroyaltwegg2015}     \\
Protoplanetary disk    & v4 &  100 &   99 &   32 &    4 &   10 & \cite{2005ApJ...622..463P}  \\ \hline
\end{tabular}
\caption{\label{tab:volatiles_composition} Explored models of volatile composition.
The number densities of the main volatile species are given relative to water ($\equiv$\cm100).}
\end{table*}

\subsubsection{Volatile material and disk gas composition}
\label{sec:methods:volatilesI}
For the composition of the volatile material we explore 5 different options, which are outlined in Table\ct\ref{tab:volatiles_composition}.
We investigate the case of a pure H$_2$O volatile composition (model \verb v0 ), a case suggested by as a possible condensation model for the
solar nebula (model \verb v1 ), 2 models based on cometary abundance measurements (models \verb|v2| and \verb|v3|) and a
model based on protoplanetary disk measurements (model \verb v4 ).
In contrast to the refractories which only occur in solid form in the planetesimals, volatiles can be present in the planetesimals in the form of ice, as well as in gaseous form in the protoplanetary disk mixed with \HHe.

However, the simplest assumption for the composition of the gas accreted into the planetary envelope is that of a \HHe \ mixture that does not contain other molecules in appreciable amounts. In our calculation we approximate this situation with a pure \HHe \ gas. The physical rationale behind such a scenario is the following: because the disk viscous timescale at the radii where the considered planets form is only of order $10^5$\cu years, the disk moves inward to be accreted onto the central star on a timescale that is short compared to that of planetary core formation. By the time the forming planets start to accrete substantial amounts of gas, the gas in the relevant disk regions has been ``flushed'' a number of times, continuously being replenished by gas from the cold, more distant disk regions. If in these cold regions all refractory and volatile materials have been put into planetesimals, then only hydrogen and helium will be left in the gaseous phase for the planet to accrete. \rch{This behavior can be seen in the model of \citet{thiabaudmarboeuf2015} where water vapor is removed from the entire disk due to viscous evolution in less than $10^5$\cu years.}

On the other hand, some parts of the planet-forming region in a protoplanetary disk may be occupied by a ``dead zone'' where the magneto-rotational instability \citep{balbushawley1991} is not operating in the disk midplane \citep{gammie1996}, so that the gas there might not have been ``flushed'' efficiently.  Therefore, a second possibility that we explore is that the gas contains, in addition to \HHe,  all those volatile species whose sublimation temperature is below the midplane temperature at the time and location where the gas is accreted (see Section\cs\ref{sec:methods:volatilesII}). This represents the other extreme scenario. Strictly speaking, it is not self-consistent with the disk evolution model which does not include a ``dead zone''.

\subsubsection{Envelope elemental abundance post-processing}
\label{sec:methods:volatilesII}
From the planetesimal mass accreted onto the planet it is possible to obtain the planets enrichment in elements other than \HHe. As described at the start of Section \ref{sec:methods:stoichiometry}, in the calculation of the planet formation and accretion history there is no accounting of individual atomic composition of  the refractory or volatile species.
Instead, for each time step the model yields the relative amounts of H ($X$),  He ($Y$), total refractory material ($Z_{\rm{r,p}}$) and total volatile material ($Z_{\rm{v,p}}$) accreted in the form of planetesimals \emph{in the envelope}.
The relative contributions of the various constituents sum up to unity.

In the formation calculation the accreted gas is assumed to consist of a pure \HHe \ mixture everywhere in the disk \rch{\citep[see][for a discussion]{mordasiniklahr2014}}. Furthermore, it is 
implicitly assumed that \emph{all} volatile species are frozen out and incorporated into planetesimals beyond the water iceline at the 
time of planetesimal formation, and no volatiles are contained in the planetesimals that formed closer to the central star than the water 
iceline. In reality, the planetesimals formed beyond the water iceline may not contain all volatile species because the condensation 
temperatures of  CO, CO$_2$, and CH$_4$ are much lower than that of H$_2$O, and hence their respective icelines lie at larger 
distances from the central star.
{Some of the highly volatile material may be captured in the form of ``clathrates'' \rch{\citep[e.g.,][]{1970P&SS...18..717D}}, such that we may have e.g. CO or CH$_4$ \ inclusions in the water ice at locations interior to the nominal icelines of the respective species.}
Furthermore, the gaseous component may contain volatile species, in addition to, \HHe, although this is not necessarily the case.
The mass budget needs to be corrected in order to account for these effects.
Firstly, $Z_{\rm{v,p}}$ is reduced accordingly by removing those volatile species from the planetesimal mass budget that are not frozen 
out during planetesimal formation, that is: the disk midplane temperature during planetesimal formation is above the sublimation 
temperature of the respective species. This implicitly means that the planetesimal surface density used to form the planet outside
of the water iceline should in reality have been somewhat lower and, therefore, introduces as slight inconsistency.
Secondly, the volatiles in gaseous form ($Z_{\rm{v,g}}$) are, optionally, introduced. These consist 
of those species whose sublimation temperature is below the midplane temperature during gas accretion.
They are added to the accreted gas according to the fraction of the total ``dust'' mass they represent, and the bulk disk dust/gas mass ratio $f_{\rm{dg}}$:
\begin{equation}
\label{eq:ZvolGas}
Z_{\rm{v,g}} = (X+Y) f_{\rm{dg}} (1-f_{\rm{r}}) \sum_{\rm{i,gas}}{g_{\rm{i}}}
\end{equation}
Here, ``dust'' is taken to be all material in the disk that is not hydrogen or helium, $f_{\rm{r}}$ represents the mass fraction of refractory material in the dust, and $(1-f_{\rm{r}})$ is the mass fraction of volatiles in the dust. The composition of the volatiles is denoted by the relative mass contributions ${g_{\rm{i}}}$ of the various species to the total mass in volatiles, where $\sum{g_{\rm{i}}}=1$ when summing over both the gaseous and icy volatiles. The sum in Equation\cs\ref{eq:ZvolGas} includes only the gaseous species, and thus denotes the fraction of the volatile material present in gaseous form. The values of ${g_{\rm{i}}}$ for the various volatile species follow from the assumed compositions that are summarized in Table\ct\ref{tab:volatiles_composition}. The refinement of the volatile mass budget necessitates a small correction factor of $\approx$\ca1 in the overall mass budget to ensure that the relative contributions to the total accreted envelope mass still add up to unity:
\begin{equation}
X+Y+Z_{\rm{r,p}}+Z_{\rm{v,p}}+Z_{\rm{v,g}}=1
\end{equation}

In total, together with the 5 volatile composition models described in Section \ref{sec:methods:volatilesI}, this introduces 3 more cases: volatiles in clathrates, volatiles not in the clathrates but also not in the gas phase, volatiles not in the clathrates but in the gas phase, leading to a total of 15 volatile abundance models. Considered together with the 8 refractory abundance models we obtain a total number of 120 different models.

Finally, if $Z_{\rm{r,p}}$, $Z_{\rm{v,p}}$ and $Z_{\rm{v,g}}$ are known, together with the compositional models for the volatiles
and refractories, this yields the atomic elemental composition in the planetary envelope and atmosphere.

Summarizing the above points, the following parameters were varied for the compositional post-processing of the envelopes:
\begin{itemize}
\item \emph{Mass fraction of refractories with respect to all metal $f_{\rm{r}}$}: 
We perform calculations for $f_{\rm{r}}$\cm$=$ 0.25, 0.32, and 0.46, corresponding, respectively, to the values found or adopted by \cite{1981IAUS...93..113H}, \cite{1989GeCoA..53..197A}, and \cite{2003ApJ...591.1220L}. \textit{This leads to 3 $f_{\rm{r}}$ model options.} 
\item \emph{Refractory composition}: We explore 4 possibilities for the relative amounts of carbon and silicates in refractories: a C/silicates mass ratio of 0.2, 0.5, and 1.0 or refractories of 46.2 weight percent ``CHON''-composition.
By default the inner parts of the disk are assumed to be carbon poor to take into account the observed the carbon depletion in the Solar System and WD atmospheres. But the inner disk carbon depletion can also be turned off. \textit{This leads to 8 refractory model options.}

\item \emph{Volatile composition}: We explore 5 different volatile compositional models, which are outlined in Table\ct\ref{tab:volatiles_composition}. We also consider clathrate formation and volatiles in the gas phase leading to 4 more options
for all volatile compositions. This leads, in total, \textit{to 20 volatile model options.} Depending on the formation location
some of these options are degenerate in their outcomes, e.g. for a planet forming inside the iceline the clathrate option is meaningless.
For planets forming outside the iceline with clathration of all volatiles the option of whether or not volatiles can be found in the
gas phase is meaningless as well. For the case without clathrates we use the iceline temperatures given in Table \ref{tab:sublimation_temperatures}.
\end{itemize}

\begin{table}[t]
\centering
\begin{tabular}{c|cc}
Species & $T_{\rm{c}}$ & Reference \\
\hline \hline
H$_2$O   &  \Tice\cu K &   \cite{2003ApJ...591.1220L}\\
CO      &    20\cu K & \cite{obergmurray-clay2011} \\
CO$_2$   &    47\cu K & \cite{obergmurray-clay2011} \\
CH$_4$   &    41\cu K & \cite{2003ApJ...591.1220L} \\
NH$_3$   &   160\cu K & \cite{2009Icar..200..672D} \\
\hline
\end{tabular}
\caption{\label{tab:sublimation_temperatures}Adopted sublimation temperatures $T_{\rm{c}}$ for the volatile species.}
\end{table}

In total there are 3 $\times$ 8 $\times$ 20 = 480 different compositional models, some of which are, as said
before, redundant. For planets which form inside the iceline the clathrate option is meaningless, leading to 240 models. For 120 of these
240 models volatiles are not in the gas phase, rendering the $f_{\rm r}$-parameter meaningless, leaving 40 of the 120 models. Further, the volatile composition cannot be important for these cases, therefore we are left with 8 of the 40 models, \textit{leading to 128 relevant models for planets forming inside the iceline}.
For planets which form outside the iceline and outside the carbon depletion region there are 4 meaningful refractory compositional models left, and the volatiles-in-gas option only makes sense if clathrate formation is turned off, leading to 3 $\times$ 4 $\times$ 15 = \textit{180 different models for planets forming outside the iceline}.

We identify different explored options uniquely using the following naming scheme:

\begin{verbatim}
    dry_r0.25_Csil0.5_Cdef_v0_noclath_gXY
    wet  0.32     0.2       1 clath   gVol 
         0.46     1.0       2
                  CHO       3
                            4  
\end{verbatim}

Here, ``wet'' and ``dry'' indicate the planet that formed completely inside or completely outside of the water iceline, respectively. The fraction of the refractory material to the total ``dust'' mass is indicated by \verb r0.XX , and \verb CsilX.X \ denotes the C/silicates mass ratio in the refractory material. If ``CHON'' material is assumed, we replace \verb CsilX.X \ with \verb CsilCHO . If a carbon deficiency in the inner disk is applied, as described in Section\cs\ref{sec:methods:refractories}, this is indicated with the \verb Cdef \ switch. \verb vX \ denotes the composition of the volatiles as detailed in Table\ct\ref{tab:volatiles_composition}. Whether or not all volatile species are frozen out during planetesimal formation along with water in the form of clathrates is indicated by \verb clath \ (they do) and \verb noclath \ (they do not, and instead each freeze out at their own sublimation temperature, see Table\ct\ref{tab:sublimation_temperatures}), respectively. The composition of the accreted gas is indicated by the last term, where \verb gXY \ indicates a pure \HHe \ gas, and \verb|gVol| indicates a gas that contains also volatile species. Thus, for example, model \verb wet_r0.32_Csil0.5_v2_clath_gXY \ indicates a planet that formed completely outside the water iceline in a disk without carbon depletion where refractory material makes up 32\% of the "dust'' mass, carbon and silicates comprise 2/9 and 4/9 of the mass in refractories, respectively, the volatiles have a composition according to \cite{2004come.book..391B}, the planetesimals contain ices with all volatile species captured in the water ice as clathrates, and the gas that is accreted after core formation contains only hydrogen and helium.
 
\subsection{Importance of planetesimal enrichment}\label{sect:importanceplanetesimals}
\begin{figure*}[ht!]
\includegraphics[angle=0,width=0.465\textwidth]{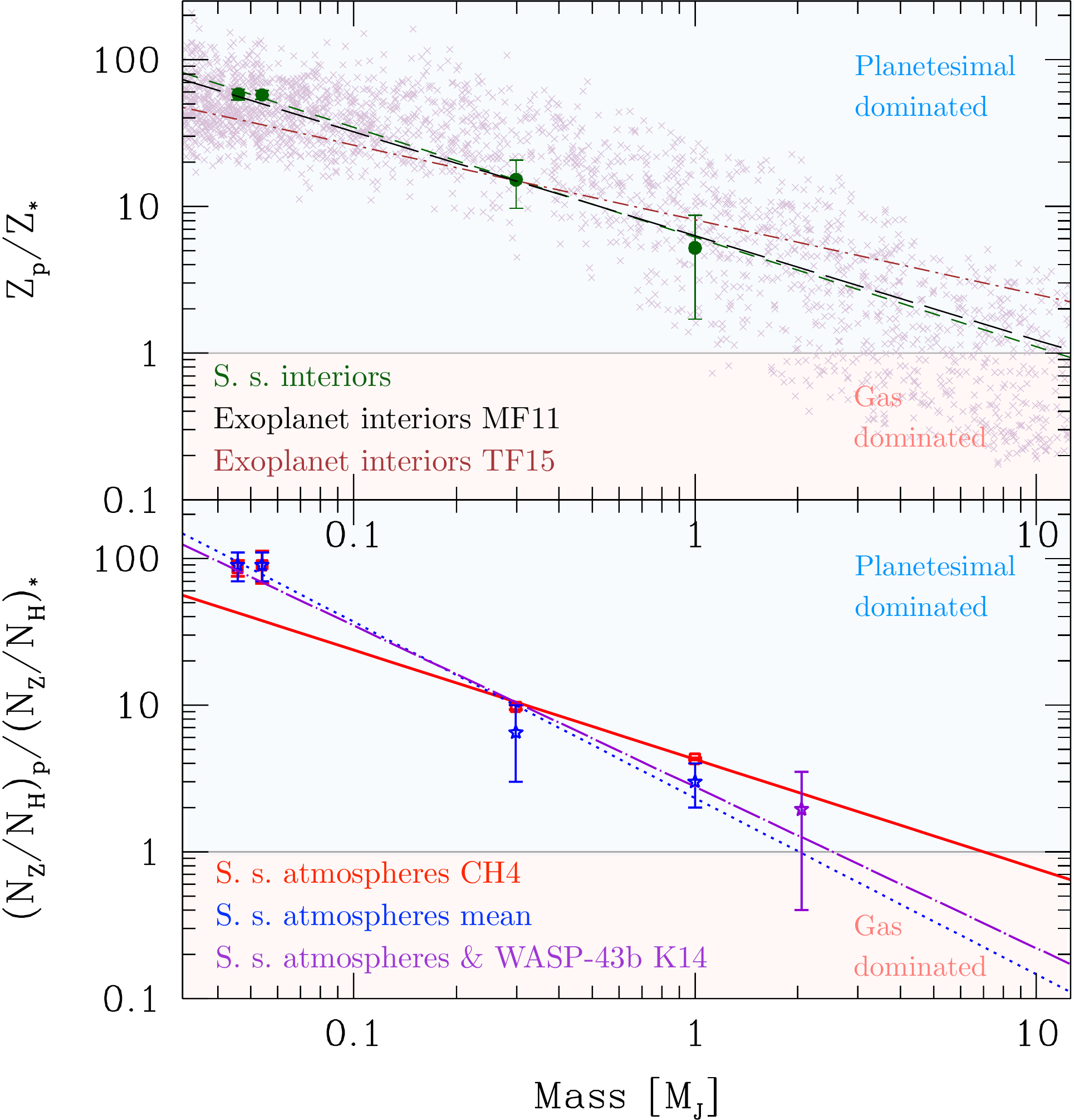}
          \hfill
\includegraphics[angle=0,width=0.487\textwidth]{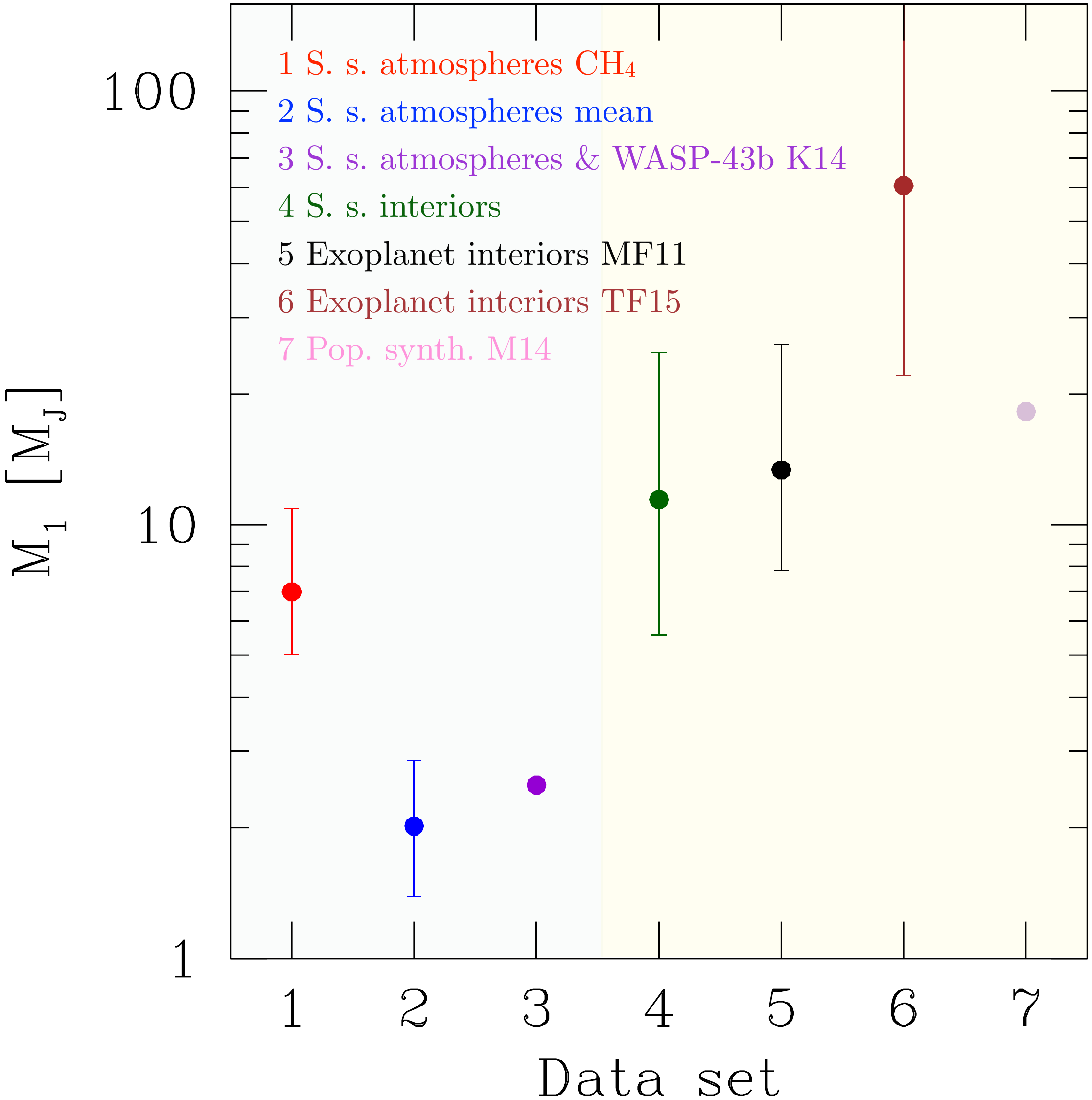}
\caption{\rch{{\it Left panel}:  Heavy element enrichment of planetary interiors (top)  and  atmospheres (bottom) relative to the host star as a function of planetary mass for seven data sets. The giants of the Solar System (Uranus, Neptune, Saturn and Jupiter) are shown. The purple star in the lower part depicts  WASP-43b \citep{kreidbergbean2014}. The values for the bulk (interior) enrichment have been taken from the models of \citet{saumonguillot2004,helledanderson2011} (green short-dashed line and green symbols) for Solar System planets, and \citet{millerfortney2011} (black long-dashed) and \citet{thorngrenfortney2015} (brown short-dashed-dotted) for extrasolar planets \rt{with an equilibrium temperature of less than $\sim$1000 K}. The many small crosses in the background show the bulk enrichment of  synthetic planets in  \citet{mordasiniklahr2014}.  The atmospheric enrichment measurements for  Solar System giant planets are taken from \citet{guillotgautier2014}, where the red solid line and red symbols are based on the CH$_4$ abundance in the atmospheres alone, whereas the blue dotted line and symbols show a mean over the measured atmospheric abundances of all heavy elements. The purple long-dashed-dotted line is from \citet{kreidbergbean2014} and additionally takes into account the atmospheric water abundance in WASP-43b. The lines shows least square fits to the various data sets. The locus where the fitting lines meet the gray horizontal line at a relative enrichment of unity defines the parity mass $M_{1}$. {\it Right panel}: Parity mass $M_{1}$  where the planetary interior or atmospheric enrichment relative to the host equals 1. The parity mass values and error bars were derived from the least square fits of the various datasets shown in the {\it left panel}.}}
\label{fig:hot_jupiters_histo}
\end{figure*}
\rch{
In the context of the two sources for the heavy elements in a planetary atmosphere which are planetesimals (or solids in general) on one hand, and nebular gas on the other hand, it is important to analyze the enrichment levels of the planets in- and outside of the Solar System. This allows to estimate the relative importance of the two sources as we see next. Past studies linking planet formation and atmospheric composition have often focussed on the enrichment by gas because the composition of the gas is more readily obtained from a disk model alone, whereas the contribution by the planetesimals can only be obtained with a proper planet formation model.}

\rch{In the left panel of Fig. \ref{fig:hot_jupiters_histo} we show the enrichment of planets relative to their host star $e_{\rm Z,rel}$ as a function of their mass. A relative enrichment clearly higher than unity can only be obtained by the accretion of solids, such that we can then speak of a planetesimal-dominated composition. Heavy elements accreted with the gas can still contribute, but cannot be dominant for  $e_{\rm Z,rel}$ clearly larger than unity. Planets which are dominated by gas accretion can in contrast have a $e_{\rm Z,rel}$ of unity or less, because the metals which are locked into the planetesimals are no longer present in the gas and therefore the accreted gas will have a sub-stellar metallicity. For example, if we assume that inside of the water iceline about half of all heavy elements are condensed and half of them in gaseous form \citep{lodders2003}, then a planet's atmosphere forming there that is enriched solely by the accreted gas will have a relative enrichment of $e_{\rm Z,rel}$=0.5. An enrichment less than unity does however still not mean that the heavy elements in a planet's atmosphere were necessarily accreted with the gas: in principle a planet could also have accreted (nearly) pure H/He, and a small amount of planetesimals, but only so little that $e_{\rm Z,rel}<$1. A relative enrichment bigger than 1 shows in contrast that solids were indeed important. Therefore, a relative enrichment of unity is a natural dividing line between a composition where solids were necessarily  important (or even dominant at higher values), and a potentially gas-dominated composition. It is not a sharp boundary, as for  $e_{\rm Z,rel}$ around unity, both sources can potentially be important.}

\rch{In the figure, we study both the relative enrichment in the interior $e_{\rm Z,rel,int}$ and atmosphere $e_{\rm Z,rel,atmo}$ for seven different data sets.  They are described in detail in Appendix \ref{appendix:relativenerichmentzpzs} together with the numerical parameters of the fits. The relative interior (bulk) enrichment is shown in the upper part. It is the mass fraction of heavy elements in a planet relative to the mass fraction of heavy elements in its host star, i.e., $e_{\rm Z,rel,int}=Z_{\rm Pl}/Z_*$ where $Z_{\rm Pl}=M_{\rm Z}/M$ with $M_{\rm Z}$ the mass of heavy elements (both in the solid core and dissolved in the envelope) in the planet of a total mass $M$, and $Z_*$ the mass fraction of heavy elements in the host star (0.0142 for the solar primordial composition, \citealt{asplundgrevesse2009}). The relative atmospheric enrichment is shown in the lower part. It is given as $e_{\rm Z,rel,atmo}=(N_{\rm Z}/N_{\rm H})_{\rm p}/(N_{\rm Z}/N_{\rm H})_{\rm *}$, where  $(N_{\rm Z}/N_{\rm H})_{\rm p}$ and $(N_{\rm Z}/N_{\rm H})_{\rm *}$ is the number of a heavy element atoms (e.g., carbon) relative to the number of H atoms in the planet and its host star, respectively. The  bulk  and atmospheric enrichments are shown separately as the former approaches $1/Z_*$ for very high planetary $Z_{\rm Pl}\rightarrow$1, whereas the latter approaches infinity. It would in principle be possible to convert them into one another and show them on one plot. But  this would involve assumptions about the (unknown) elemental stellar and planetary composition, such that we prefer to show them here separately. Both of them can be used to estimate whether planets fall into the planetesimal- or gas-dominated regimes.}

\subsubsection{Bulk enrichment}
We first discuss the relative bulk enrichment.  From internal structure models one can derive the $M_{\rm Z}$ necessary to reproduce the observed mass and radius, and - for the Solar System planets - the gravitational moments. Studies inferring in this way  $e_{\rm Z,rel,int}$ of transiting exoplanets have found that $e_{\rm Z,rel,int}$ decreases with increasing mass (left panel of Fig.  \ref{fig:hot_jupiters_histo}). The planetary mass where $e_{\rm Z,rel,int}=1$ defines the parity mass $M_{1}$ (see Appendix \ref{appendix:relativenerichmentzpzs}). It is shown in the right panel of Fig. \ref{fig:hot_jupiters_histo}. It is extrapolated to be between $\sim$13 and 60 M$_{\rm Jup}$ \citep{millerfortney2011,thorngrenfortney2015}. \rt{The planets analyzed in these studies have equilibrium temperatures of less than $\sim$1000 K (corresponding to an orbital distance of about 0.08 AU for a solar-like star) so that they are not affected by the aforementioned bloating mechanisms.} A similar \rt{decrease of $e_{\rm Z,rel,int}$ with increasing mass} is found for the bulk metal content of Solar System giants \citep{saumonguillot2004,helledanderson2011}, where the  mass where $e_{\rm Z,rel,int}$ reaches 1 is extrapolated to be at about 11 M$_{\rm Jup}$. From theoretical planet population syntheses based on the core accretion theory one finally finds that the parity mass is at about 10 to 18 M$_{\rm Jup}$ \citep{mordasiniklahr2014}. Considering that of the 255 extrasolar giant planets ($M\sin i > 0.1\mj$)  inside of 0.1 AU currently listed on \url{www.exoplanets.org} \citep{HanWang2014} only 4 have a mass exceeding 10 $\mj$ (which is not an observational bias). \rt{W}e thus deduce that \rt{at least based on their masses} regarding the bulk composition of hot Jupiters, it appears that almost all of them \rt{should be} dominated by planetesimal enrichment. \rt{We add the caveat that the bulk heavy element content cannot  be inferred directly for typical hot Jupiters at equilibrium temperatures of $T_{\rm eq}\gtrsim$1500 K because of bloating mechanisms. But the fact that both the planets analyzed by \citet{millerfortney2011,thorngrenfortney2015} ($a=$0.03-1 AU, $T_{\rm eq}\lesssim$1000 K) and the solar system planets ($a\approx$5-30 AU) follow the same trend, makes it appear unlikely -even though in principle not excluded- that the hot Jupiters at $a\sim0.04$ AU do not follow the same enrichment pattern.}

\subsubsection{Atmospheric enrichment}
\rch{An interesting question is whether the planetesimal enrichment is also visible in the planetary atmosphere. If the atmosphere is not sufficiently well mixed the heavier species might slowly settle to the central regions of the planet and are therefore no more visible in the atmosphere. This effect has been looked at in our paper (see last paragraph of Section \ref{sec:methods:atmosphere}) and at least from these simple estimates it is found that it is not important. Another effect could be important, however as discussed in the Sections \ref{sec:methods:interior}, \ref{sec:metallicity_evolution}: the planets accrete most of the planetesimals enriching their envelopes before the phase of runaway gas accretion. If one traces where the mass of the disintegrating planetesimals is deposited in the envelope's deep layers, and if mixing is inhibited due to semi-convection, then this could lead to planets where only a small amount of the planetesimal enrichment reaches the envelope's upper layers and, therefore, its atmosphere. Whether the onset of semi-convection occurs during the formation of the planets is currently not tested in our model and currently not known from other formation models.}

\rch{However, looking at the relative {\it atmospheric} heavy element enrichment in Solar System and extrasolar gas planets may indicate that the mass where $e_{\rm Z,rel,atmo}$=1 is at around $M_{1}\sim$ 2 to 7 M$_{\rm Jup}$ \citep[][Appendix \ref{appendix:relativenerichmentzpzs}]{guillotgautier2014,kreidbergbean2014}, lower than inferred for the interiors. Thus the aforementioned discrepancy between the interior and atmospheric enrichment may indeed exist, meaning that compositional gradients and semiconvection could play a role, but the  $e_{\rm Z,rel,atmo}$=1 (which does not mean necessarily gas-dominated yet, see above) is still only occurring at rather large masses $M_{1}$ \rt{compared to typical hot Jupiter masses}.}

\rch{In summary, given that the enrichment of interiors of planets appear to be planetesimal-dominated up to at least 10 $\mj$ and up to a few Jupiter masses for their atmospheres, we \rt{think} that gas-dominated enrichment of hot Jupiters should \rt{probably} be rare because most hot Jupiters have rather low masses when compared to $M_{1}\approx$2-10$\mj$: the mass distribution of giant planets ($M> 0.1\mj$)  within 0.1 AU from their host star peaks at about 0.9 $\mj$, and 81 \% have masses below 2 $\mj$, and 98 \% have masses below 10 $\mj$ (see, e.g., \url{exoplanets.org}). Given these observations, it seems that hot Jupiters with envelopes and atmospheres with a composition dominated by planetesimal accretion should \rt{likely} be the rule, whereas hot Jupiters with a composition dominated by gas accretion should be an exception. In that sense considering planetesimal-dominated planets is \rt{probably} quite general \rt{for hot Jupiters, unless they form in a completely different way than envisioned here}. The studies focussing on the C/O ratio of the disk gas and its implication for the planetary C/O ratio, like done in \citet{obergmurray-clay2011,ali-dipmousis2014}, are therefore likely only relevant for planets heavier than a typical hot Jupiter. Examples could be massive directly imaged planets like $\beta$ Pictoris b \citep{lagrangebonnefoy2010} or around HR 8799 \citep{maroismacintosh2008}. Future precise measurements of  planetary atmospheric abundances may show whether there is indeed such a transition from planetesimal to gas-dominated compositions.}

\newcommand{\mum}{$\mu$m}

\begin{table*}
\begin{center}
\begin{tabular}{l|ccc|cc}

                   & \multicolumn{3}{|c|}{JWST} & \multicolumn{2}{c}{ARIEL} \\
\hline
                   & NIRISS SOSS I            &  NIRSPEC MRS III    &  MIRI LRS     &   Ch0       & Ch1\\
\hline
\hline
wavelength range    & 0.8-2.8~\mum             &  2.9-5.0~\mum       & 5.5-13.5~\mum   & 1.95-3.9~\mum   & 3.9-7.8~\mum  \\
quantum Efficiency  & 0.8                      & 0.8                 & 0.6             &  0.8            &  0.8   \\
full well capacity  & 60\,000 e-               & 60\,000 e-          & 250\,000 e-     &  40\,000 e-     &   40\,000 e- \\
readout noise       & 23 e-                    & 6 e-                & 14 e-           &  20 e-          &  20 e-    \\
dark current        & 0.02 e- s$^{-1}$         & 0.01 e- s$^{-1}$     & 0.17 e- s$^{-1}$ &  16 e- s$^{-1}$ &  16 e- s$^{-1}$  \\
total system transmission &  0.15              & 0.54                & 0.35            &  0.30           &  0.30    \\
systematics noise floor &  50~ppm              &  75~ppm             &  100~ppm        &  20~ppm         &   20~ppm  \\
\hline
\end{tabular}
\end{center}
\caption{\label{tab:JWST_parameters}Basic instrument parameters assumed for the simulated JWST observations for the three adopted instruments and configurations. We assume the JWST to have a collecting area of 24\,m$^2$ and the ``warm'' mirrors to be at 35\,K. For ARIEL we assume a collecting area of 0.81\,m$^2$ and the ``warm'' mirrors to be at 70\,K. We adopt a ``noise floor'' due to uncorrected systematic effects as indicated in the last row, in units of parts per million.}
\end{table*}

\subsection{Atmospheric model for spectral calculations}\label{sec:methods:atmo_models}
As described in Section \ref{sec:methods:atmosphere} the atmospheric model used during the evolution of the planets is the analytic double-gray model by \citet{guillot2010}, which takes the planet's enrichment into account in an approximative fashion, by using scaled solar abundance Rosseland mean opacities. While this atmospheric description might be sufficient for the planetary evolution, the emission and transmission spectra of the planets are very sensitive to the actual atomic compositions derived from the formation and post-processing modules. 

Furthermore the emission and transmission spectra of the planets need to be calculated, for which a wavelength dependent treatment of the planetary radiation field is necessary. Therefore, in the fourth chain link, we couple the planets' quantities such as mass, radius, insolation and atomic abundances to a self-consistent fully non-gray 1-D atmospheric model. We use this model to calculate the planet's emission and transmission spectra. A short description of the code is given next.

\subsubsection{\emph{PETIT} code}
The \emph{PETIT} code is a 1-D plane-parallel atmospheric code, which solves the radiative-convective equilibrium
structure of the atmosphere under the assumptions of LTE and equilibrium chemistry. The code models the wavelength
dependent radiative transfer making use of the correlated-k assumption. It considers molecular opacities for CH$_4$, C$_2$H$_2$,
CO, CO$_2$, H$_2$S, H$_2$, HCN, H$_2$O, K, Na, NH$_3$, OH, PH$_3$, TiO and VO, as well as H$_2$--H$_2$ and H$_2$--He
collision induced absorption (CIA).
The results calculated by the code are the atmosphere's self-consistent pressure-temperature structure, the atomic and molecular
abundances throughout the atmosphere and the planet's emission and transmission spectra.
A detailed description of the code can be found in \cite{molliereboekel2015}.
The code was only recently extended to also calculate transmission spectra. To this end we directly calculate the transmission through planetary annuli as seen by the observer during a transit. We then combine the annuli's individual transmissions to obtain an effective planetary radius. For the transmission calculation we include Rayleigh scattering of H$_2$ molecules and He atoms, using cross-sections from \citet{dalgarnowilliams1962} and \citet{chandalgarno1965}, respectively. In order to verify our implementation of the transmission spectra calculations we carried out a comparison to the 1-d transmission spectra shown in Figures 2 and 3 in \citet{fortneyshabram2010}. We found a very good agreement.

\subsection{Simulated observations}\label{sec:methods:observations}
We simulate secondary eclipse observations using the performance model of \citet{vanboekelbenneke2012}. It employs a library of stellar model atmospheres alongside the planetary model spectra calculated with PETIT to generate realistic astrophysical signals. These are then propagated through parameterized descriptions of telescope, spectrograph, and detector properties in order to estimate the achieved SNR on an eclipse observation of a given system with a given facility. In this paper we perform calculations for observations with the James Webb Space Telescope (JWST) as well as the proposed dedicated eclipse spectroscopy mission ARIEL \citep[formerly Thesis, then EChO, e.g.][]{2010SPIE.7731E..25S,2012SPIE.8442E..1HK,2012ExA....34..311T,2012SPIE.8442E..1GS,2013JAI.....250004G}. The assumed telescope and instrument parameters are summarized in Table~\ref{tab:JWST_parameters}.

\begin{figure*}[t]
\begin{minipage}{0.48\textwidth}
	      \centering
         \includegraphics[width=1.02\textwidth]{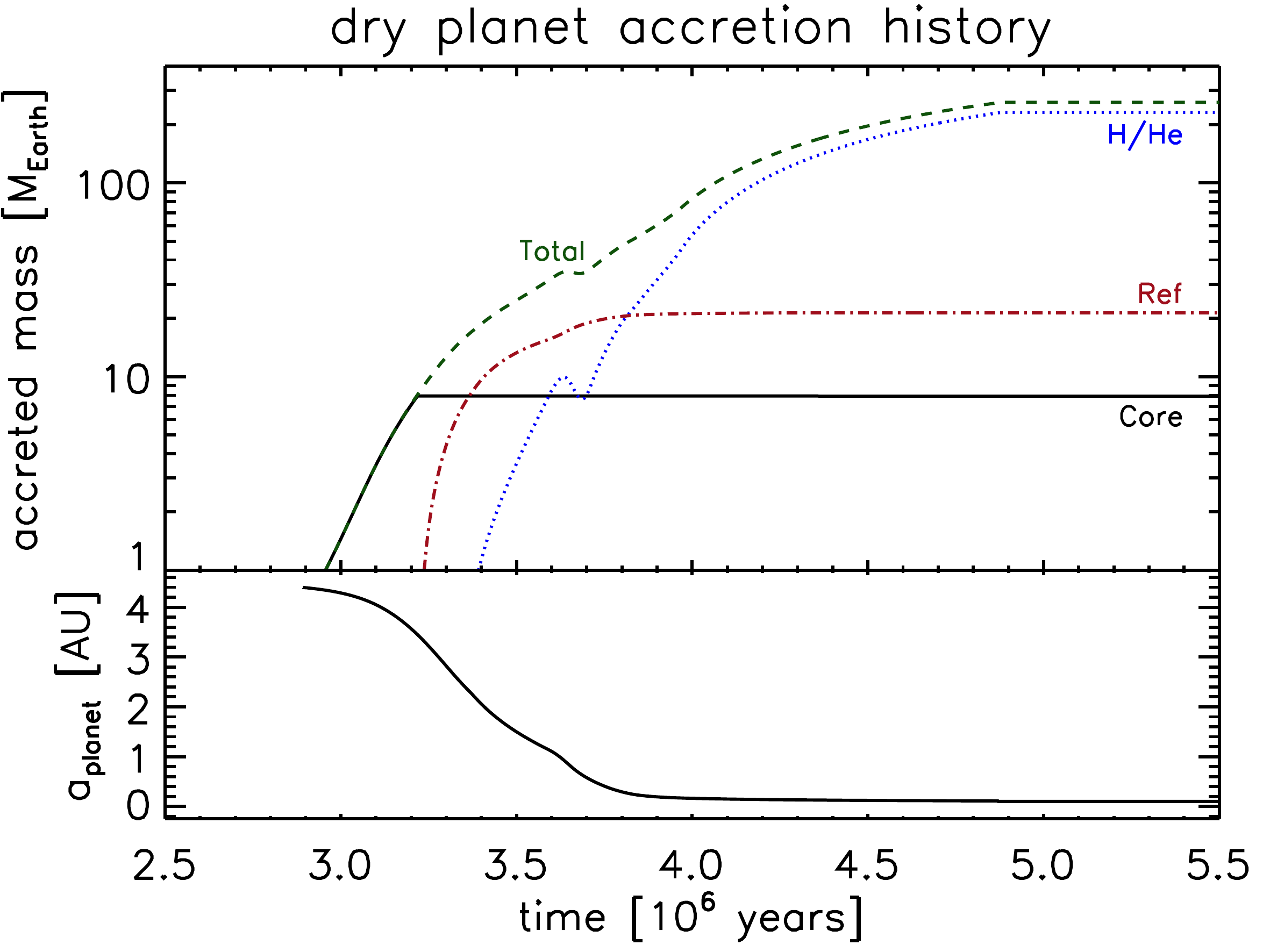}
     \end{minipage}
          \hfill
     \begin{minipage}{0.48\textwidth}
      \centering
         \includegraphics[width=1.02\textwidth]{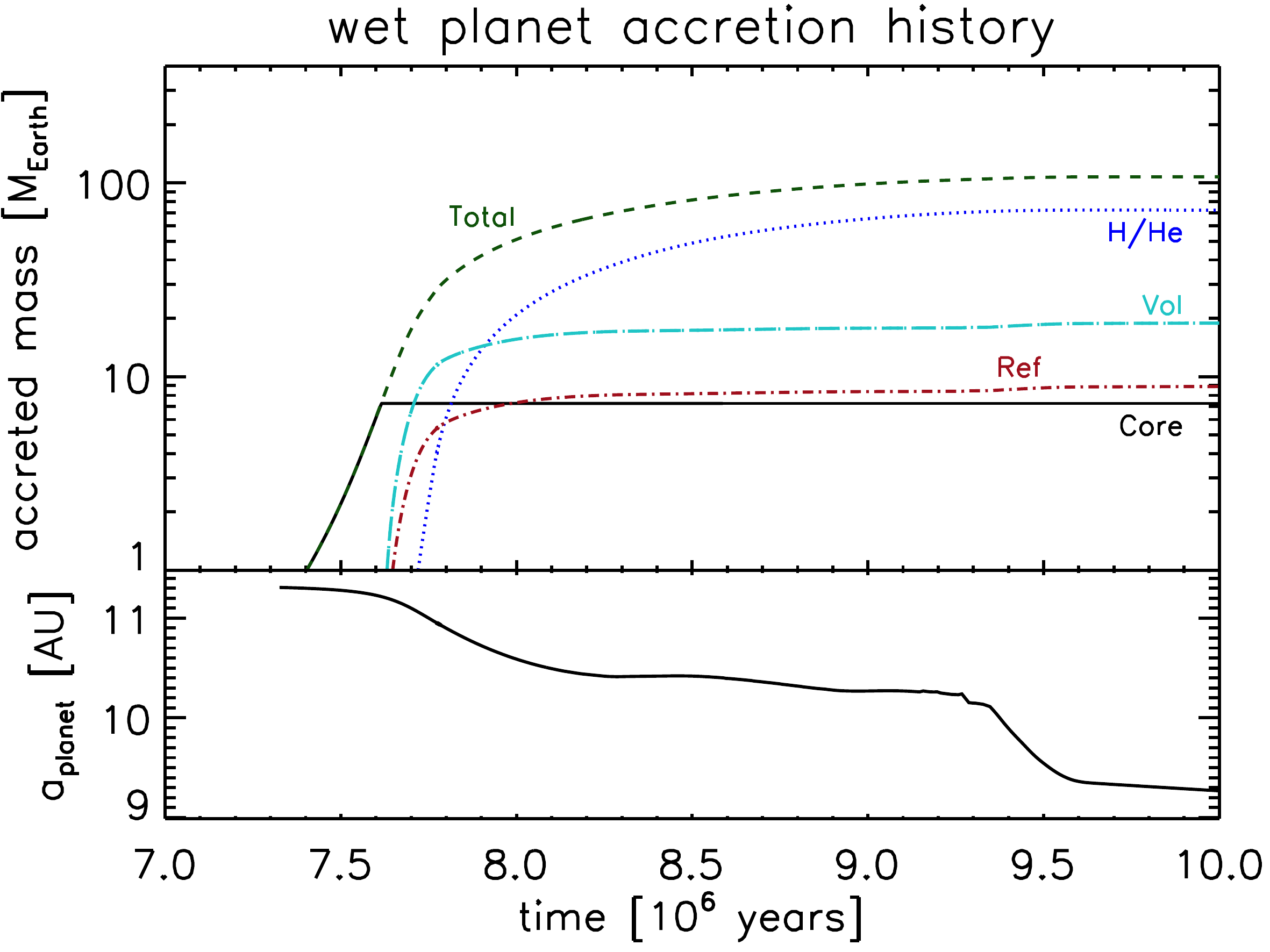}
     \end{minipage}
\caption{Formation phase: the upper panels show the mass accretion as a function of time for the ``dry Jupiter'' ({\it left}) and ``wet Saturn'' ({\it right}). The mass of the central solid core, the \HHe \ in the envelope, and the mass of refractories and volatiles delivered by planetesimals and mixed mixed into the \HHe \ is shown. These simulations assume an $f_{r}$=0.32 and that the accreted gas is pure \HHe.  The lower panels show the semi-major axis of the planet. \ColorJournal}
\label{fig:mass_vs_Time} 
\end{figure*}

\section{Calculations \& Results}
\label{sec:results}
\begin{table}[t]
\centering
\begin{tabular}{c|cc}
&     ``dry Jupiter''       &     ``wet Saturn'' \\
\hline \hline
Initial disk mass [\Msun] &	0.092	&	0.077   \\
Initial disk mass [MMSN]	& $\approx$\ca7$\times$ & $\approx$\ca6$\times$  \\
Disk (and star) [Fe/H]			&    -0.05	&	-0.40  \\
Location of iceline [AU]	&	6.9		&	6.2     \\
Initial planet location [AU] &	4.4		&	11.3     \\
\hline
\end{tabular}
\caption{\label{tab:initial_conditions}Disk initial conditions.}
\end{table}

In this study we concentrate on results obtained for the formation and evolution of two prototypical planets, a ``dry Jupiter'' and a ``wet Saturn''. These simulations were taken from a population synthesis calculation of  \citet{mordasinialibert2012c} and thus have initial conditions expected from the observed distributions of disk properties in terms of mass, metallicity, and lifetime. The most important initial conditions for the two cases are  given in Table\ct\ref{tab:initial_conditions}.
For comparison, the mass of the MMSN of \citet{hayashi1981} is about 0.013\cm\Msun. Both disks are therefore rather massive in terms of gas mass. But given the low [Fe/H] in the ``wet Saturn'' case, the surface density of planetesimals is only moderately larger than in the MMSN case. The disk in the ``dry Jupiter'' case has an approximately solar metallicity, therefore its surface density of planetesimals is significantly higher than in the MMSN.

\begin{figure*}[t]
\begin{minipage}{0.48\textwidth}
	      \centering
       \includegraphics[width=0.95\textwidth]{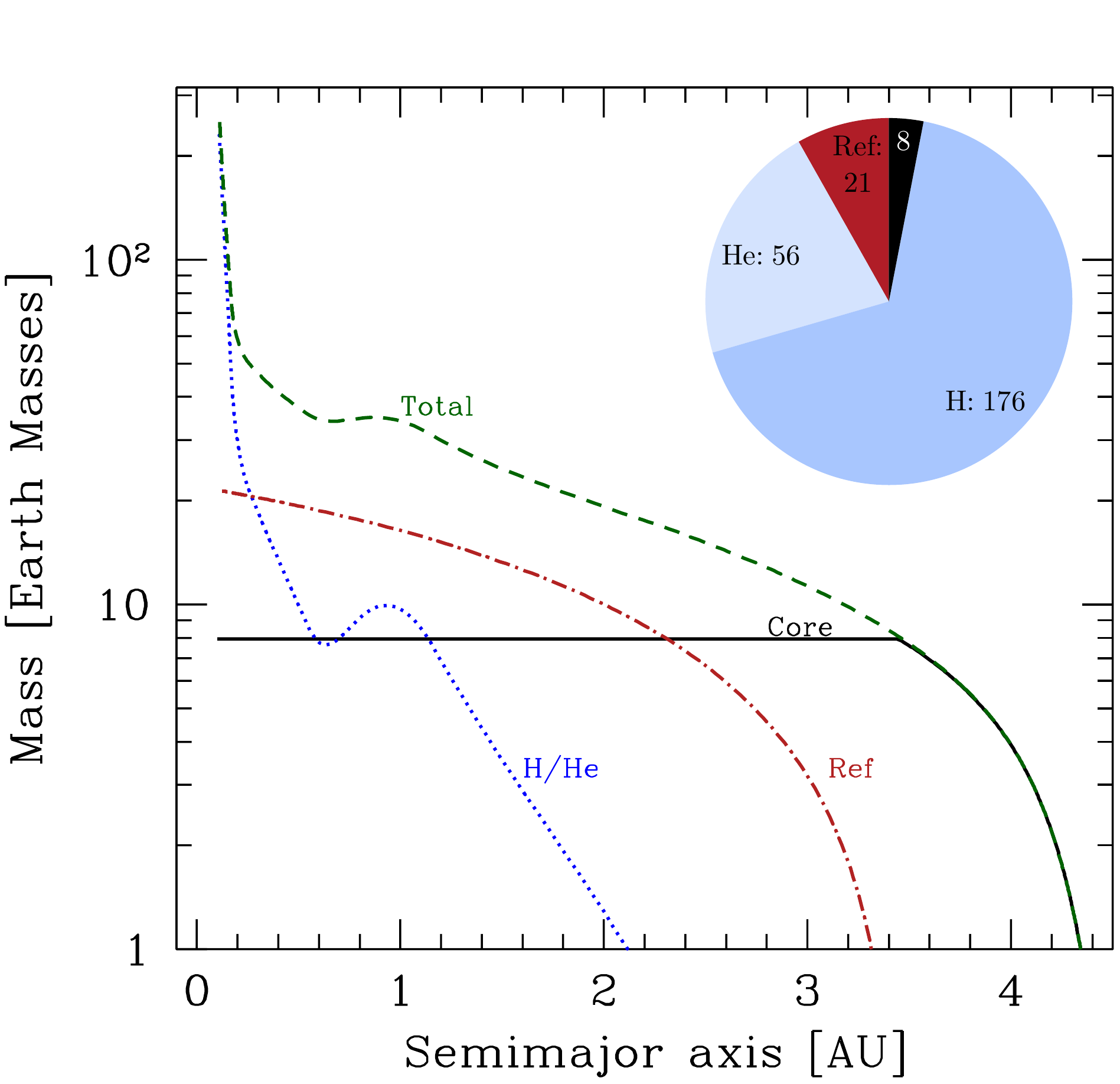}
     \end{minipage}
          \hfill
     \begin{minipage}{0.48\textwidth}
      \centering
                    \includegraphics[width=1.04\textwidth]{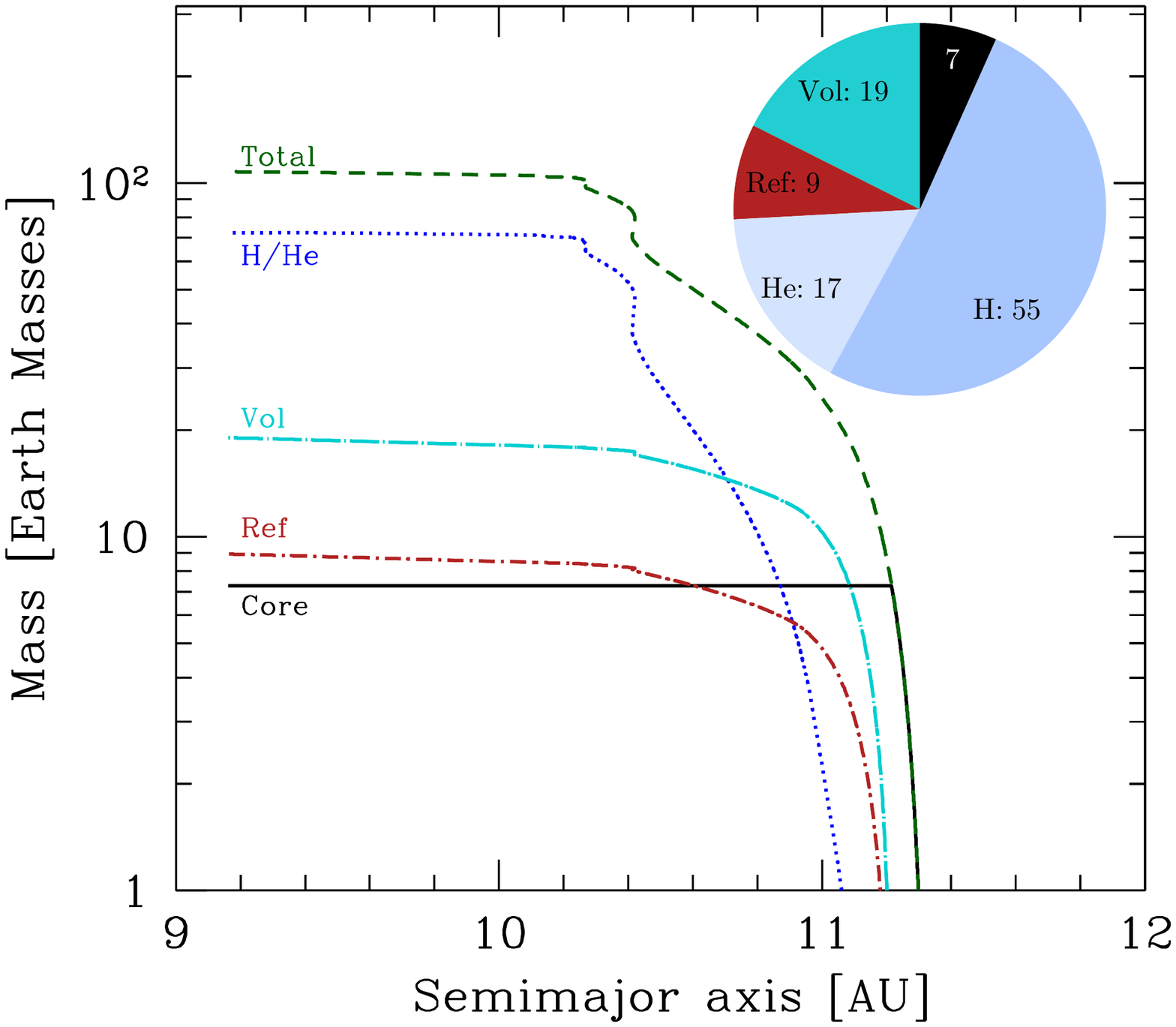}
     \end{minipage}
\caption{Formation phase: mass as a function of semimajor axis for the ``dry Jupiter'' ({\it left}) and ``wet Saturn'' ({\it right}). The mass of the central solid core, the \HHe \ in the envelope, and the mass of refractories and volatiles delivered by planetesimals and mixed mixed into the \HHe \ is shown. The pie charts show the final bulk composition in \Mearth \ at the end of formation. These simulations assume an $f_{r}$=0.32 and that the accreted gas is pure \HHe. \ColorJournal}
\label{fig:mass_vs_aPlanet} 
\end{figure*}
This allows a giant planet to form completely inside of the water iceline (see \citealt{idalin2004a} and \citealt{mordasinialibert2012a}) because the refractories alone are sufficient to allow a formation of a supercritical core that can trigger gas runaway accretion during the lifetime of the disk. For disks with a low surface density of planetesimals the increase of the surface density due to water condensation is, in contrast, necessary for giant plant formation. For the interpretation of our results, it is important that the planetesimal surface density in the disk and the stellar [Fe/H] are correlated (see Sect. \ref{sec:methods:disk_model}), a view that is supported by the well-know positive correlation of stellar [Fe/H] and the frequency of giant planets \citep[e.g.][]{santosisraelian2001,fischervalenti2005}.

The location of the water iceline at 6.9 and 6.2\cu\AU \ may appear relatively large. However, it is a direct consequence of viscous heating in such massive disks. For comparison, for a MMSN disk, our model predicts an iceline location of about 3.4\cu\AU, which is not very different from the classical 2.7\cu\AU \ in a disk that is only heated by the central star \citep{hayashi1981}. Since we assume that planetesimals of 100\cu km in size form instantaneously, it is consistent to assume that the location of the iceline does not further evolve in time.

\subsection{Formation phase}\label{sec:results:formation_phase}
Figures\cs\ref{fig:mass_vs_Time} and\cs\ref{fig:mass_vs_aPlanet} show the accretion and migration history and the resulting bulk compositions of both the ``dry'' planet, which formed inside the water iceline, and the ``wet'' planet, which formed outside of it.
 The calculations start at the moment in time that is necessary to build up the initial seed of 0.6 $\mearth$ at its starting position, given the properties (mass and metallicity) of the protoplanetary disk \citep[see][]{mordasinialibert2009a}. This leads to a significant time delay between the start of disk evolution (i.e., $t$=0), and the start of the calculations, especially for the   ``wet'' planet because of its large initial semimajor axis.

\subsubsection{Accretion history}\label{sec:results:accretion_history} 
The ``dry'' planet starts assembling its core around 4.4\cu\AU \ and migrates inward. At $\approx$\ca3.4\cu\AU \ it has acquired a gaseous envelope of $\approx$\ca0.08\cu\Mearth, which is sufficiently thick for infalling \rch{100 km} planetesimals to be evaporated before they reach the core. Hence, the core stops growing, and any planetesimals accreted from here on will enrich the envelope with heavy elements. \rch{For smaller planetesimals, the core would stop growing already at a lower mass, resulting in a correspondingly higher final enrichment of the envelope. For example, the core would stop growing already at $\sim$$2\mearth$  for  1 km planetesimals  as discussed in Section \ref{sec:methods:planetesimalimpacts}. This would lead to a final envelope enrichment that is about 29\% higher. We thus see that 100 times smaller planetesimals indeed lead to a higher enrichment, but the difference is not very large. Even if all solids would be mixed into the envelope as it could occur for very small bodies, an enrichment that is higher by 38\% would result, again not a very large change. The reason for this moderate increase is that even for 100 km sized planetesimals, already most of them are destroyed in the envelope. The reason for this is that envelopes of protoplanets embedded in the protoplanetary nebula with core masses of  a few $\mearth$ are very massive compared to secondary atmospheres of the terrestrial planets in the Solar System. But even Venus' atmosphere that is with a mass of $\sim$$10^{-4} \mearth$  very tenuous in comparison, is able to shield Venus' surface from $\sim$$1$ km asteroids. This is shown by the  crater size distribution that falls off steeply for craters $\lesssim$20 km. Smaller craters do not form as smaller impactors are destroyed in the atmosphere as indicated by radar dark ``shadows'' \citep{zahnle1992}.} 

The planet continues to migrate inward, accreting both gas and planetesimals. The envelope mass is initially dominated by the accreted and dissolved planetesimals. Only in the innermost disk, at $<$\cm0.3\cu\AU, runaway gas accretion occurs and the envelope mass becomes dominated by the accreted gas\footnote{At about 3.6\cu Myrs, at $\approx$\ca1\cu\AU, there is a phase where the envelope mass decreases. This is a consequence of the following: At this time, the libration timescale of gas on horseshoe orbits becomes longer than the viscous timescale across the horseshoe region. Therefore, the positive corotation torque (which slows down the planet's inward migration) saturates, and the planet migrates more rapidly inwards (only negative Lindblad torques are left; see, e.g., \citealt{paardekooperbaruteau2010,dittkristmordasini2014}). It therefore accretes more planetesimals, because it migrates faster into regions with a high planetesimal surface density. This means that the luminosity in the envelope increases because of more impacts, and so does the pressure support in the gas. The envelope expands, pressing some previously bound gas out of the Hill sphere, so that the envelope mass decreases temporary.}. In total, the planet envelope consists of 232\cu\Mearth \ of material accreted in gaseous form, and 21\cu\Mearth \ of material originating from dissolved planetesimals. Eventually, the planet migrates to the inner border of the computational disk. We assume that the planet then stops at a orbital distance of 0.04 AU due to the stellar magnetospheric cavity \citep[e.g.][]{linbodenheimer1996,benitez-llambaymasset2011}.

The core of the ``wet'' planet starts forming around 11.6\cu\AU, and also starts a slow inward migration. Due to its large starting distance, this planet forms only in the final phases of the lifetime of the disk, when the disk's gas mass has already much decreased. At $\approx$\ca11.5\cu\AU \ core formation is complete and further accreted planetesimals evaporate and enrich the envelope. A short runaway accretion phase ensues when the planet is at $\approx$\ca10.4\cu\AU.  At the end of the simulation the planet's envelope has accreted $\approx$\ca72\cu\Mearth \ of gaseous material and $\approx$\ca28\cu\Mearth \ in planetesimals. We assume that also the ``wet'' planet ends up close to the central star as a hot Jupiter on a time scale that is short compared to the 5\cu Gyr assumed age of the mature planet, via a mechanism that is not explicitly modeled. It is assumed that this happens due to few-body interactions (Kozai migration, planet-planet scattering) and tidal circularization, without the accretion of further material. This is a missing chain link in our model. 

\subsubsection{Envelope metallicity evolution}\label{sec:metallicity_evolution} 

\begin{figure}[t]
\includegraphics[angle=0,width=\columnwidth]{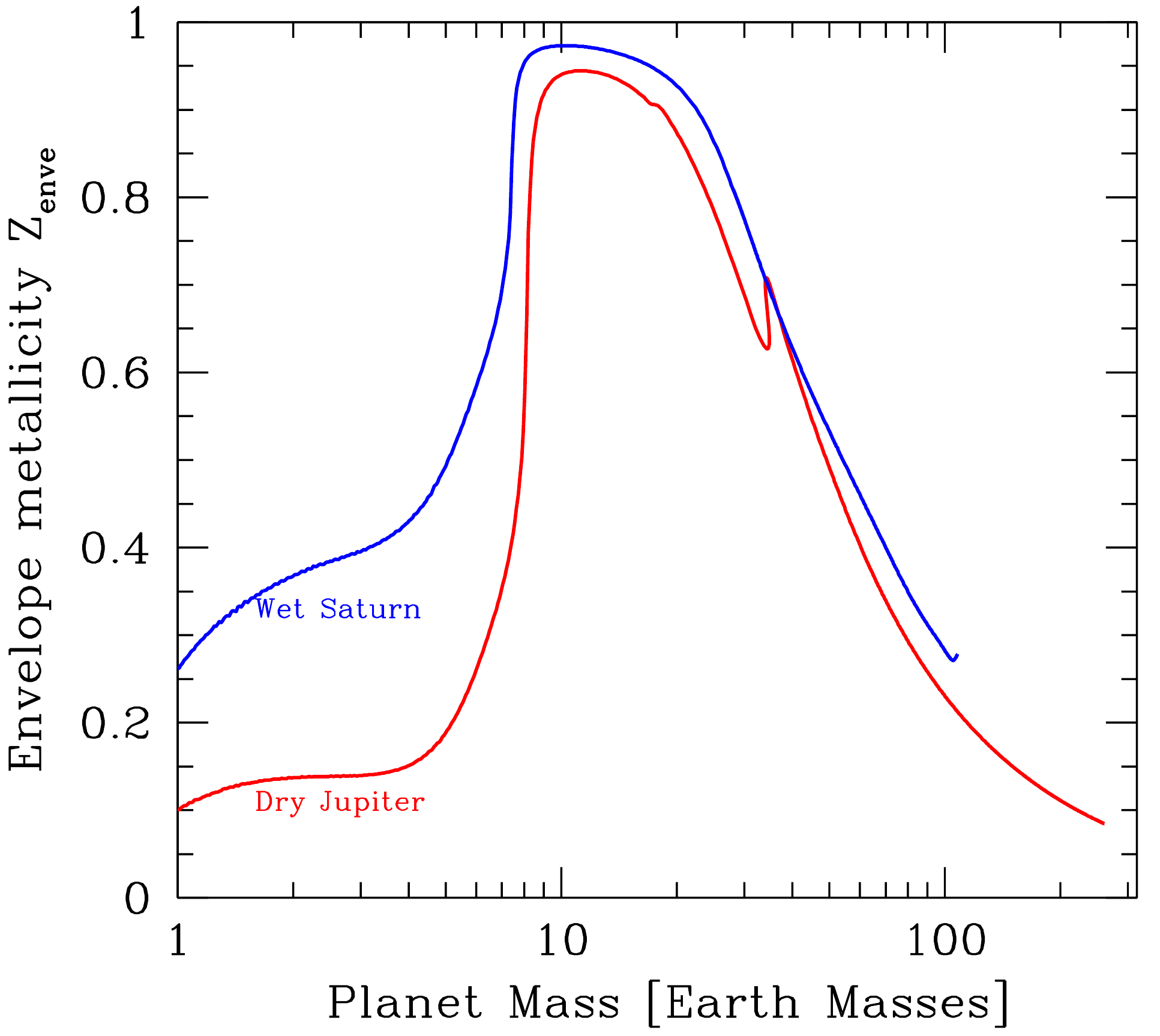}
\caption{\label{fig:zenvejupsat}Formation phase: envelope metallicity (metal mass fraction) as a function of total planet mass for the ``dry Jupiter'' and ``wet Saturn''.}
\end{figure}

From Figures~\ref{fig:mass_vs_Time} and \ref{fig:mass_vs_aPlanet} it becomes  clear that the accretion history of the solids and icy volatiles is quite different from that of the gas: much of the heavy elements accreted in the form of planetesimals are deposited in the envelope at \emph{earlier} times than most of the gas. This is reflected in the evolution of the envelope metallicity (mass fraction of dissolved metals) during the formation process, that is shown in Figure\cf\ref{fig:zenvejupsat}. At early times, when the planet has less than $\approx$\ca5\cu\Mearth, our assumed 100\cu km planetesimals loose only  little mass during their flight through the tenuous envelope and essentially all their mass ends up in the core, keeping $Z_{\rm enve}$ initially low. Once the planet reaches $\approx$\ca5\cu\Mearth \ the envelope becomes thick enough for the planetesimals to disintegrate and evaporate. This leads to very strong enrichment of the envelopes, with metallicities peaking above 90\%. As runaway gas accretion ensues, the envelope is diluted again and at the end of formation the ``dry'' and ``wet'' planets' envelopes contain, by mass $\approx$\ca8\% and $\approx$\ca28\% heavy elements, respectively.  The accretion of planetesimals which are smaller than 100\cu km or tiny bodies like pebbles would increase the metallicity \citep[][Sect. \ref{sec:methods:planetesimalimpacts}]{fortneymordasini2013}. Besides its lower H/He content (due to its late formation), the metallicity for the ``wet Saturn'' is further increased because icy planetesimals are more susceptible to mass loss during their flight through the protoplanetary envelope \rch{(see \citealt{mordasini2014} for a discussion of the effect of the planetesimal composition on the atmospheric destruction)}.

\subsubsection{(Proto-)planet radius}\label{sec:results:proto_planet_radius}

\begin{figure}[t]
\includegraphics[angle=0,width=\columnwidth]{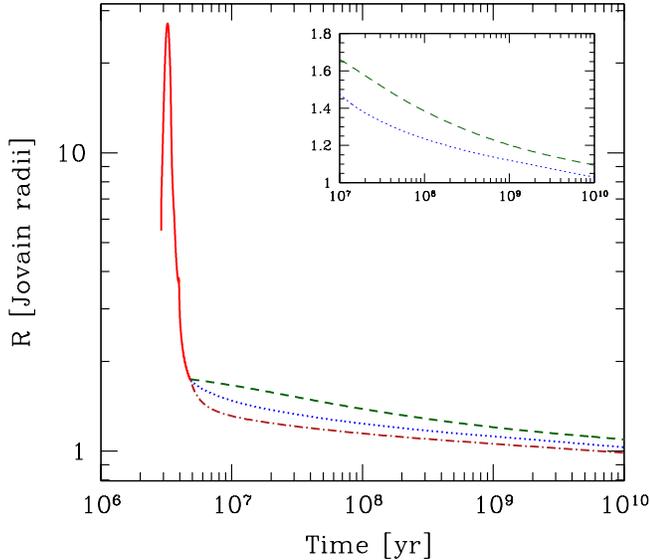}
\caption{\label{fig:tRjupi} Formation and evolution phase: radius of the ``dry Jupiter'' as a function of time. The red solid line shows the radius during the formation phase when the planet is growing in mass. During the subsequent evolutionary phase, the radius ($\tau$=2/3) is shown for three different models: at 0.04 AU, double-gray atmosphere, solar composition (blue dotted); at 0.04 AU, double-gray atmosphere, enriched (green dashed), and, for comparison, at 0.1 AU, gray atmosphere, solar (brown dashed-dotted). The inset figure is a zoom on the evolution phase. At 5 Gyr, the radius is 1.12 and 1.06 $R_{\jupiter}$ for an enriched and solar composition opacity, respectively.}
\end{figure}

In Figure\cf\ref{fig:tRjupi} we show the outer radius of the ``dry Jupiter'' during the formation stage as well as during the subsequent evolution.   
During the early formation the proto-planet is in the attached phase, so that its radius is approximately the smaller of the accretion radius and one third of the Hill sphere radius \citep{lissauerhubickyj2009}.  
As the planet grows, its Hill sphere initially expands, but after about 3.3 Myrs, the Hill sphere shrinks again due to inward migration. At about 4 Myrs and a radius of 3.7 \cu\Rjup, the planet detaches from the disk and contracts further. The gas accretion rate is now limited by the supply from the protoplanetary disk rather than the planet's Kelvin-Helmholtz contraction. The accretion of gas is now assumed to cause a supercritical shock on the planet surface such that the kinetic energy of the infalling gas is instantly radiated away. Hence, due to the accretion of low-entropy material, the radius of the  protoplanet decreases  during the disk-limited accretion phase to about 1.7 \cu\Rjup \ at the moment when the planet reaches the inner border of the disk.

\subsection{Evolution phase}\label{sec:results:evolution}

\subsubsection{Cooling and contraction of the ``dry Jupiter''}
At the end of the formation phase, the ``dry Jupiter'' has accreted a solid core of about 8\cu\Mearth, about 231\cu\Mearth \ of H/He, and 21\cu\Mearth \ of heavy elements mixed into the envelope. As mentioned in Section\cs\ref{sec:methods:interior}, for the evolution at an orbital distance of 0.04\cu\AU \ from a solar-like star, we assume that all solids are concentrated in the computational core. This inconsistency should lead to an overestimation of the planetary radius of $\sim5$ \% at late times \citep{baraffechabrier2008}. However, for the nominal evolution we take the enrichment for the opacity into account in an approximative way by using opacities for a scaled solar-composition gas with [M/H]=0.92 taken from \citet{freedmanlustig-yaeger2014}. As shown in Figure \cf\ref{fig:tRjupi} the $\tau=2/3$ radius at 5 Gyrs is 1.12 and 1.05\cu\Rjup \ for the  [M/H]=0.92  and solar composition opacity case, respectively.

The entropy in the deep convective zone at the end of formation is 9.73\cu k$_{\rm B}$ per baryon, which is an important diagnostic for the formation history of a giant planet \citep[cf.][]{marleyfortney2007}. This entropy is lower than  in \citet{mordasini2013} for the corresponding core and total mass. The lower entropy is likely due to the lower gas accretion rate in the disk-limited phase: in \citet{mordasini2013}, a rate of $10^{-2}$ \Mearth/yr was assumed, while here the typical gas accretion rate is only of order $2\times10^{-4}$ \Mearth/yr, which is known to lead to lower post-formation entropies \citep{spiegelburrows2012}. In any case, for this comparatively low-mass planet, by the time the planet's spectrum is typically observed (here we assume 5\cu Gyrs), the initial entropy only has an indirect impact via its influence on the planet's radius at early times, which in turn enters into the calculation of the atmospheric escape rate, and therefore the envelope mass at late times. However, for both giant planets considered here,   atmospheric escape is also not of high importance, as we will see below.

Figure\cf\ref{fig:pTdryjupi} shows the long-term evolution of the ``dry Jupiter's'' interior in the $p$\cm-\cm$T$ plane. The upper ends of the lines correspond to the planet's outer atmosphere, while the lower ends are  at the core-envelope boundary. At early times after formation, the radiative-convective boundary is at about  3 bar. Subsequently, a deep radiative zone forms as expected \citep{guillotshowman2002}, and the radiative zone extends down to several $10^{2}$ bars. One can also see the decrease of the central temperature. The temperature at the outer boundary in contrast remains virtually fixed as the change of the stellar luminosity in time is neglected, but strongly dominates the temperature structure in the upper parts of the planetary atmosphere, as it is typical for hot Jupiters. In the figure, the evolution for both the nominal increased opacity, and the solar-composition opacity is shown. The  first post-formation structure is identical in the two cases for the deep interior, as both start with the same entropy. At a higher opacity, the subsequent cooling is as expected delayed \citep[e.g.][]{burrowshubeny2007}. One also sees that the temperature in the deep isothermal zone (between about 1 to 100 bars) is identical in both models. This is expected for the simple semi-gray atmosphere, where the ratio of the opacities in the infrared and visual is assumed to depend only on the planet's temperature, but not its composition. In the detailed atmospheric models discussed in sections \ref{sec:methods:atmo_models} and \ref{sec:results:atmospheric_structure_chemistry} this is different. In future work, we will calculate the planet's evolution with an atmospheric model that directly uses the composition given from formation, in an approach as demonstrated by \citet{fortneyikoma2011} for the giant planets of the Solar System. 

\begin{figure}[]
\includegraphics[angle=0,width=\columnwidth]{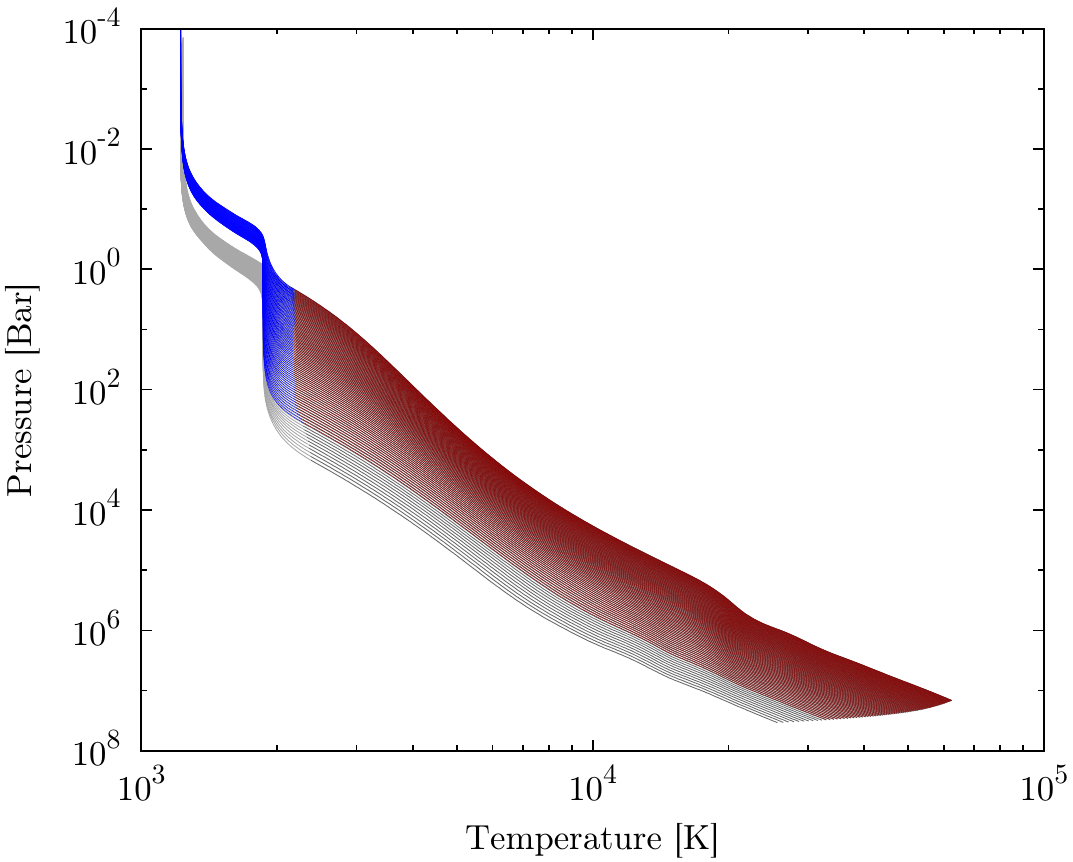}
\caption{\label{fig:pTdryjupi} Evolution phase: temporal sequence of  $p$\cm-\cm$T$ structures in the interior and atmosphere of the ``dry Jupiter'' ($a$\cm$=$\cm0.04\cu AU). The uppermost line corresponds to a moment in time shortly after the beginning of the evolutionary phase, while the bottom profile is at an age of 5\cu Gyrs. The structures are calculated with a opacity corresponding to [M/H]\cm$=$\cm0.92 with radiative and convective parts shown in blue and brown, respectively. The evolution for a solar opacity is also shown for comparison (light/dark gray lines).  The initial structure is identical at large pressures. \ColorJournal}
\end{figure}

\subsubsection{Cooling and contraction of the ``wet Saturn''}
\begin{figure}[t]
\includegraphics[angle=0,width=\columnwidth]{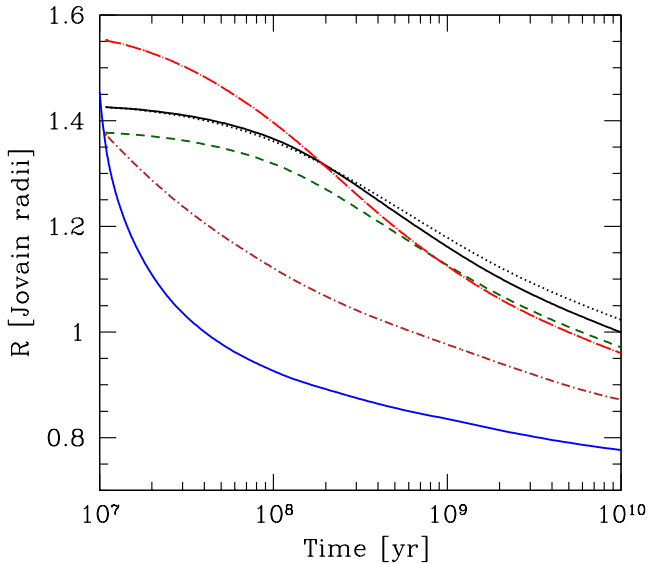}
\caption{\label{fig:tRsaturn} Evolution phase: radius of the ``wet Saturn'' as a function of time for different model settings. The nominal case assuming an opacity corresponding to [M/H]\cm$=$\cm1.4 is shown by the black solid line. The dotted black line shows the evolution without envelope evaporation. The brown dashed-dotted line corresponds to a solar composition opacity. The green dashed line assumes a purely rocky core, while the red long dashed-dotted curves uses $Y$\cm$=$\cm0.3. All the simulations use the double-gray atmospheric model and $a$\cm$=$\cm0.04\cu AU. For comparison, the blue line shows the evolution at $a$\cm$=$\cm9.4\cu AU with a simple gray atmosphere and solar composition opacities.}
\end{figure}

Figure\cf\ref{fig:tRsaturn} shows the contraction of the ``wet Saturn'', testing the impact of several parameters. All models use the post-formation entropy of 8.86\cu k$_{\rm B}$ per baryon, a computational core mass of 35.2\cu\Mearth, and an envelope mass of pure \HHe \ of 72.5\cu\Mearth. As shown in Sect. \ref{sec:metallicity_evolution}, if the effect of the 28 $\mearth$ of dissolved planetesimals would be self-consistently taken into account, the envelope would actually have a Z of about 0.28. This is a significant value, therefore our simplification of putting all solids in the computational core instead of mixing them with the H/He in the evolutionary module here has stronger consequences than for the less enriched ``dry Jupiter''.  Putting all metals in the core could result in an overestimation of the radius by up to $\sim$15 \% at late times \citep{baraffechabrier2008}.  For the opacity, we include in contrast the effect of the enrichment by assuming an opacity that corresponds to a scaled solar composition with [M/H]=1.4 for the nominal cooling curve. It predicts a radius of 1.04 $\rj$ at 5 Gyrs. Besides the nominal case, Figure \ref{fig:tRsaturn} also shows the radius as a function of time for five other simulations: Neglecting the atmospheric escape leads to a radius that is only slightly larger (difference of about 0.03 $\rj$ at 5 Gyrs). Changing the composition of the solid core, or increasing the helium mass fraction to 0.3 instead of the nominal 0.24 also leads to small differences at late times of about 0.05 $\rj$. Assuming a solar composition opacity has a larger impact, as it leads to a radius of about 0.9 $\rj$ at 5 Gyr. For comparison, we also show in the figure the radius for the same planet, but evolved at a semi-major axis of 9.4 AU which is the orbital distance at the end of the disk lifetime, before the assumed few-body interaction occurs, leading to the scattering of the planet close to the star. It is calculated with a simple gray atmosphere, and solar composition. In this comparison case, the radius at 5 Gyrs is about 0.78 Jovian radii. This is less than Saturn's current radius (about 0.84 $\rj$), and is caused by the high Z value of the planet (maybe a factor 1.2-2.2 higher than in Saturn, \citealt{guillotgautier2014}), and the too low  opacity.

We thus find that in the absence of special bloating mechanisms, the radius of the planet at 5 Gyrs should lie between 0.9 and 1.1 $\rj$, but this value is affected by the various simplifications like the double-gray atmospheric model or the scaled solar-composition opacities. As outlined in Section \ref{sec:methods:interior} the effect of radius uncertainties of the magnitude stated above on the emission spectra should be minor. The differences in the planetary evolution, and therefore its intrinsic luminosity, are not expected to be of high importance, as a hot Jupiter's atmospheric temperature structure in the regions where the emission spectrum stems from is determined by the insolation (and the atmospheric abundances). There exists a log$(g)$ dependence as well, but the uncertainty in log$(g)$ resulting from the radius uncertainty stated above is too small. For planetary transmission spectra the situation might be different, as the scale height in the planets should vary non-negligibly, being visible in the value of the planet's transit radius and the height of the spectral transmission features.

\subsubsection{Atmospheric escape}
\label{sec:results:atmopheric_escape}

\begin{figure}[t]
\includegraphics[angle=0,width=\columnwidth]{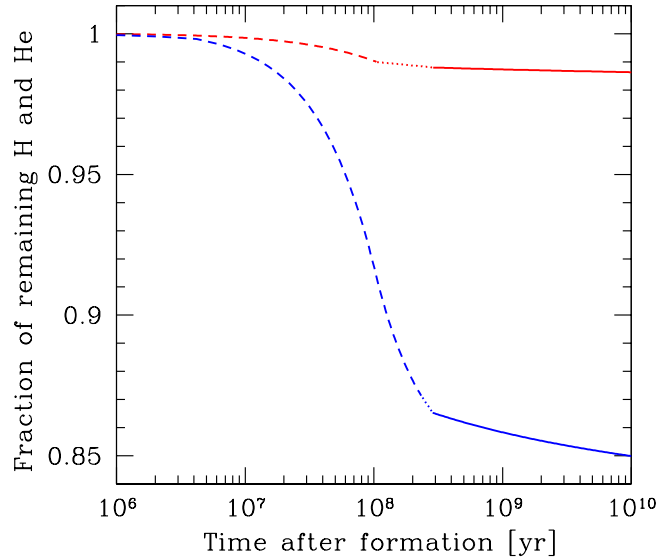}
\caption{\label{fig:tMxy} Evolution phase: evaporation of the gaseous envelope. The plot shows the fraction of the remaining envelope material as a function of time after formation, normalized to the initial value for the ``dry Jupiter'' (red) and ``wet Saturn'' (blue), respectively. The dashed, dotted, and solid lines stand for X-ray driven, EUV radiation-recombination, and EUV energy limited evaporation. }
\end{figure}

Figure\cf\ref{fig:tMxy} illustrates the hydrodynamic mass loss driven by photoionization heating from stellar ultraviolet and X-ray radiation during the evolutionary phase for the nominal models. The plot shows the fraction of the remaining envelope material as a function of time, normalized to the post-formation value. For these relatively massive planets at 0.04\cu\AU \ (i.e., not extremely close to the host star), mass loss is, as expected \citep[e.g.][]{owenwu2013,lopezfortney2013,jinmordasini2014}, not very important: the ``dry Jupiter'' looses only $\approx$\ca1.3\% of its initial envelope.

Due to its lower mass (but similar radius), the ``wet Saturn'' is more vulnerable to evaporation,
and looses $\approx$\ca15\% of the initial envelope. Note that only \HHe \ escape is calculated explicitly by the evaporation model, but because  mass is lost in the hydrodynamic regime, we assume that the heavier elements are dragged along with the \HHe \ and the composition of the material lost to evaporation is equal to the bulk envelope composition. The actual envelope composition, in particular the significant enrichment in heavy elements, is not considered in detail the evaporation model but it could affect the loss rate via modified heating and cooling rates, and a different mean molecular weight. 

The plot also shows that the evaporation first occurs in the X-ray driven regime  as found by \citet{owenjackson2012a}, then in the EUV-driven radiation-recombination regime, and finally in the EUV-driven energy limited regime \citep[e.g.][]{murray-claychiang2009}. Most of the loss occurs during the first $\sim$\ca100 Myrs after formation, when the stellar X-ray and EUV luminosity is high, and the planetary radii are large. Due to the dependency of the evaporation rate on the radius we find that the amount of material that is lost depends on the assumed opacity. For the ``wet Saturn'', for example, only 8\% of the envelope is lost at solar opacity instead of the 15\% found in our nominal model with [M/H]=1.4. 

\begin{figure*}
\includegraphics[angle=0,width=\textwidth]{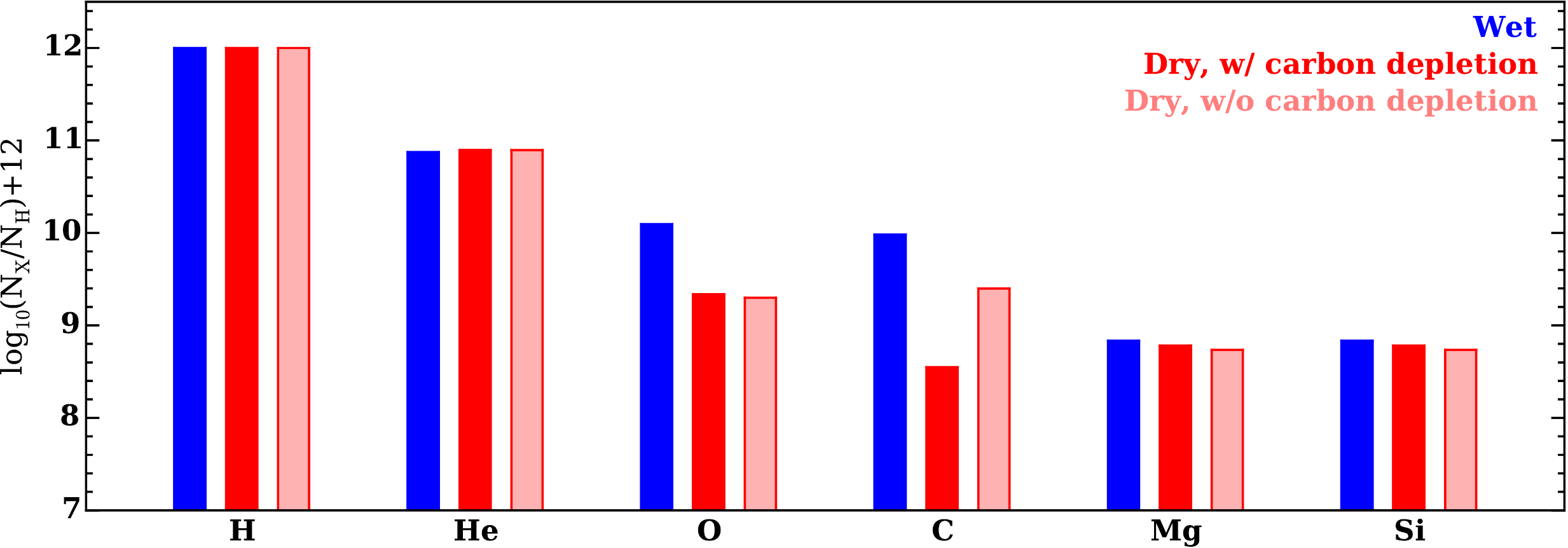}
\cprotect\caption{\label{fig:stoichiometry} Logarithmic elemental number densities normalized by the hydrogen number density log$_{10}$(N$_{\rm X}$/N$_{\rm H}$)+12. The number densities are shown for the ``wet'' planet (blue bars) and ``dry'' planet (dark red bars)  for the \verb|r0.32_Csil0.50_Cdef_v1_clath_gVol| case for the elemental abundances, meaning in particular that refractory carbon is depleted in the inner disk.  The effect of turning off the \verb|Cdef| switch, i.e., assuming that the planetesimals in the inner disk contain 0.5 refractory carbon in mass leads to the light red bars with a much higher carbon content. \ColorJournal}
\end{figure*}

Despite the rather low mass loss at 0.04\cu\AU, the ``wet Saturn'' is actually quite close to the threshold where catastrophic runaway evaporation occurs \citep{baraffeselsis2004,kurokawakaltenegger2013,jinmordasini2014}: if this planet is evolved at 0.03\cu\AU \ instead of 0.04\cu\AU \ for otherwise nominal parameters, it undergoes a phase of runaway gas loss about 40\cu Myrs after the end of formation. In this phase, the envelope expands, and most of the gas is evaporated. The nature of the planet would then be completely different at 5\cu Gyr, showing the possibly strong impact of evaporation, and the necessity of it to be included as a chain link, at least for certain planet types. This vicinity of the planet to the catastrophic evaporation threshold is in good agreement with the findings of \citet{kurokawanakamoto2014}. 

In Table\ct\ref{table:physprops5gyrs} we list the physical properties of the two planets at 5\cu Gyrs. They are used together with the elemental composition in the final chain link, where the atmospheric $P$$-$$T$ structure, chemistry, and the resulting spectra are calculated.

\begin{table}
\centering
\begin{tabular}{l|cc}
Quantity & ``dry Jupiter'' & ``wet Saturn''  \\ 
\hline \hline
Radius [$\rj$] & 1.12 & 1.04 \\
Mass [\Mearth] & 257.3 & 97.0 \\
$T_{\rm int}$  [K] & 98.0 & 66.4 \\
Orbital radius  [AU] & 0.04& 0.04 \\
Core mass [\Mearth] & 7.9 & 7.3 \\
\hline
\end{tabular}
\caption{Physical properties at an age of 5 Gyr.}
\label{table:physprops5gyrs}
\end{table}

\subsection{Elemental abundances}
\label{sec:results:stoichiometry}
The resulting elemental abundances for the ``dry'' and ``wet'' planet cases depend on the various assumptions on the composition of the accreted refractories, volatiles, and gas, described in Section\cs\ref{sec:methods:stoichiometry}. We visualize the elemental abundances in Figure\cf\ref{fig:stoichiometry} for a specific set of assumptions, using the \verb|r0.32_Csil_0.50_Cdef_v1_clath_gVol| option, i.e., a refractory mass fraction of $f_{\rm r}$ = 0.32 in the metals (rest ices), a carbon to silicate mass ratio of 1/2  with the default carbon depletion in the inner disk. The light red bars show the same abundance model with the difference that no carbon depletion is assumed for the inner disk. Furthermore we assume a solar nebula model for the disk volatile composition and consider the formation of clathrates outside the disk and that the volatiles are found in the gas inside the iceline. 

The most important difference between all models is the carbon to oxygen ratio C/O which varies quite significantly for the different
models. First, the ``wet'' planet which forms outside the iceline contains a high amount of oxygen, as the oxygen can be found in
both the refractory and volatile material accreted in the planetesimals. It also contains a considerable amount of carbon, as carbon can
also be found in both the refractories and volatile ices. Nonetheless, the refractory and volatile composition and their ratio in the planetesimals result in carbon to be less common than oxygen, leading to a C/O ratio of 0.77.

For the ``dry Jupiter'' the resulting C/O ratio 0.16, i.e. it is richer in oxygen relative to carbon than the ``wet 
Saturn''. This can easily be understood by the fact that the planet formed in the carbon depleted inner part of the disk, where the 
refractories contain very little carbon, but are rich in oxygen due to the high amount of silicates.
The volatiles being accreted by the envelope in gaseous form are oxygen rich as well {for all explored compositions (Table~\ref{tab:volatiles_composition}). Here it should be noted that the carbon that has been removed from the solid phase has not been added to the gaseous phase, i.e. it is assumed that since the early phase of carbon combustion and subsequence planetesimal formation the gaseous disk has evolved viscously. However, even if this carbon were all locally retained in the gas phase to be partially accreted onto the planet later on, it would never lead to C/O ratios near unity for our planetary envelopes, as they are always heavily enriched and get most of their heavy elements from planetesimals \rch{(see Sect. \ref{sect:importanceplanetesimals} ).}

Therefore, in the example case studied here, both planets forming in- and outside of the iceline have a small C/O ratio, with the ``dry''
planet being depleted in carbon relative to oxygen more strongly than the ``wet'' planet.
 
\rch{While the result that no carbon-rich ``dry'' and ``wet'' planets can form in our standard carbon depletion model is robust, the fact that the ``wet'' planet has a larger C/O ratio than the ``dry'' planet is specific to the example shown here. It critically depends on two assumptions for the example studied here: The maximum carbon-depletion factor is very low (we assumed 10$^{-4}$) and all non-water volatiles freeze out with the water as clathrates.
If one runs calculations with different maximum carbon-depletion factors (10$^{-4}$ to 1, i.e.  non-depleted, in steps of 1 dex), one sees that the statement that the ``dry'' planet has a lower C/O ratio than the ``wet'' one is not generally true; if the depletion factor is a factor $\sim$100 or less then the ``dry'' planets generally have more carbon than the equivalent ``wet'' cases, even if one assumes all volatiles to be in clathrates. If one considers cases without clathrate formation then the ``wet'' planets all fall in a narrow range with C/O=0.10-0.12. It is known that clathrate formation occurs and that a non-negligible fraction of non-water volatiles freezes out with the water \citep{1992IAUS..150..437B,marboeufthiabaud2014a,2016SciA....2E1781L}. However, assuming this fraction is close to 1 is not necessarily realistic and was adopted only to test the most extreme case of clathration, knowing that the actual clathration fraction lies somewhere in between. Therefore, a better quantitative understanding of the carbon depletion and clathrate formation is clearly needed.}

If we consider the ``dry Jupiter'' once more, but in the case without carbon depletion, we find a very different result, namely that the planet's C/O ratio would be 1.26, i.e., clearly bigger than one. In this case we would find that planets formed inside the iceline would have a C/O ratio
$>$ 1, while planets formed outside the iceline would have a C/O ratio $<$ 1. As we will see in Section \ref{sec:results:simulated_observations},
this kind of dichotomy of the C/O ratio, related to the formation location, would  be easily distinguishable in the planetary spectra.
The reason for this is that the spectra of either oxygen-dominated (C/O $<$ 1) or carbon-dominated (C/O $>$ 1) atmospheres are very
different in their appearance \citep{madhusudhan2012,molliereboekel2015}. 

\subsection{C/O number ratios}
\label{sec:results:CtoO}

In order to get a better overview over the global characteristics of our compositional models connected to the ``dry'' and ``wet'' formation scenarios we show a histogram of all resulting C/O ratios in Figure \ref{fig:COratios_Cdepleted}. For the left panel of this plot we used the standard refractory model, i.e., a carbon to silicate ratio of 1:2 and assumed a carbon depletion in the inner disk, but varied all other chemistry options (the refractory to all metals fraction $f_{\rm r}$, the volatile model, the presence or absence of clathrates and of volatiles in the gas).
Therefore, every count in the histogram stands for one of our compositional models applied to the ``wet'' and ``dry'' planet formation case.
The histograms should not be confused with probability distributions for the C/O ratios for planets forming inside or outside the iceline, as we did not assign any prior probability to any of the compositional models. They merely give an graphical impression of the impact of different disk chemistry models on the resulting planetary C/O. 

\begin{figure*}[t]
\begin{center}
\includegraphics[angle=0,width=1.03\columnwidth]{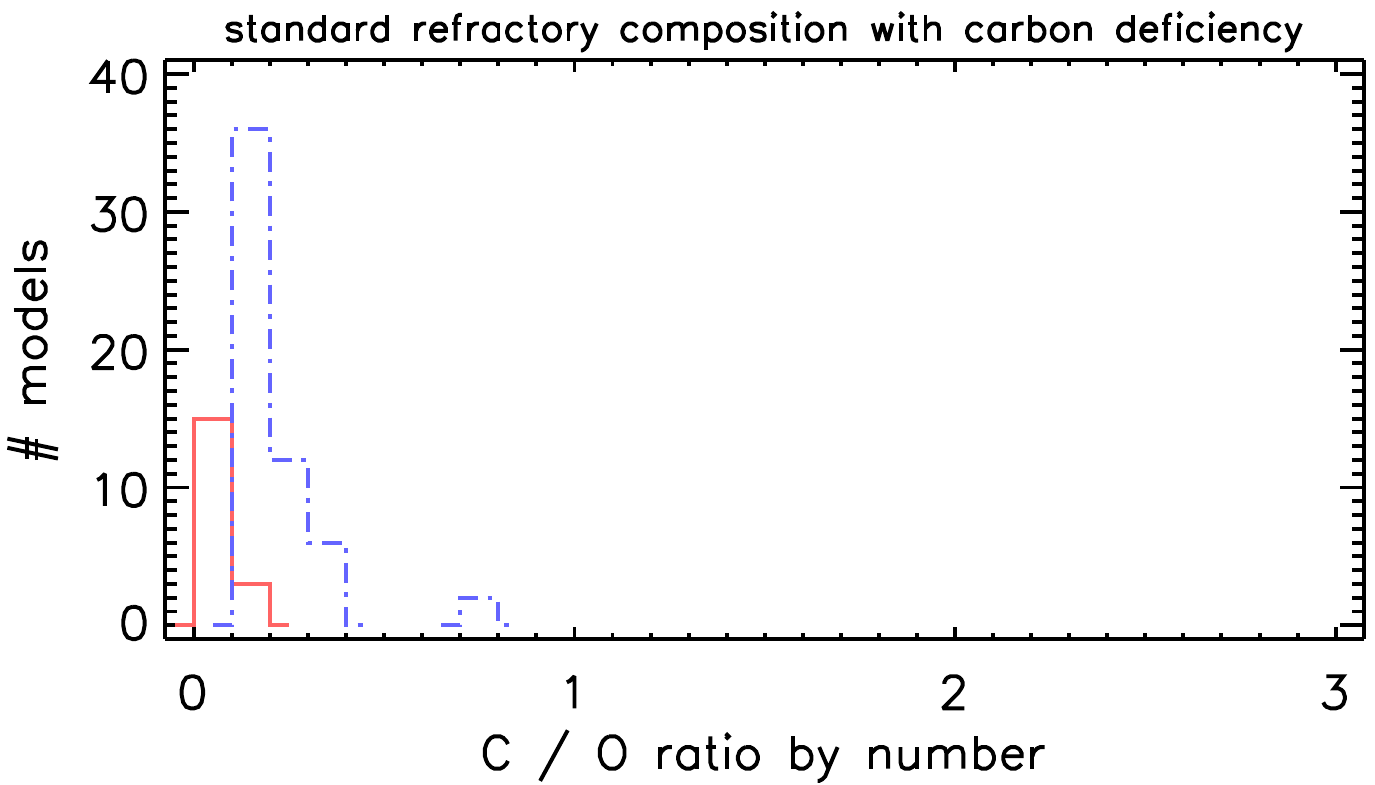} 
\hspace{0.cm}
\includegraphics[angle=0,width=1.03\columnwidth]{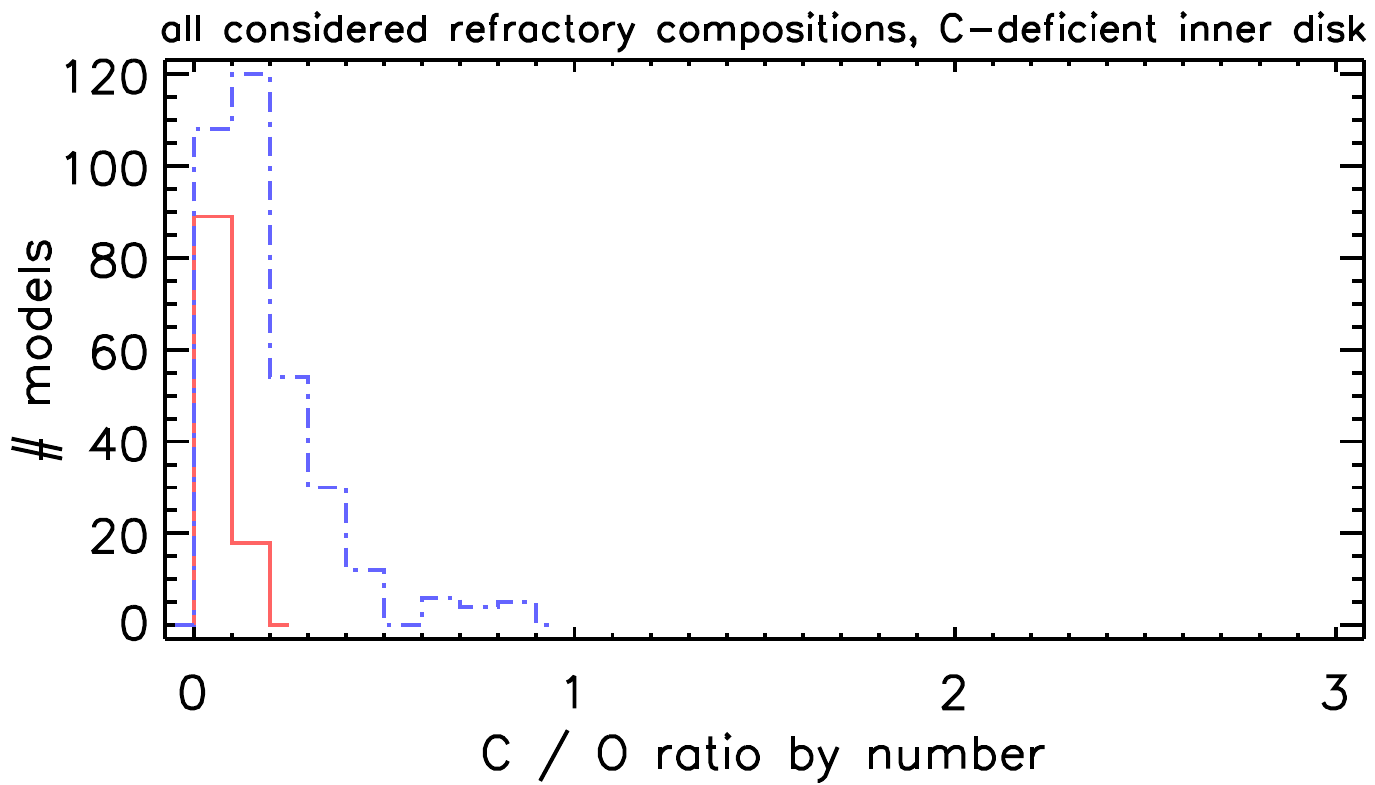}
\end{center}
\caption{\label{fig:COratios_Cdepleted} {\it Left panel:} Resulting C/O number ratios with carbon depletion in the inner disk and the ``standard'' compositional model for the refractories (i.e., a C:silicates mass ratio of 1:2), but varying all other compositional parameters. {\it Right panel:}  as on the left, but allowing now also the C:silicates mass ratio to vary. \ColorJournal }
\hspace{-0.3cm}
\begin{tabular}{cc}
\includegraphics[angle=0,width=1.03\columnwidth]{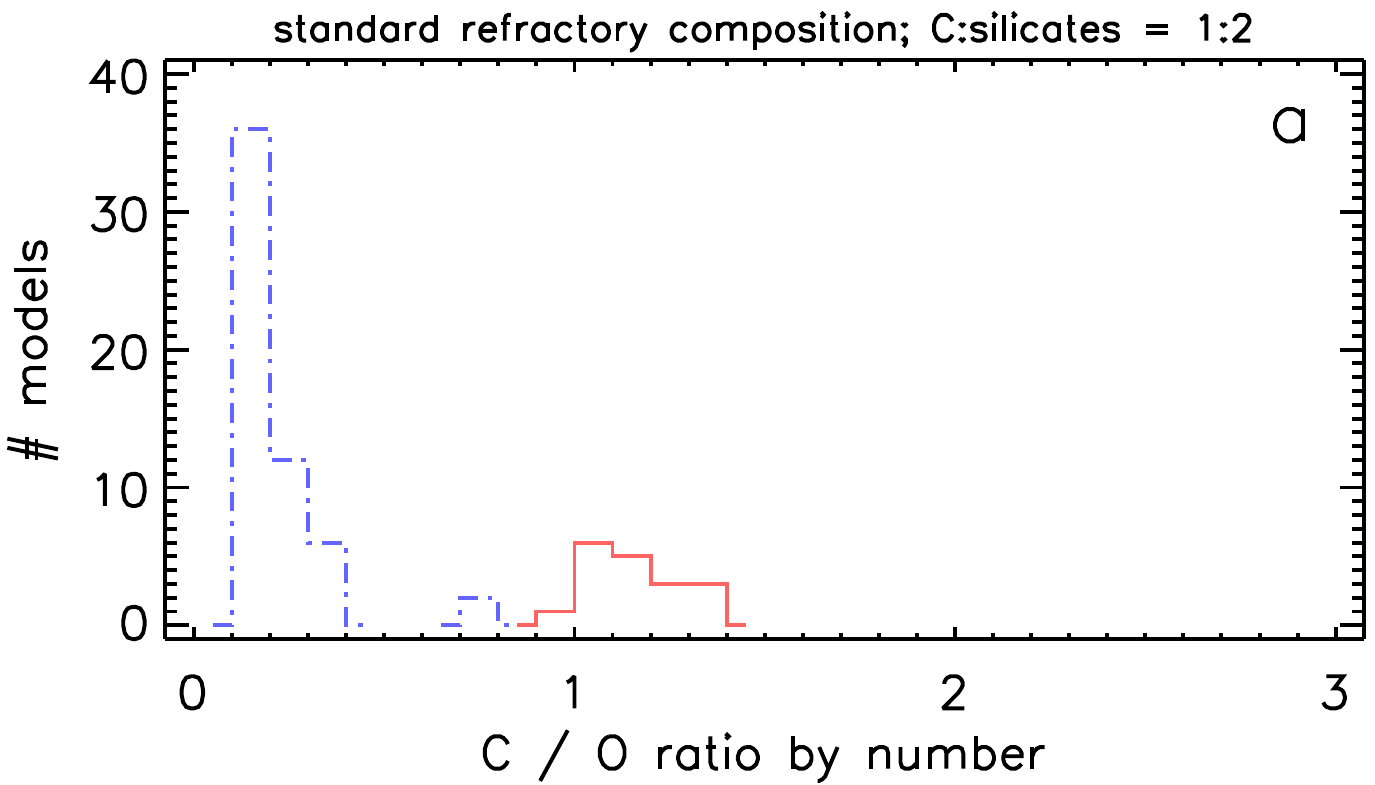} &
\includegraphics[angle=0,width=1.03\columnwidth]{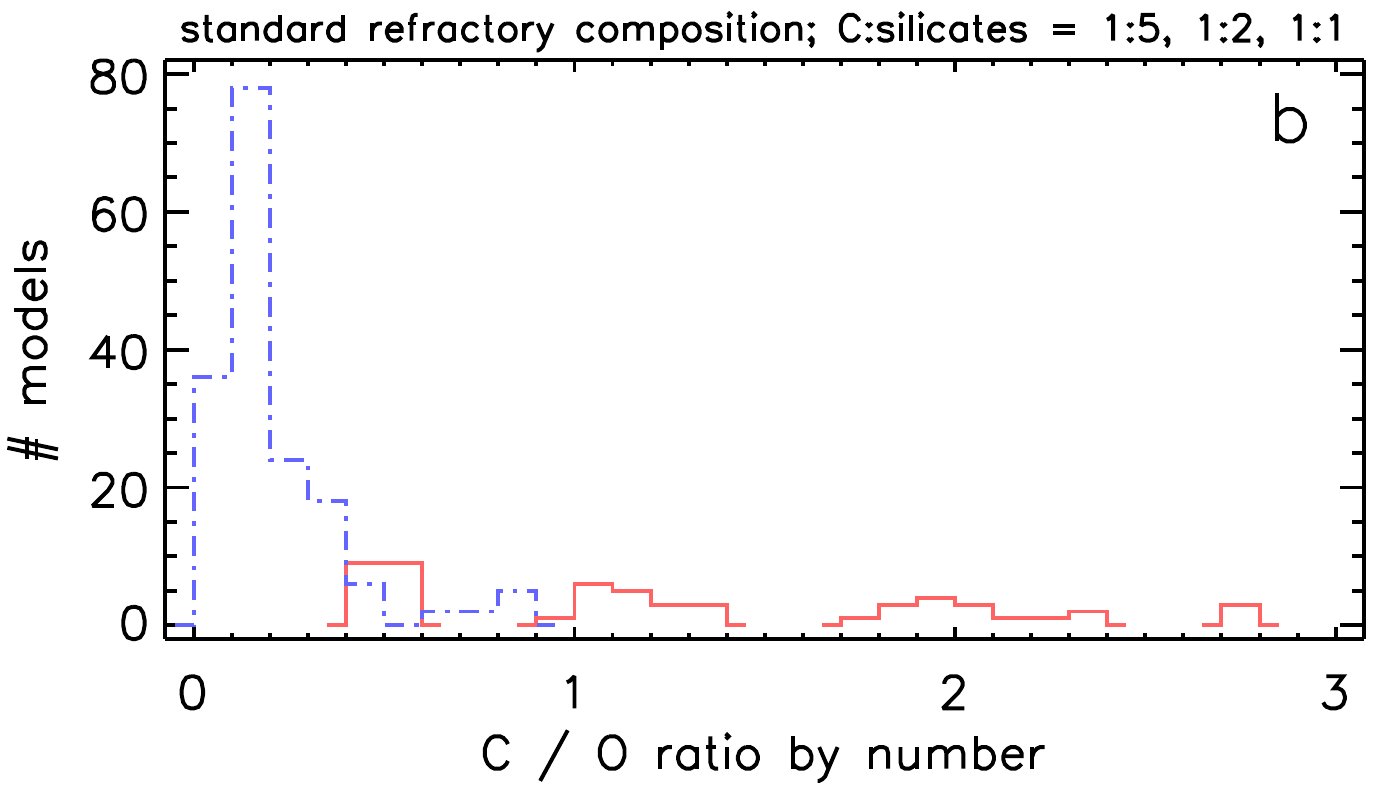}

\end{tabular}
\caption{\label{fig:COratios_simple} Resulting C/O number ratios without carbon depletion. In the {\it left panel} (a) the C:silicates mass ratio in the refractory material is 1:2. In the {\it right panel} (b) 3 values of the C:silicates mass ratio are included: 1:5, 1:2, and 1:1. The other compositional parameters are varied. \ColorJournal}
\hspace{-0.3cm}
\begin{tabular}{cc}
\includegraphics[angle=0,width=1.03\columnwidth]{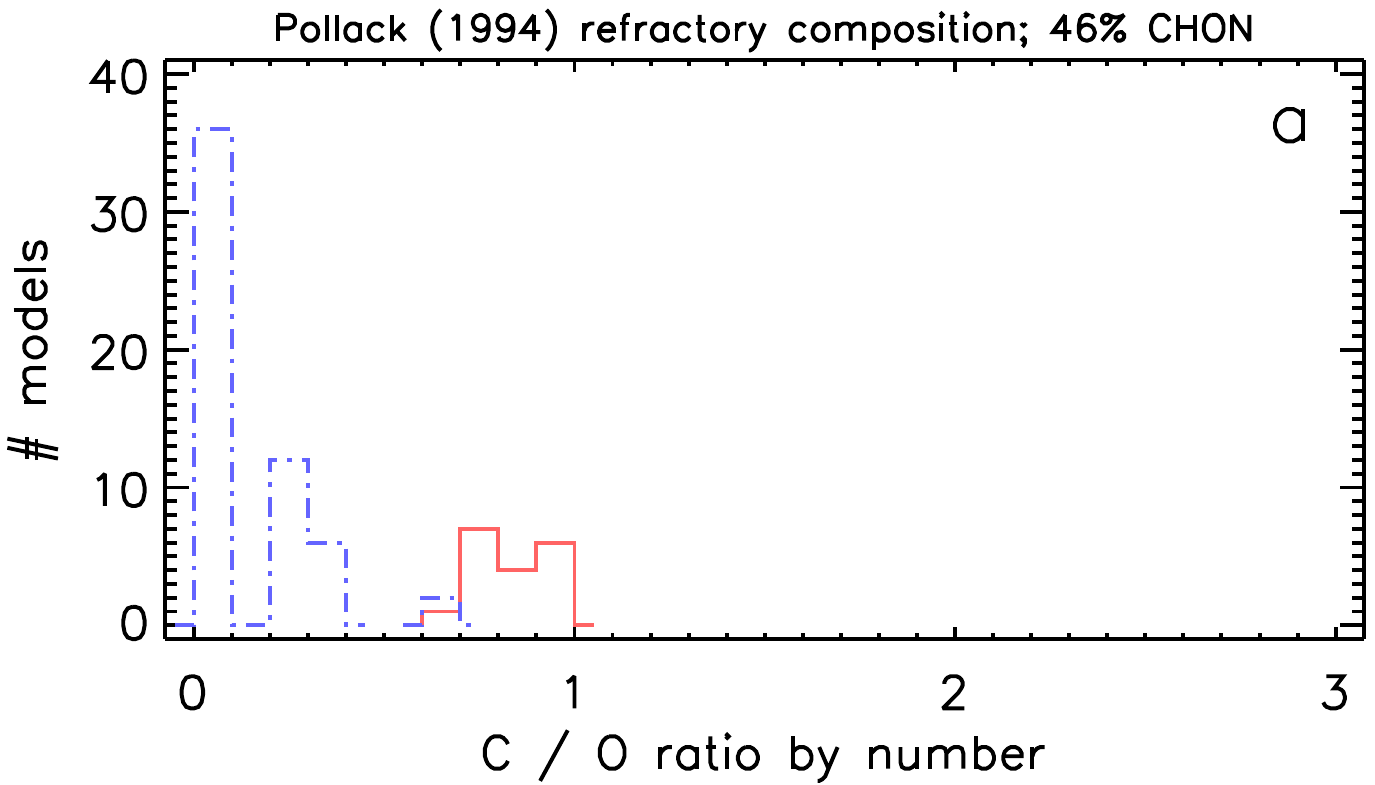} &
\includegraphics[angle=0,width=1.03\columnwidth]{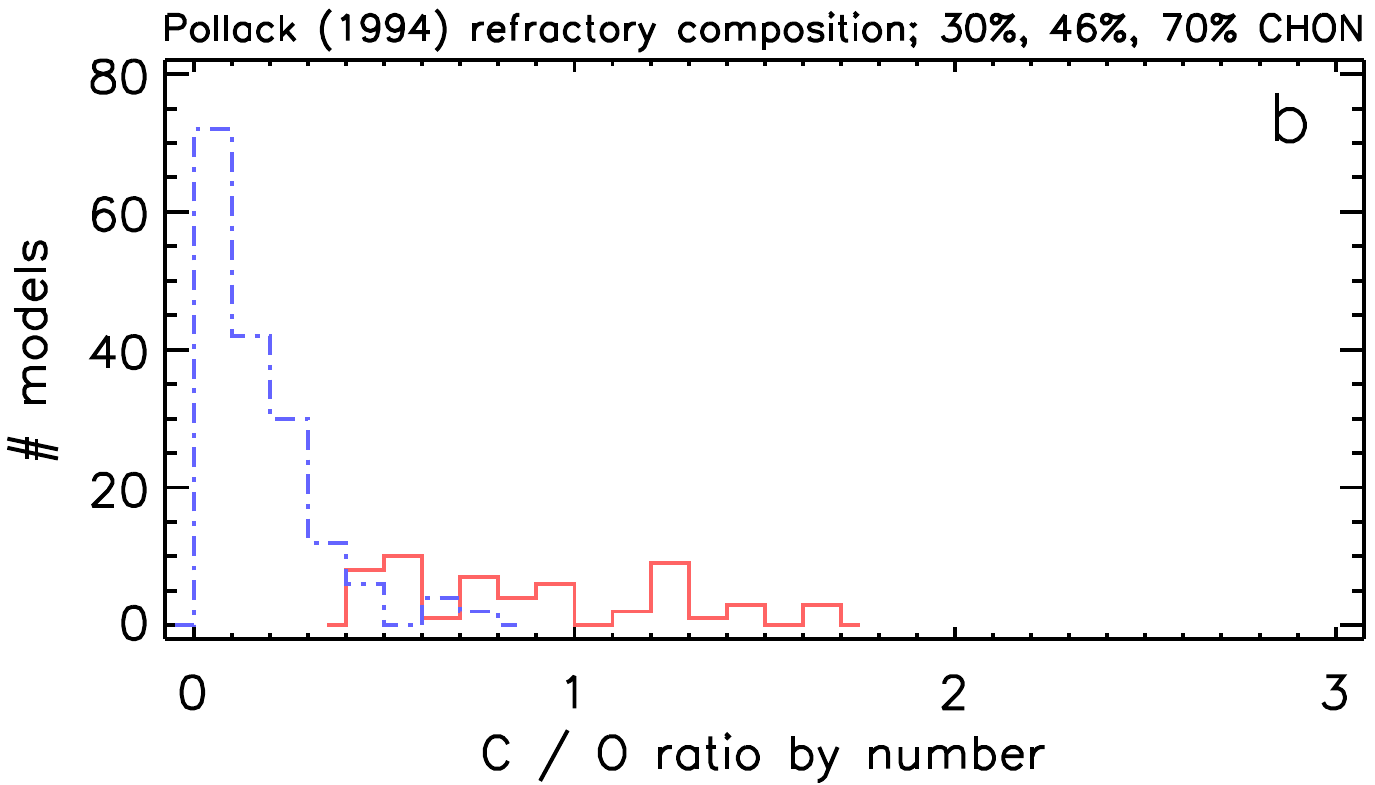}

\end{tabular}
\caption{\label{fig:COratios_Pollack} Resulting C/O number ratios without carbon depletion for a refractory composition following \cite{1994ApJ...421..615P}, as described in Section\cs\ref{sec:methods:refractories}. In the {\it left panel} (a) the CHON material comprises 46\% of the refractory mass as in the Pollack et al. model, in the {\it right panel} (b) we include 3 values of the CHON mass fraction: 30\%, 46\%, and 70\%. \ColorJournal}
\end{figure*}

It is straightforward to see, however, that no planet with a C/O ratio bigger than 1 forms in both the ``dry'' and the ``wet'' case.
Further, the ``dry'' case is even more enriched in oxygen with respect to carbon than the ``wet'' case. Note, however, that the absolute
oxygen enrichment of the ``wet Saturn'' is still higher, as it is formed outside of the iceline and because it has a smaller total mass
it is strongly enriched by water-rich planetesimals. The right panel of Fig. \ref{fig:COratios_Cdepleted} shows the effect of varying also the C:silicates mass ratio (Sect. \ref{sec:methods:refractories}). We see that this does not change the conclusion that both planets have C/O$<$1. \rch{The histogram shows that the ``dry'' planet's C/O ratio is found to be $<$0.2, whereas the ``wet'' planet can have C/O values ranging typically between 0.1 and 0.5, depending on the assumed disk chemistry model.} {In some cases, ``wet'' planets with  C/O$>$0.5 occur; this happens when we adopt the volatile composition \verb|v1| and assume all volatiles are trapped as clathrates. This is a rather extreme assumption but falls within the parameter space we explore. Most of explored possibilities, however, result in planets with a sub-solar C/O ratio \rch{$<$ 0.5.} }

Consequently we conclude that in our standard assumption on the composition, in particular with the carbon depletion in the inner disk, hot Jupiters should be oxygen-rich. {This follows directly from the fact that in our model the planetary envelopes obtain the majority of their heavy elements from planetesimals, which contain more oxygen than carbon.}

\rch{In the non-carbon-depleted abundance model the highest value of the carbon-to-oxygen ratio in the refractory planetesimals occurs if the mass ratio of carbon to silicates is equal to 1. As we assume that the silicates are composed of MgSiO$_3$ this means that the maximum C/O ratio in the refractories is $\sim 2.7$. Therefore, if one uses a minimum carbon depletion value of 10$^{-1}$, instead of 10$^{-4}$, there would still no carbon-rich planets be formed.}

\subsubsection{Could there be a carbon sweet spot?}\label{sec:results:carbonsweetspot}
Regarding carbon depletion it is worthwhile to mention the following point. Considering the position of the iceline and the  dependency of the carbon depletion with distance from the star (Fig. \ref{fig:Cdef}), we see that there there is in principle a sweet spot for carbon-rich planet formation in our model: Outside of 5 AU our ad hoc carbon depletion model does not yet decrease the carbon, but this  still lies inside of the water iceline used for the planetesimal formation, which is at 6.9 AU for the ``dry'' planet's disk. Taken at face value, planets forming in this semimajor axis range would therefore end up with high C/O as for the cases without carbon depletion. 

It is clear that quantitatively, this finding is a direct consequence of the ad hoc way carbon is depleted. But it shows that it might not be  impossible to form carbon rich planets even in disks with (partial) carbon depletion in their inner regions. Whether such regions exist, and over which orbital distance they extend will depend on the mechanism that destroys carbon in the inner disk like ion-induced erosion of solid carbon \citep{sabribaratta2015}. In an alternative model of \citet{leebergin2010}, hot atomic oxygen produced by photodissociation of O bearing species erode carbon grains and release the carbon into the gas in the upper layers of the disk. The efficiency of this process decreases with orbital distance, and finally drops  outside of the iceline, as the oxygen is there locked on grain surfaces as water ice. Whether this quantitatively leads to significant C rich regions inside the ice line must be investigated with detailed disk chemistry models including  chemical-kinetic pathways, ion irradiation and photochemistry, as purely thermodynamic condensation models are insufficient \citep{jurayoung2014,sabribaratta2015}. 

Such a carbon rich region inside of the iceline, but outside of a ``tar'' line is reminiscent of the one proposed by \citet{lodders2004} to explain the high  C/O ratio in Jupiter. According to this paper, the carbon rich region would start at around 2.2 AU. This point illustrates the importance of a good understanding of the chemistry in protoplanetary disks \citep[e.g.,][]{henningsemenov2013}. Observationally, an extrasolar planetesimal with a carbon-rich and water-poor composition that is potentially compatible with a carbon sweet spot was found by \citet{juradufour2015}. If such a region does indeed exist, and is sufficiently wide, it would be a good tracker of the planet's formation location. 

In light of this, and as our carbon depletion model is largely based on observational evidence in the Solar System and white dwarf atmospheres, rather than a physically robust model, it is worthwhile to also study the opposite extreme for the inner disk, i.e., the case where
all carbon in the refractories is retained. Adopting again the ``standard'' carbon to silicate mass ratio of 1:2, the resulting C/O ratios now yield a clear-cut behavior: the ``wet'' planet that formed completely outside the water iceline always ends up with an oxygen-rich envelope (nothing changed for it), whereas the ``dry'' planet which formed completely inside the water iceline now attains a carbon-rich envelope with C/O $>$ 1. The reason is for this is that in the accreted planetesimals which eventually determine the atmospheric C/O, the refractory carbon outnumbers the oxygen atoms in the silicates. There is one exception to this: one ``dry'' model yields C/O=0.975 by number, this is the \verb|dry_r0.25_Csil0.50_v0_noclath_gVol| model where the volatiles consist only of water and the rock/volatiles mass ratio is 1/3, the lowest explored value. In this model, the oxygen atoms accreted in the form of gaseous water just barely outnumber the C atoms accreted in the form of planetesimals, leading to C/O $<$ 1. The corresponding histogram for this case can be seen in Figure\cs\ref{fig:COratios_simple}a.

In the absence of carbon depletion in the inner disk, if we explore a larger range of carbon to silicate mass ratios, the resulting distributions show a  more or less pronounced difference between the ``wet'' an ``dry'' case, i.e., a formation in- or outside of the water iceline, as is illustrated in Figure\cs\ref{fig:COratios_simple}b. The ``wet'' planet accretes an oxygen-rich envelope. For it, the highest C/O values are obtained with those chemical models with the highest assumed rock/volatiles mass ratio and the highest assumed carbon/silicates mass ratio in the refractories. They result in C/O number ratios of  up to $\approx$\ca0.9.

The C/O number ratio in the envelope of the ``dry'' planet directly reflects the assumed carbon to silicate mass ratio in the refractory material. Carbon-poor refractories (C:silicates $=$ 1:5) yield oxygen-rich planets with C/O number ratios between $\approx$\ca0.4 and $\approx$\ca0.6; the nominal ratio of C:silicates $=$ 1:2 yields C/O number ratios between $\approx$\ca1.0 and $\approx$\ca1.4; if we assume very carbon-rich refractory material (C:silicates $=$ 1:1 by mass) we get high C/O number ratios between $\approx$\ca1.7 and $\approx$\ca2.8 for the explored range of assumptions.

Still without carbon depletion, but when adopting the refractory composition by \cite{1994ApJ...421..615P}, a somewhat different picture emerges. In Figure\cf\ref{fig:COratios_Pollack}a we show the resulting C/O number ratios for their standard refractory composition, where ``CHON'' material makes up 46\% of the mass in refractories. Interestingly, both the ``wet'' and the ``dry'' planet always attain an oxygen-dominated envelope, though in the ``dry'' case the C/O number ratios are close to unity. If we vary the relative contribution of CHON material to the total refractory mass, the resulting C/O number ratios in the planetary envelope show a correspondingly larger spread, with carbon-rich atmospheres occurring when we adopt refractories that are very rich in CHON material (70\% by mass), see Figure\cf\ref{fig:COratios_Pollack}b. 

In summary, for giant planets where the accreted planetesimals determine the atmospheric composition (not too high mass \rch{of less than $\sim$2-10$\mj$}, mixing of interior and atmosphere), without carbon depletion a formation inside of the water iceline leads to a high C/O (between 0.4 and 2.8, with typical values around 1), while a formation outside of it leads to a low C/O$<$0.5, with typical values around 0.2. This would allow to make a direct link between atmospheric spectra and the formation location. However, for the nominal and likely case of carbon depletion, there is unfortunately no such clear-cut distinction: both a formation inside and outside of the water iceline leads to a  C/O$<$1. 

\rch{\subsubsection{C/O: Comparison with previous work}\label{sec:comparisonpreviouswork}}
\rch{Several previous studies with a similar direction as the work here have predicted that under certain conditions, planets with high C/O could form. This is in contrast to the low C/O we find here for the nominal disk chemistry for both the ``wet'' and ``dry'' planet, raising questions about the generality of our result. These previous studies have shown that important ways to get high C/O in giant planets are (1) to form them farther out in the disk, in particular beyond the CO$_{2}$ snow line \citep{obergmurray-clay2011,madhusudhanamin2014} or (2) because of a preferential depletion of H$_{2}$O vapor in comparison to CO vapor at smaller distances \citep{ali-dipmousis2014}. The crucial point is that both these mechanisms can only be important if there is very little ``pollution'' by planetesimal accretion, i.e., if the  atmospheric composition is dominated by the composition of the accreted gas and not planetesimal-dominated. As discussed by \citet{obergmurray-clay2011}, already small additions of planetesimals (about 1-2\% in mass, corresponding to an enrichment level equal to the stellar one) are sufficient to get into the planetesimal-dominated regime.  In view of the result outlined in Sect. \ref{sect:importanceplanetesimals} that several theoretical and observational findings indicate that planets with masses below $\sim$2-10 $\mj$ have enrichment levels higher than the star, and thus are in the planetesimal-dominated regime, we conclude that these pathways towards high C/O are unlikely to apply for typical hot Jupiters because of their lower masses. The dominance of planetesimal enrichment for Jovian mass planets is further backed up by the study of \citet{MousisMarboeuf2009} who showed that for the Solar System giant planets the enrichment by planetesimals  allows to fulfill several detailed constraints on both the atmospheric and interior composition.}


\rch{The behavior that the enrichment by planetesimals is of paramount importance is clearly illustrated in the study of \citet{thiabaudmarboeuf2015}.  Thanks to their population synthesis approach they cover a wide range of planetary and disk properties. If they assume that only gas determines the atmospheric composition of giant planets, they find a wide spread of C/O covering sub- and supersolar values. If they assume on the contrary that planetesimals fully dissolve in the envelopes, all giant planets end up with a low subsolar C/O. We argue that only the latter case is relevant for most hot Jupiters because of their mass usually below 2-10 $\mj$. Their population-wide results furthermore mean that our results of water-dominated hot Jupiters should be of general validity and not be a consequence of our specific initial conditions.}

\rch{\rt{Regarding the} planetesimal enrichment, it is positive to see that our results are consistent with what all these previous studies have found, i.e., that substantial planetesimal accretion leads to O-rich envelope compositions. The exception to that is when the solids have themselves a carbon-rich composition. This is the case in our model if we assume for the solids  inside of the water iceline a complete inheritance of the ISM-like carbon-rich composition (the non-nominal models without carbon depletion). This result in turn is in good agreement with \citet{thiabaudmarboeuf2014}. Other studies have proposed  that carbon-rich planets could form around stars that have themselves intrinsically C/O$>$1 \citep[e.g.,][]{carter-bondobrien2010,2012ApJ...759L..40M}. In view of several recent observational and theoretical results that stars with such high C/O are probably very rare \citep{fortney2012, TeskeCunha2014,gaidos2015}, also this pathway towards carbon-rich hot Jupiters now appears rather unlikely. This paucity is also the reason why we did not investigate planet formation around stars with intrinsically C/O$>$1. This gives further support to the generality of water-rich compositions of hot Jupiter with low C/O as found in our study. }

\subsection{Atmospheric structure and chemistry}\label{sec:results:atmospheric_structure_chemistry}
\begin{figure*}[t]
\hspace{-0.4cm}
\includegraphics[angle=0,width=1.03\textwidth]{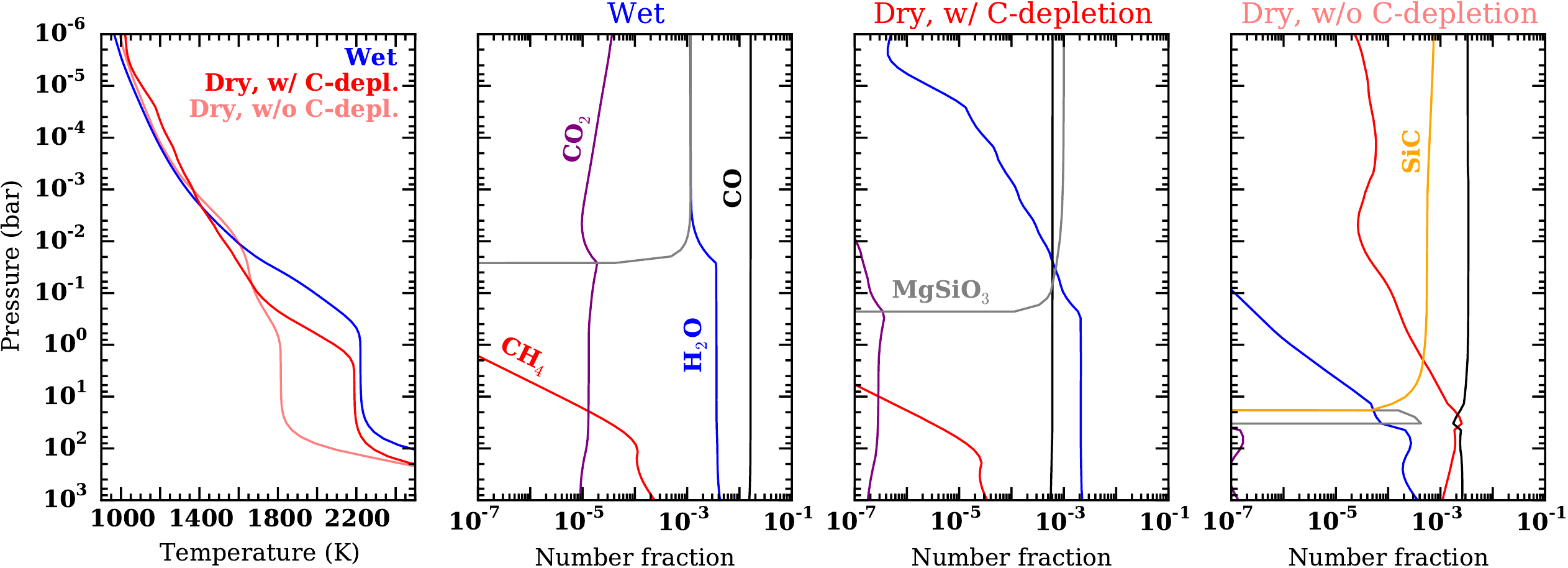}
\cprotect\caption{\label{fig:chemistry} Atmospheric structure and element abundances obtained with the \emph{PETIT} code. {\it Left-most panel}: $PT$-structures for the \verb|wet_r0.32_Csil0.50_Cdef_v1_clath_gVol| (blue line), \verb|dry_r0.32_Csil0.50_Cdef_v1_clath_gVol| (red line) and \verb|dry_r0.32_Csil0.50_v1_clath_gVol| (light red line) compositional models. In the {\it center-left},
{\it center-right} and {\it right-most} panel we show the corresponding molecular number fractions in the atmospheres of the three cases. The number fractions are shown for CH$_4$ (red line), H$_2$O (blue line), CO$_2$ (purple line), CO (black line), MgSiO$_3$ (gray line), and SiC (orange line). \ColorJournal}
\end{figure*}

We use the  atmospheric code \emph{PETIT} to calculate self-consistent structures and spectra for the ``wet Saturn'' and ``dry Jupiter'' whose abundances we showed in Figure \ref{fig:stoichiometry}.
In addition to the atomic abundances for H, He, C, N, O, Fe, Mg and Si, which result directly from our chemistry model, we considered also K, Fe, P, Al, Ca, Na, S, V, Ti, Ni and Cl. The abundances for these additional species were calculated by multiplying their relative solar abundance with respect to Si by the Si abundance obtained from our formation and compositional model.

For the planetary parameters of the ``wet Saturn'' and ``dry Jupiter'' we used the values given in Table \ref{table:physprops5gyrs}. 
The star was assumed to be a sun-like main sequence star with $T_*$~=~5777~K and a radius of 1~$\rsun$.
For the planet these parameters result in an effective day-side averaged temperature of 1656 K.

\subsubsection{\emph{PETIT} code results}\label{sect:petit_atmos_struct}
For the insolation of the atmosphere we used \emph{PHOENIX} spectra as described in \citet{vanboekelbenneke2012}.
We then calculated the self-consistent structure of the planet's atmosphere using the ``dry'' and ``wet'' abundances which we obtained for the
\verb|r0.32_Csil0.50_Cdef_v1_clath_gVol| model. We also calculated the structure of the ``dry'' case with carbon depletion turned off as a third option.
For the calculation we divided the atmosphere in 150 layers spaced equidistantly in log-space between 10$^{-14}$ and 10$^{5}$ bar
The corresponding $PT$-structures and molecular and atomic abundances for all 3 models can be seen in Figure \ref{fig:chemistry}.
We plot the molecular abundances for CH$_4$, H$_2$O, CO$_2$, CO, MgSiO$_3$ and SiC.

For the ``wet'' case one clearly sees that the planet is oxygen-rich, with H$_2$O being the most abundant molecule in terms of its impact
on the spectrum.\footnote{CO is more abundant, and has some visible spectral features, but in its entirety its impact on the general shape
of the SED is small when compared to water.} The condensation of MgSiO$_3$ decreases the abundance of H$_2$O above the 10$^2$~bar~altitude.
As the ``wet'' planet is depleted in carbon with respect to oxygen, methane is much less common in the planet's atmosphere
and will not leave a spectral imprint (see Section \ref{sec:results:spectra}). For C/O $<$ 1 the carbon is preferentially put into CO, at least
at the high temperatures considered here. Additionally there is quite a lot of CO$_2$ formed in the ``wet'' atmosphere, which is first and foremost a consequence of the planet's high metallicity, rather than its C/O ratio \citep[see, e.g.][]{mosesline2013}.

In the standard ``dry'' case (i.e., with carbon depletion) we see again that water is the most common molecule in terms of the spectral impact.
CO is much less common than in the ``wet'' case because there is much less carbon present in this planet, almost all of which was accreted in
the form of volatile gases inside the iceline. Due to the planet's lower enrichment (in comparison to the ``wet'' planet), CO$_2$ is much less abundant than in the ``wet'' case.
Water becomes depleted above the 10$^{-1}$ bar altitude due to the condensation of MgSiO$_3$, decreasing the atmospheres ability to
cool somewhat, but not significantly enough to produce an isothermal layer or even an inversion in the planet's atmosphere.

In the ``dry'' case without carbon depletion the most common molecule for the spectral signature is CH$_4$, as it should be expected for a carbon-rich atmosphere. At $\sim$20~bar the condensation of MgSiO$_3$ increases the CH$_4$ abundance somewhat, as MgSiO$_3$ takes
away oxygen which can then not be stored in CO anymore. Above the $\sim$20 bar altitude SiC condenses, decreasing the CH$_4$ abundance significantly, thereby decreasing the atmosphere's ability to cool away absorbed stellar radiation. This leads to an approximately isothermal region between the 10$^{-1}$ and 10$^{-2}$ bar altitude. The CO$_2$ abundance is very low, as this planet has a C/O $>$ 1.

\subsection{Emission spectra}\label{sec:results:spectra}

\subsubsection{\emph{PETIT} code results}\label{sec:results:spectra:petit}

The emission spectra for the planets studied in Section \ref{sect:petit_atmos_struct} can be seen in Figure \ref{fig:spectra_nominal_petit}.
As expected the nominal ``wet'' planet  and ``dry'' planet with carbon depletion show clear water signatures in their spectra, while the non-nominal ``dry'' planet without carbon depletion shows signatures of methane absorption \rch{and makes it evident that in a disk without carbon depletion a distinction between the formation inside and outside the would be possible}. However, due to the approximately isothermal region found in this planet's atmosphere the overall appearance of this atmosphere is closer to a blackbody spectrum than the two nominal cases. If this planet would be even hotter, then SiC condensation would not occur, which would lead to even clearer methane absorption features.

For the nominal chemistry model with carbon depletion,  both planets forming either in or outside the water iceline are, as seen earlier, poor in carbon with respect to oxygen, and thus the spectra of both planets are dominated by water features,  and therefore similar. There are, however, also differences: as the ``dry'' planet contains less water than the ``wet'' one, the water absorption troughs between $\sim$ 1 - 10 $\mu$m are less deep when compared to the ``wet'' case. The reason for this is that the ``dry'' case has in total less water than the ``wet'' case, and the lower enrichment causes the planetary photosphere to sit a larger pressures, where the absorption minima are more strongly affected by pressure broadening of line wings \citep[see, e.g.,][]{molliereboekel2015}.

One spectral signature which looks as if it might enable to discriminate between the ``dry'' and the ``wet'' planet,
i.e., a formation inside or outside the water iceline, is the CO$_2$ feature at 4.3 $\mu$m. CO$_2$ is a molecule which is abundant in atmospheres with C/O $\lesssim$ 1, i.e.,  it is formed in atmospheres which are oxygen-rich when compared to the carbon abundance. The ``dry'' planet is very oxygen-rich. On the other hand it is so carbon-poor that one may think that this hampers the formation of CO$_2$, even though the planet has C/O $\lesssim$ 1. Consequently there is no CO$_2$ feature present in the spectrum of the ``dry'' planet. However, as said in Section \ref{sect:petit_atmos_struct}, the CO$_2$ abundance is rather connected to the total enrichment, and not so much the C/O ratio. This can also be seen in Figure 12 in \citet{molliereboekel2015}, where there is a weakening of the CO$_2$ feature and
abundance associated with a decreasing metallicity in the atmospheres.
The ''dry'' planet is less strongly enriched when compared to the ``wet'' planet. Testing this further we found that the CO$_2$ feature vanished when we decreased all metal abundances of the ``wet'' planet by a factor 10, leaving the relative atomic abundances constant. An increase of the ``dry'' planet to a C/O = 0.5 (where we increased the carbon fraction) did not generate a CO$_2$ feature in the atmosphere's spectrum. And a decrease of the ``wet'' planet's carbon abundance to a C/O ratio of 0.15 did not significantly weaken the CO$_2$. Therefore this feature is not well suited to discriminate between the two formation locations (in- or outside the iceline).


\rch{We note that while we consistently find that C/O$<$1 with carbon depletion, there is a substantial range of possible C/O values, depending on the disk chemistry model. The ``dry'' planet typically has C/O$<$0.2, while for the ``wet'' planet values range between 0.1 and 0.5. For this range, the efficiency of inclusion of carbon-bearing volatiles as clathrates in water ice is the most critical parameter. If no clathrates are included all, the resulting C/O value for the ``wet'' planet is close to 0.1 for all explored chemistry models. Such differences in the C/O value may be constrained using retrieval methods on, e.g., JWST observations \citep[see][]{greeneline2016}: the main carbon- and oxygen-carrying molecular species water and CO can be retrieved with a high SNR. However, as shown in Section \ref{sec:results:stoichiometry}, without a detailed quantitative understanding of the carbon depletion in the refractory material and other processes such as clathrate formation, directly linking an observed C/O ratio to a specific formation scenario or location is not yet possible.}

\subsection{Transmission spectra}\label{sec:results:spectra_transm}\subsubsection{\emph{PETIT} code results}
The transmission spectra obtained for the three planets studied in emission in Section \ref{sec:results:spectra:petit} can be seen in the right panel of Figure \ref{fig:spectra_nominal_petit}.
The clear dichotomy between oxygen-dominated and carbon-dominated planets persists, with the ``wet'' and the carbon-deficient ``dry'' planet showing strong water features and the ``dry'' planet without carbon deficiency being dominated by methane absorption. The strong CO$_2$ feature seen in the ``wet'' planet's emission also leads to an increased planetary radius in the transmission spectrum at $\sim$4.2~$\mu$m. In total, as the ``wet'' planet is less massive than the ``dry'' planets its contrast between transmission maxima and minima is larger than when compared to the carbon-deficient ``dry'' planet. The ``dry'' planet without carbon deficiency has in general a smaller radius than the carbon-deficient ``dry'' planet because it's methane is decreased in the higher atmospheric layers by the condensation of SiC which reduces the atmospheric opacity in these layers. This allows for the emergence of relatively strong CO features in the transmission spectrum of this planet, as the relative importance of CO increases.

\rch{The distinction between the ``dry'' and ``wet'' planet in the carbon depleted scenario based on the ``dry'' planet's lower C/O ratio as compared to the ``wet'' planet for the specific chemistry model shown here may be possible from the transmission spectra. Note, however, that the presence of clouds could potentially inhibit the retrieval of the C/O ratio based on the water and CO abundances \citep{greeneline2016}. The reason for this is the clouds' higher optical depth when probed under transit geometry which obscure the molecular features \citep{fortney2005}. Further the aforementioned need for a better quantitative understanding of the relevant disk chemistry processes persists.}

\subsection{Simulated observations}\label{sec:results:simulated_observations}

\begin{figure*}[ht!]
\includegraphics[angle=0,width=0.48\textwidth]{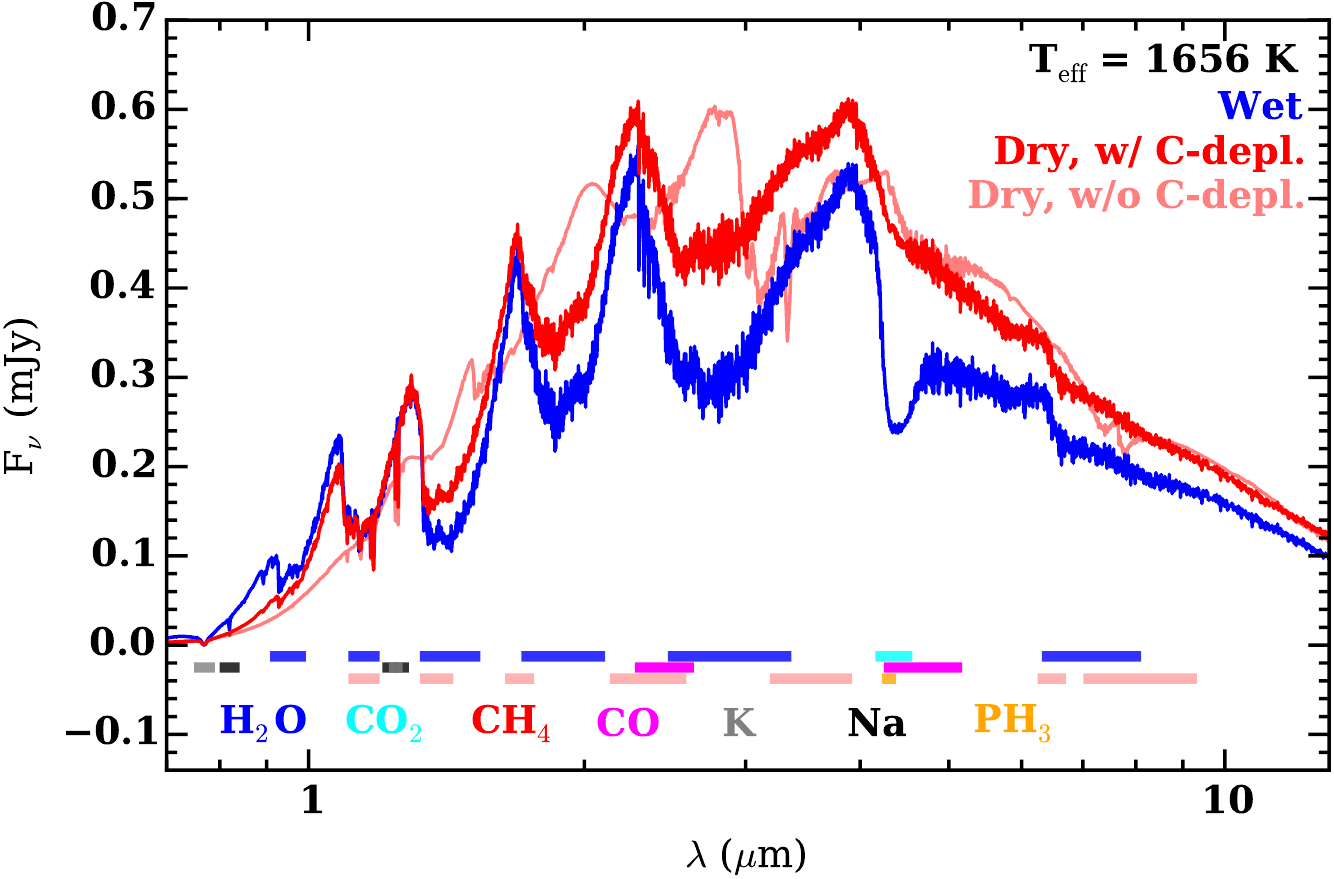}
\includegraphics[angle=0,width=0.48\textwidth]{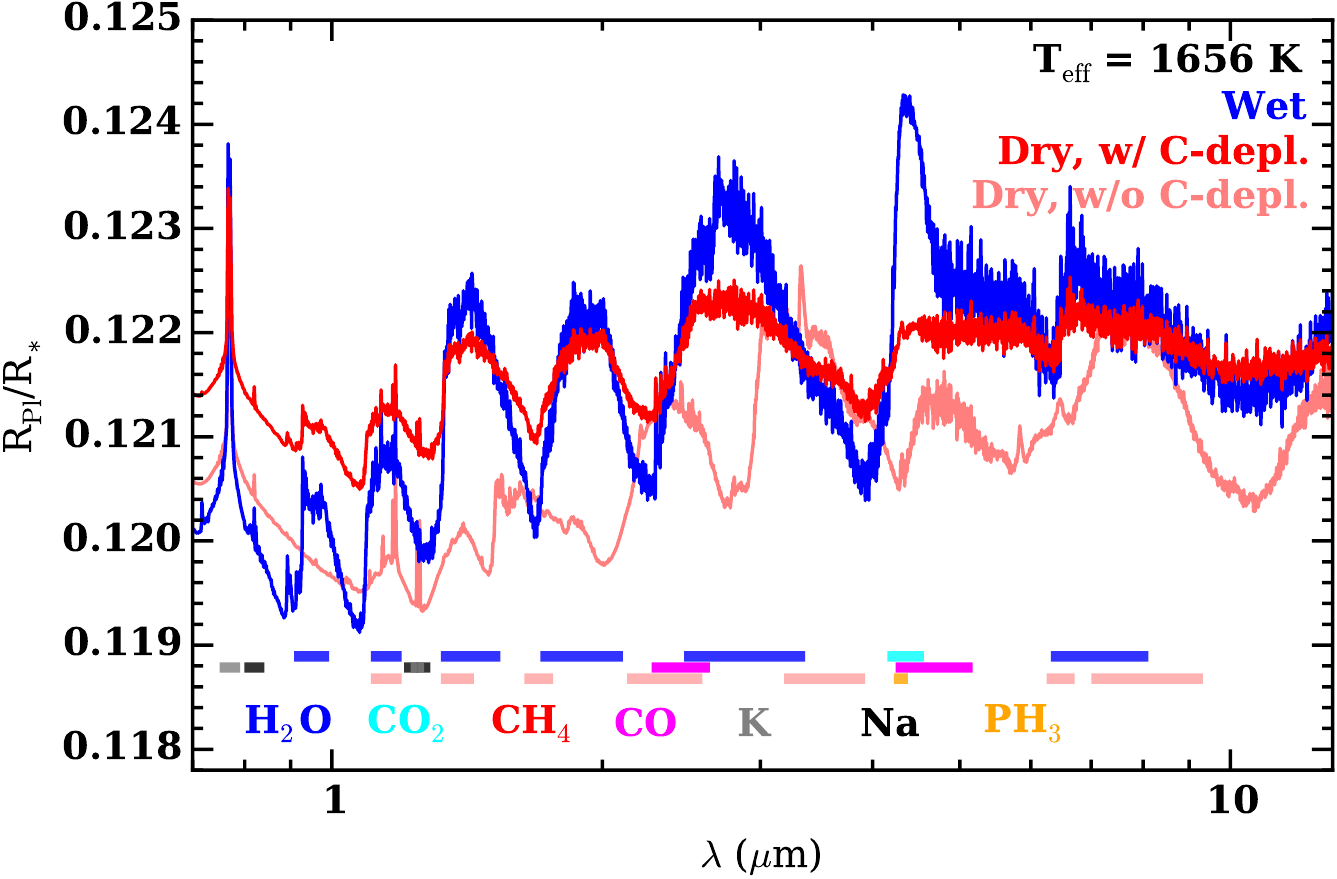}
\cprotect\caption{\label{fig:spectra_nominal_petit} Emission ({\it left panel}) and transmission ({\it right panel}) spectra of the 3 planets studied in Figure \ref{fig:chemistry}. The spectra are shown for the ``wet'', ``dry'' (\verb|Cdef|) and ``dry'' (without \verb|Cdef|) planet as a blue, red and light red line, respectively. We indicate the position of absorption bands of the most important absorbers at the bottom of the figure. The planets have been put to a distance of 55.6 pc for the emission spectra \ColorJournal}

\includegraphics[angle=0,width=0.48\textwidth]{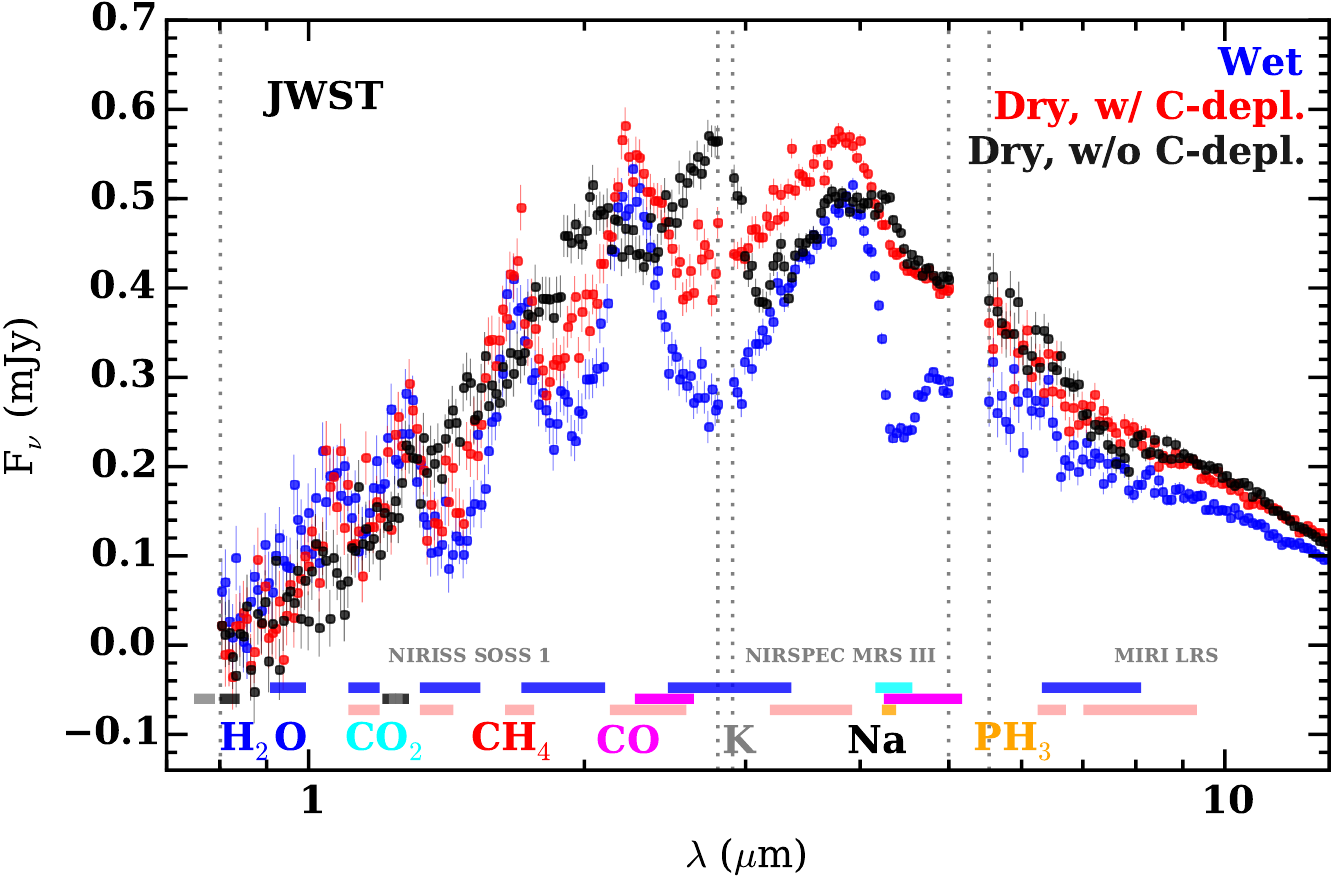}
\includegraphics[angle=0,width=0.48\textwidth]{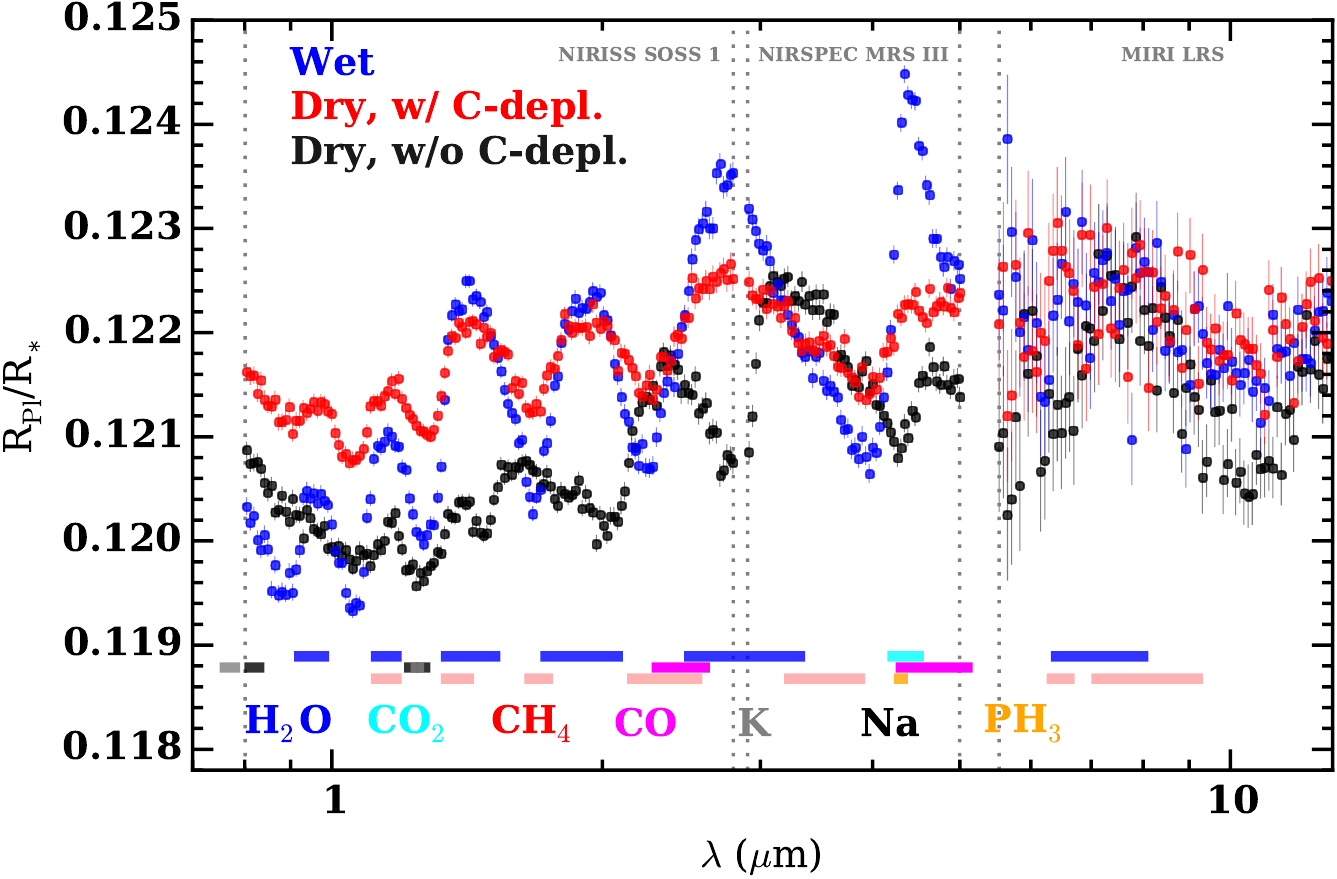}
\cprotect\caption{\label{fig:simulated_observations}Simulated observations of the dayside emission spectra (secondary eclipse) and day-night terminator transmission spectra of selected models, as observed with the JWST for a system with a G2-type host star and an apparent brightness of K\,$=$\,7.0\,mag. \ColorJournal}

\includegraphics[angle=0,width=0.48\textwidth]{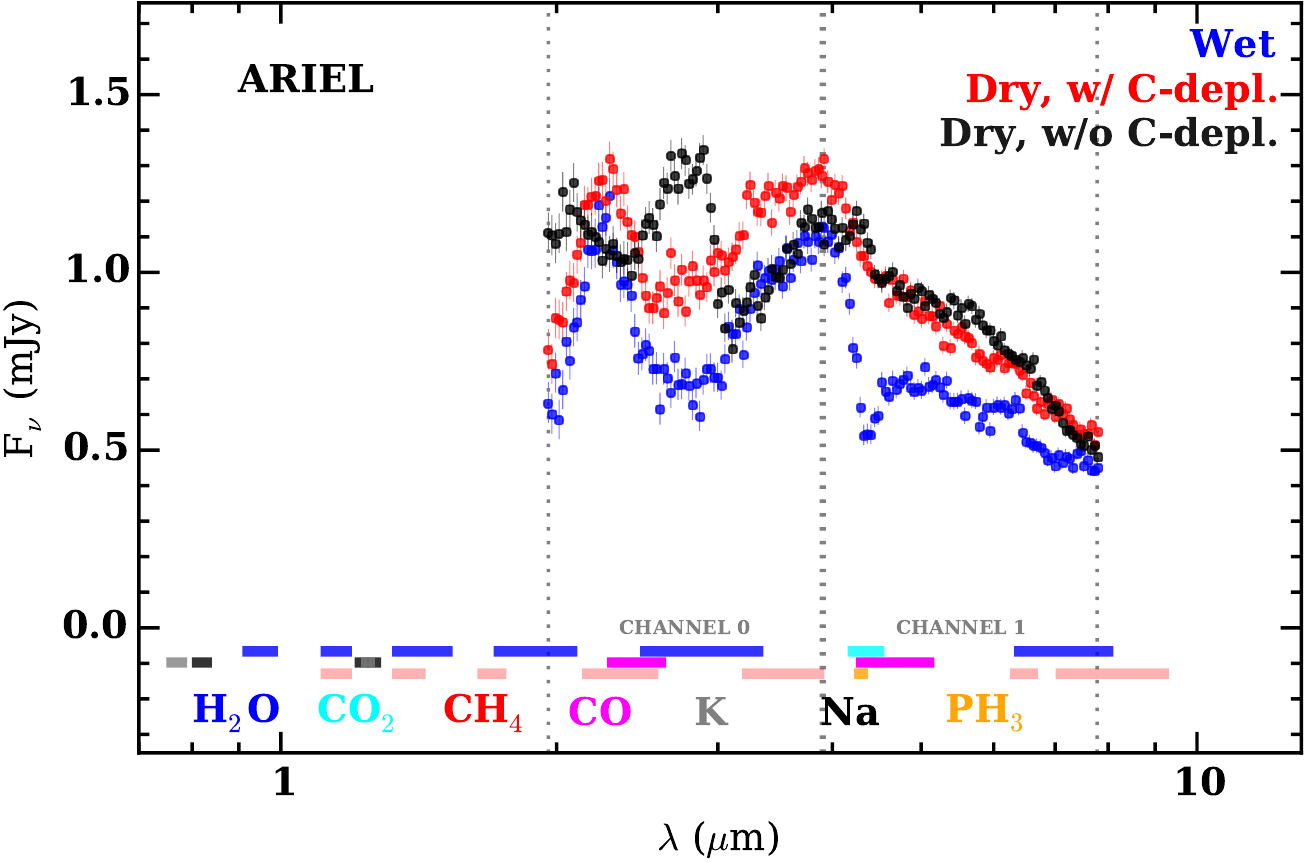}
\includegraphics[angle=0,width=0.48\textwidth]{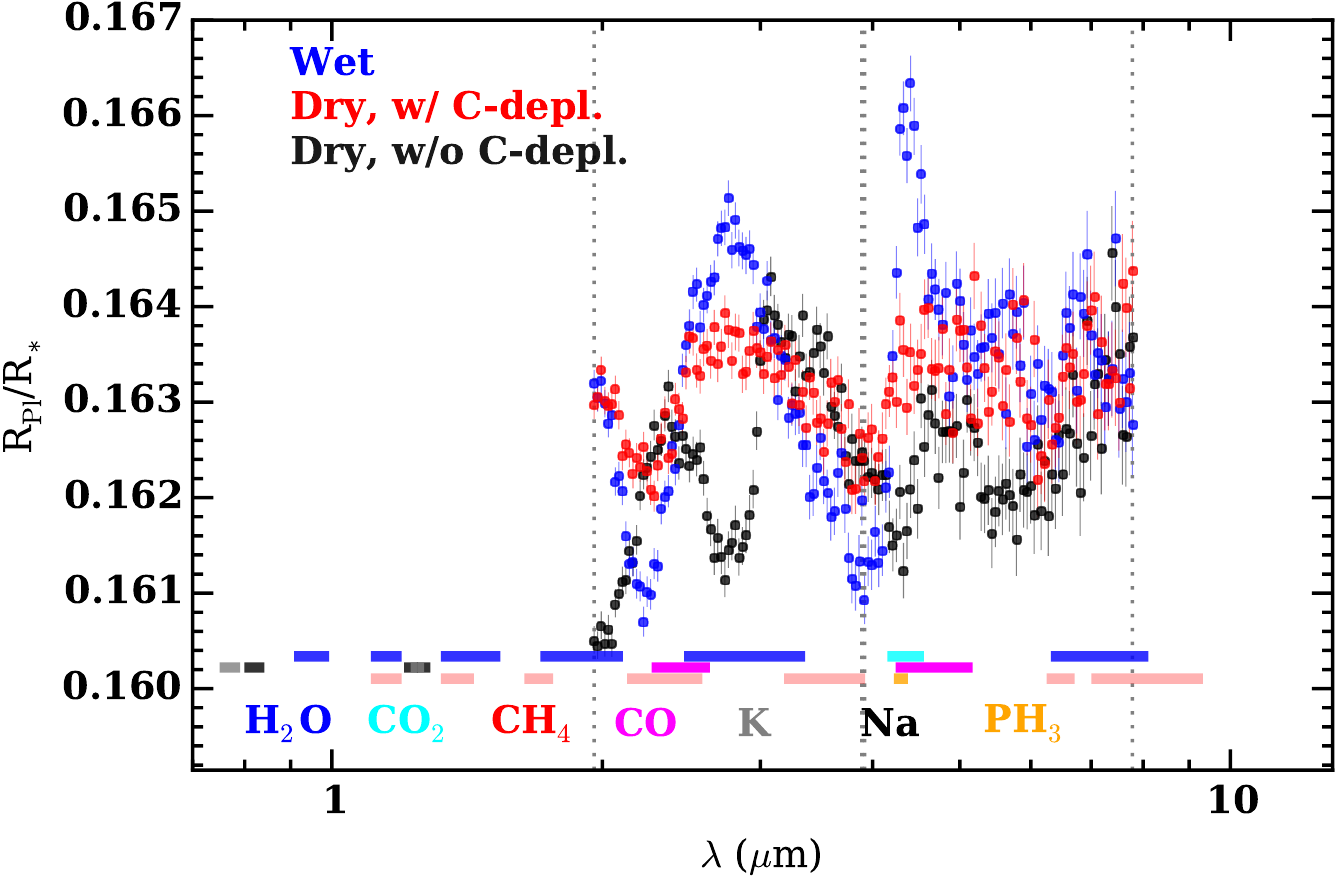}
\cprotect\caption{\label{fig:simulated_observations_ARIEL}Simulated observations of the dayside emission spectra (secondary eclipse) and day-night terminator transmission spectra of selected models, as observed with the proposed mission ARIEL for a system with a K2-type host star and an apparent brightness of K\,$=$\,8.0\,mag, where we have averaged over 5 eclipses. \ColorJournal}
\end{figure*}

{In Figure~\ref{fig:simulated_observations} we show simulated observations of secondary eclipse events of the systems as described in Section\cs\ref{sec:results:spectra}, which are placed at a distance of 55.6\cu pc, yielding an apparent stellar brightness of $K$\cm$=$\cm7.0\cu mag. We simulate eclipse observations with the JWST using the \emph{EclipseSim} package \citep{vanboekelbenneke2012}. We observe the system during 1 eclipse in each using the NIRISS \citep{doyonhutchings2012}, NIRSPEC \citep{ferruitbagnasco2012}, and MIRI instruments \citep{wrightrieke2010}. The length of each observation is taken to be the full eclipse duration T$_{14}$ bracketed by a ``baseline'' of length T$_{23}$ before as well as after the eclipse\footnote{T$_{14}$ is the duration between the moment the planet starts to disappear behind the star and when the planet has completely reappeared from behind the star. T$_{23}$ is the length of time the planet is completely behind the star. For the chosen stellar parameters and orbit, T$_{14}$\ca$\approx$180\cu min and T$_{23}$\ca$\approx$147\cu min, assuming a zero impact parameter.}, yielding a total time of 8\cu h per observation. Since the system can only be observed by one instrument at a time, observations of 3 separate eclipses are required to obtain the complete spectral coverage as shown in Figure~\ref{fig:simulated_observations}, which requires approximately 24h of observing time.}

{The planets considered here are favorable targets for eclipse spectroscopy observations and could be observed with substantially smaller telescopes. In Figure~\ref{fig:simulated_observations_ARIEL} we illustrate what the planet spectra would look like if we observe a similar system with ARIEL. This is a space telescope dedicated to exoplanet eclipse spectroscopy with an effective diameter of 0.9m that is a candidate mission for the M4 slot in ESA's cosmic vision program. The planets are now placed around a star of the same apparent brightness as in the previous example ($K$\cm$=$\cm7.0\cu mag), but with an early K spectral type and a distance of 37\,pc. To keep also the equilibrium temperature identical the orbital separation is reduced to 0.021\cu\AU. Due to the somewhat redder spectrum of the stellar irradiation compared to the G2-type host star, the resulting $p-T$ structures are somewhat closer to isothermal \citep{molliereboekel2015} and the contrast of the molecular absorption features in the dayside emission spectra is slightly lower.}

 {We observe the system during 5 eclipses, which yields the same total observing time of $\approx$\ca25h, and average the measurements to build up SNR. The resulting spectra are of very similar quality to those obtained with the JWST for the system with a solar-type host star, and the wavelengths of the strongest molecular features are covered.}

\section{Summary and conclusions}\label{sec:discussion}
In this work we present a ``chain'' of models linking directly the formation of a planet to its spectrum. The spectrum of a planet represents a window into its composition. This composition depends on the planet's formation history, its subsequent evolution, and the present-day irradiation. This opens the possibility to use spectra of extrasolar planets as a novel way to better understand planetary formation. However, due to the multitude of physical processes affecting the outcome, the link between formation and spectrum is complex. 

To make it tractable,  we construct  a  chain of simple but linked models where the output of one model serves self-consistently as the input for the next one. Our chain consists of five chain links: (1) a core accretion formation model that describes the accretion history of a giant planet, its interaction with the disk, the evolution of the disk, and the internal structure of the forming planet. It follows in particular which materials are accreted at what time and location in the disk, and whether the accreted refractory and icy material is added to the solid core or gets mixed into the gaseous H/He envelope. This is achieved by simulating \rch{explicitly} the planetesimal impacts into the protoplanetary envelope\rch{, a new aspect relative to previous similar studies}. This yields the bulk composition of the planetary envelope as inherited from the formation process. (2) In the second chain link, we use a planet evolution model to calculate the thermodynamic evolution of the planetary structure including cooling, contraction, and atmospheric evaporation. This yields the planet's mass, luminosity, and radius at an age of 5\cu Gyrs. (3) In the third link we assign a range of elemental compositions to the refractory, icy, and gaseous components that result from the formation model. We use a large number of models for the disk chemistry to obtain the associated elemental composition of the planet's atmosphere, assuming that the bulk elemental composition of the envelope is representative for the atmospheric composition. We explore various assumption on the composition of the gaseous, refractory, and volatile material that represent extremes of the plausible parameter space. (4) Using a fully-non gray radiative-convective model of the atmosphere and given the planet's physical properties and elemental composition, we calculate in the fourth chain link the atmospheric $p$-$T$ structure and molecular composition yielding also the planet's emission and transmission spectrum. (5) In the fifth chain link, we simulate  spectroscopic observations of the planet's eclipse \rch{and transit with JWST and ARIEL}.

We apply this chain of models to two hot Jupiters with very different formation histories to investigate whether this leads to visible spectral imprints: (1) a ``dry'' planet that formed completely in the warm inner disk inside of the water iceline, and (2) a ``wet'' planet that formed completely in the colder outer disk, outside of the water iceline. \rch{Because we find that the enrichment of hot Jupiters is dominated by  planetesimal accretion, these two planet formation pathways  represent two extremes of the possible scenarios, intermediate cases could occur when a planet crosses the iceline during its formation.} The first planet becomes a hot Jupiter by disk migration. We assume that also the second planet is moved close to the central star, but this time due to dynamical interactions (Kozai migration or planet-planet scattering), without accreting ``dry'' material in the inner disk. This  leads to the following main results:
\\
\\
(1) \textit{Planetesimals play a dominant role for the planetary atmospheric composition \rch{of  hot Jupiters}.} An important difference between our model and some previous efforts to predict the planetary composition from the parent disk properties (e.g., \citealt[][]{obergmurray-clay2011,hellingwoitke2014}\rch{; \citealt{ali-dipmousis2014}}) is that in our model the planetesimals, for which we explicitly calculate their atmospheric dissolution, form the prime source of heavy elements in the planetary envelope\rt{. They are dominant over the heavy elements accreted with the gas}, at least for the giant planets of  low\rch{er} mass that we study \rch{here} (Saturnian to Jovian mass)\rch{, in agreement with \citet{MousisMarboeuf2009}}.  Core accretion models predict  that the planetary enrichment due to planetesimal accretion is a decreasing function of planet mass \citep{mordasiniklahr2014}, in good agreement with observations \rch{of the interior and atmospheric enrichment of Solar System and extrasolar giant planets \rt{with an equilibrium temperature of less than $\sim$1000 K}} \rch{\citep{millerfortney2011,kreidbergbean2014,guillotgautier2014,thorngrenfortney2015}}. Thus, there should be a transition from a planetesimal-dominated composition for lower mass planets like the ones considered here to a composition that is dominated by the composition of the accreted gas at large masses (above \rch{2}-10 Jovian masses, \rch{as estimated in Sect. \ref{sect:importanceplanetesimals}). Since the large majority of  known transiting hot Jupiters have lower masses than these values,  planetesimal-dominated compositions should \rt{likely} apply to most hot Jupiters. More massive planets like those detected by direct imaging may in contrast have different, gas-dominated abundances}. 

In this work, the bulk elemental composition of the envelope is taken to be representative for the atmosphere of the mature planet. This is valid only if the envelope remains well mixed during the formation phase where most heavy elements are accreted earlier than most gas (Figure\cs\ref{fig:mass_vs_Time} and \cs\ref{fig:mass_vs_aPlanet}) leading \rch{potentially} to compositional gradients in the interior, as well as during the subsequent evolution. As in most planet formation and evolution models, a fully convective interior is a fundamental assumption in our model, but we note that the compositional gradients   (Fig. \ref{fig:zenvejupsat}) may halt large scale convective mixing \citep{lecontechabrier2012}. If the envelope does not remain well mixed during  formation, then the final atmospheric composition will depend primarily on the heavy element abundances in the gas acquired during gas runaway accretion, rather than on material supplied by planetesimals during the early formation phase \citep{thiabaudmarboeuf2015}. A similar situation arises regarding the compositional mixing across the deep radiative zone that develops during a hot Jupiter's evolution. This could also decouple the observable atmospheric composition from the bulk composition. \rch{Simple estimates indicate (Sect. \ref{sec:methods:atmosphere}) that the latter effect should not be important, but t}he detailed mixing processes in the envelope during runaway gas accretion remain to be investigated, as well as the long-term evolution of the relation between the bulk interior and atmospheric composition. 

{\it (2) Hot Jupiters are most likely oxygen-rich, i.e., have C/O$<$1, except for non-standard disk chemistries that have no depletion of refractory carbon in the inner disk.} Our result for the planetary C/O depends critically on the assumptions made for the refractory composition in the inner  disk. We adopt a composition that is inherited from the ISM \citep{gaidos2015}, in contrast to the approach often used that initially all material is hot and gaseous, and that solids are formed along the condensation sequence as the disk cools. The actual disk composition could lie between these extremes and will most likely resemble one or the other depending on location in the disk and evolutionary stage \citep{pontoppidansalyk2014}. We allow the disk chemistry to alter the ISM refractory composition only in a single, however crucial, pathway: namely that carbon grains initially present in the ISM material can be destroyed in the inner parts of the disk by oxidizing reactions at the carbon grain-gas interface \citep{2001A&A...378..192G,leebergin2010}. \rch{This is based on the observation that the inner part of the Solar System is very carbon poor \citep{1988RSPTA.325..535W,2001E&PSL.185...49A,2015PNAS..112.8965B}, and that freshly polluted white dwarf atmospheres are also carbon poor \citep{2013MNRAS.432.1955F,Wilson01072016} which points towards a generality of carbon depletion.}

Then, the result is that all hot Jupiters are oxygen-rich (i.e. C/O $<$1):  \rch{as their composition is planetesimal dominated}, a planet forming inside of the water iceline is oxygen-rich because silicates from the dissolved rocky planetesimals add high amounts of oxygen atoms to the planet's envelope. Some carbon is accreted in the form of C-depleted planetesimals and as CO and CH$_{4}$ gas, but the amount is small compared to the oxygen that is accreted in the form of planetesimals. A planet forming outside of the water iceline is also oxygen rich. Its envelope gets enriched via the planetesimals by refractory and volatile material that contains both oxygen and carbon, but due to the oxygen-dominated composition of these building blocks containing water ice, they also end up with a C/O$<$1. Thus, both the ``dry'' planet formed inside of the water iceline and the ``wet'' one formed outside of it are dominated by oxygen, and with our assumed inner disk carbon depletion profile (Fig. \ref{fig:Cdef}) the ``dry'' planets \rch{sometimes} have an even lower C/O than the ``wet'' ones. For the former, we find C/O$<$0.2, for the latter, C/O$<$0.9 with most values clustering around 0.1-0.3. Only in disk chemistry models without carbon depletion in the inner disk (i.e., an ISM-like composition) we robustly find that planets forming inside of the water iceline can have a C/O$>$1: in the accreted planetesimals, the refractory carbon now dominates over the oxygen in the silicates, leading to planetary C/O ratios of $\sim$1.2 for an approximately ISM refractory composition with a 2:1 mass ratio of silicates:carbon, and correspondingly higher or lower values (between 0.5 and 2.5) for different assumption of this ratio. However, as outlined above, disks without carbon depletion in the inner regions appear unlikely, rendering the formation of carbon-rich hot Jupiters  \rch{via this channel} unlikely. 

\rt{In our model, we include a planet's enrichment  by both planetesimal impacts, and by the heavy elements that are accreted in gaseous form together with the H/He gas. Our finding that a planetesimal-dominated enrichment (usally) leads to O-rich compositions is in good agreement with earlier works that merely assumed planetesimal domination to be the case. The new aspect that is added by our study is that we explicitly calculate the planetesimal dissolution and then directly find that planetesimal enrichment is really the dominant enrichment pathway for hot Jupiters, at least within the fundamental assumptions of our model. The fundamental reason for this is that already relatively modest planetesimal contributions are sufficient to move from the gas to the planetesimal-dominated regime.} \rch{This probably planetesimal-dominated nature of the enrichment of most hot Jupiters furthermore means that other mechanisms that were proposed  to lead to high C/O in previous studies \citep{obergmurray-clay2011,hellingwoitke2014,ali-dipmousis2014} appear unlikely for most hot Jupiters. The reason is that they rely on the accretion of gas of different C/O ratios such that they only apply to gas-dominated enrichments. They may be applicable  to planets more massive than 2-10 $\mj$.}

We neglect the effect of a moving iceline which could condense ice on grains and planetesimals otherwise consisting of refractories. If all condensible volatiles are trapped in solids outside of the initial iceline position and the inner disk is cleared from volatiles due to the diffusive disk evolution \citep[see, e.g.,][]{ali-dipmousis2014}, this might be a viable assumption. If we would allow for ice condensation on grains, and therefore planetesimals, inside the initial iceline at later times our main result would not be changed: The planets stay oxygen-rich, as water ice only adds more oxygen.

\rch{Regarding the impact of model parameters and settings,  the carbon depletion model has the biggest impact on the possible outcome of the ``dry'' planet's C/O ratios. We found that introducing a partially ad hoc model of the actual carbon depletion function is sufficient, as already relatively modest carbon reduction factors ($10^{-1}$ instead of the nominal $10^{-4}$) do not change the result that carbon-rich hot Jupiters cannot form under carbon-depleted conditions. For the ``wet'' planet the clathrate formation can have a non-negligible impact on the C/O ratio if a significant amount of carbon-bearing volatiles is trapped in the water ice planetesimals (this requires a volatile abundance model with a non-negligible carbon-fraction). Under no circumstances do we find ``wet'' planets with C/O ratios bigger than 1, however. In the ``dry'' cases without carbon depletion, the silicate-to-carbon mass ratio has the biggest influence on the planetary C/O ratio, leading to C/O values of $\sim 2.8$ for the maximum value considered in the paper (C/Silicate mass ratio = 1).}

{\it (3)  \rch{Constraining a hot Jupiter's formation location and migration mechanism based on the spectral imprint of a C/O higher or lower than 1 alone appears difficult because hot Jupiters are expected to be oxygen-rich for a formation both inside and outside of the water iceline, at least for our nominal disk chemistries.}}
\rch{The ``dry'' and the ``wet'' planets have oxygen-rich envelopes such that their atmospheres both show strong water features and are dominated by oxygen-rich chemistry. For the example shown in sections \ref{sec:results:spectra} and \ref{sec:results:spectra_transm} the ``dry'' planet has a much lower C/O ratio than the ``wet'' planet and it has been shown that both the water and the CO abundance may be retrieved with high SNR in hot Jupiter emission spectra, and thus the C/O ratio \citep{greeneline2016}. Therefore a distinction of the formation location may be possible for this specific disk chemistry. However, in Section \ref{sec:results:stoichiometry} we show that depending on the details of the carbon depletion and clathrate formation model other scenarios may arise where the ``dry'' and ``wet'' planet have overlapping C/O$<$1 ratios, usually $<$ 0.3. Thus an important step to improve the link between planet formation and spectra would be a detailed and quantitative treatment of the carbon depletion and clathrate formation chemistry in exoplanet disks.}

\rch{Nonetheless, we find some secondary features distinguishing the two classes:} Planets forming outside of the water iceline are at a fixed total mass more enriched in C and O relative to H/He because of the larger reservoir of planetesimals in the outer disk. This can result in a higher CO$_{2}$ abundance. Next, planets forming outside of the water iceline are more strongly enriched in C and O relative to Si and Mg because of the accretion of icy planetesimals (Fig. \ref{fig:stoichiometry}). \rch{This is generally true for O; for C it is true only if efficient trapping of C-rich ices as clathrates occurs.} In the case where carbon depletion is neglected in the inner parts of the disk, albeit favored by neither observation nor theory, carbon-rich planets can form.  \rch{A complete inheritance of carbon-rich ISM-like grains into the solid building blocks of hot Jupiters forming inside of the water iceline thus represents a planetesimal-driven, but probably unlikely pathway towards high C/O$>$1. A related formation path to C-rich planets was suggested for solid planets \citep{gaidos2000,carter-bondobrien2010} around stars which have intrinsically themselves a C/O$>$1. However, such carbon-rich stars are probably very rare \citep[e.g.,][]{fortney2012,gaidos2015}.} In this case there is a clear dichotomy between planets having accreted exclusively inside or outside the iceline, leading to C/O-ratios~$>$~1 or $<$~1, respectively. Carbon-rich (C/O $>$ 1), ``dry'' hot Jupiters would be dominated by methane absorption, rather than water, leading to a distinctively different spectrum when compared to the water-rich, ``wet'' planet. 

It is interesting to link these findings to predictions by formation models. Giant planet formation models based on the core accretion paradigm predict that around low-metallicity stars, giant planets form only outside of the water iceline, while around high-metallicity stars, giant planets can form both outside and entirely inside of the water iceline (Fig.\,5 in \citealt{idalin2004}; Fig.\,8 in \citealt{mordasinialibert2012a}). The reason is that around high metallicity stars, the amount of refractories alone is high enough to form a critical core of $\sim$10 $\mearth$  triggering runaway gas accretion, while at low [Fe/H], the extra mass provided by the condensation of ice is needed to form such a massive core. Here it is implicitly assumed that the stars have a scaled solar composition. This leads to two predictions: (1) that around low [Fe/H] stars, hot Jupiters with the signs of having accreted only inside of the iceline should be rare, while around high [Fe/H] stars, hot Jupiters with the signs of an accretion inside as well as  outside of the iceline are predicted. (2) for stars where tidal interactions have not damped high obliquities, hot Jupiters that show signs of an accretion only beyond the iceline should have a wide range of obliquities including high ones, at least if a high obliquity is a sign of a dynamical interaction, and if this interaction does not lead to the accretion of solids originating from inside of the water iceline, contrarily to disk migration.

Using our chain of models we were able to predict the planets' spectra based on their formation \rt{history}. The most striking of our results described above is that the formation of carbon-rich \rch{hot Jupiters} with C/O $>$ 1 is unlikely. 
\rcp{This result is in good agreement with observations, because the \rch{hot Jupiters recently} characterized appear to be oxygen-rich \citep{lineknutson2014,benneke2015,singfortney2015}. \citet{lineknutson2014}  do not find any conclusive evidence for super-solar C/O ratios. The study by \citet{benneke2015} allows for super-solar C/O ratios, while robustly excluding cases with C/O$>$1. The latter is due to the fact that a water detection in {HST WCF3} firmly rules out a carbon-rich chemistry for the considered hot Jupiters. \citet{singfortney2015} further show that the low water abundance in some hot Jupiters is due to the presence of clouds and hazes, and not to a  water depletion during formation. Such a primordial depletion would be in contradiction to our results.} 
Tentative evidence for planets with carbon-rich atmospheres exists \rch{for types of planets other than hot Jupiters like HR8799b} \citep{lee2013} \rch{or 55 Canc e}  \citep{tsiarasrocchetto2015}. In our paper we also discuss the possibility of a carbon sweet spot in the disk which lies outside of the region of carbon depletion, but still inside of the iceline \citep{lodders2004}. Planets which would form exclusively within this region could attain carbon-rich envelopes and atmospheres. If such planets end up close to their stars they should be easily distinguishable due to their methane-dominated spectra. While the exact location and processes which give rise to this carbon sweet spot are specific to our model assumptions, the existence and formation of carbon-rich planets is therefore not downright refutable, but should be the exception, rather than the rule. At least under the assumptions made in this work, the majority of hot Jupiters should be oxygen-rich.

\acknowledgements{\emph{{Acknowledgments}} We thank K. Dullemond, U. Marboeuf, and A. Thiabaud for useful discussions. C. M. acknowledges the support of the Swiss National Science Foundation via grant BSSGI0$\_$155816 ``PlanetsInTime''. Parts of  this work have been carried out within the frame of the National Centre for Competence in Research ``PlanetS'' supported by the Swiss National Science Foundation (SNSF).}
\appendix

\section{Detailed description of some chain links}
\label{appendix:chain}
\subsection{Formation}
\label{appendix:chain_form}
\subsubsection{Viscous gas disk evolution}
\label{appendix:chain_form_gas_disk}
As mentioned in Section \ref{sec:methods:disk_model} our time dependent protoplanetary disk model is describing a
1+1D (vertical and radial) viscous disk. We include the effects of turbulent viscosity in the $\alpha$ approximation, photoevaporation by the star and from external sources, and mass accretion onto the planet.
The governing equation for the evolution of the surface density of the gas $\Sigma$ in time $t$ is given as \citep{papaloizouterquem1999}:
\beq
\frac{\partial \Sigma}{\partial t}=\frac{1}{r}\frac{\partial}{\partial r}\left[3 r^{1/2} \frac{\partial}{\partial r}\left(r^{1/2}\nu \Sigma\right)\right]-\dot{\Sigma}_w(r)-\dot{\Sigma}_{\rm pla}(r).
\eeq
In this equation $r$ is the distance from the star, $\nu$ the viscosity while $\dot{\Sigma}_w(r)$ and $\dot{\Sigma}_{\rm pla}(r)$ denote the change of the surface density due to photoevaporation and planetary gas accretion, respectively. The viscosity is written as $\nu=\alpha c_{\rm s} H$ \citep{shakurasunyaev1973} where $c_{\rm s}$ is the sound speed and $H$ the vertical pressure scale height of the disk. The parameter $\alpha$ is set to 0.007. The methods and boundary conditions that are used to solve this equation are described in \citet{alibertmordasini2005} while the planet accretion and photoevaporation term, which includes external and internal photoevaporation, and the initial surface density profile are described in \citet{mordasinialibert2012b}. For the calculation of the vertical structure, the impact of stellar irradiation is included in the equilibrium angle approximation \citep{barriere-fouchetalibert2012}.

\subsubsection{Planetesimal and gas accretion}
\label{appendix:chain_form_accretion}
The solid accretion rate of the protoplanet is obtained by considering its gravitationally enhanced cross-section
as it moves through the disk. As described in \citet{pollackhubickyj1996} the solid accretion rate can be found using
a Safronov type rate equation:
\beq
\frac{d M_{Z}}{dt}=  \Sigma_{\rm P} \Omega F_{\rm G} \pi R_{\rm capt}^{2}.
\eeq
In this equation, $\Sigma_{\rm P}$ is the surface density of planetesimals, $\Omega$ the Keplerian frequency, and $F_{G}$ the gravitational focussing factor \citep{greenzweiglissauer1992}. The planetesimal random velocities are the same as in \citet{pollackhubickyj1996}. $R_{\rm capt}$ is the protoplanet's capture radius for planetesimals. It is larger than the core radius due to the braking effect of the gaseous envelope and calculated with the planetesimal-protoplanet interaction model mentioned in Section
\ref{sec:methods:planetesimalimpacts} and Appendix \ref{appendix:chain_form_planetesimal_enve_interact}.

The accretion rate of gas is found by solving a slightly simplified set of internal structure equations of the planet's 1D radial structure in the quasi-hydrostatic approximation. 
The internal structure of the gaseous envelope is described by the equations of mass conservation, hydrostatic equilibrium, energy transfer, and energy conservation. The latter equation is simplified by assuming a radially constant luminosity. The temporal evolution of the total luminosity is found by energy conservation arguments as described in \citet{mordasinialibert2012b}. One then has \citep[e.g.][]{bodenheimerpollack1986}:
\begin{eqnarray}
\frac{\partial m}{\partial r}&=&4 \pi r^{2} \rho\\   
\frac{\partial P}{\partial r}&=&-\frac{G m}{r^{2}}\rho    \\
\frac{ \partial T}{\partial r}&=&\frac{T}{P}\frac{\partial P}{\partial r}\nabla(T,P).
\end{eqnarray}
In these equations, $G$ is the gravitational constant, $r$ the distance from the planet's center, $P, T, \rho$ the pressure, temperature, and density of the gas, $m$ the mass within $r$, and the gradient $\nabla(T,P)$ can either be the radiative gradient in the diffusion approximation, or the adiabatic gradient in convective layers as determined by the Schwarzschild criterion. Effects of a compositional gradient that could suppress convection are therefore currently neglected, but could be important, as discussed in Section \ref{sec:methods:interior}. For the radiative gradient, the grain opacity is assumed to be reduced by a factor 0.003 relative to the ISM grain opacity \citep{mordasiniklahr2014}. The boundary conditions that are necessary to solve the internal structure equations differ depending on whether the planet is in the attached phase at low core masses or in the detached phase during runaway gas accretion \citep{bodenheimerhubickyj2000,mordasinialibert2012b}. In the former phase, the outer radius is proportional to the Hill sphere radius. In the latter, the planet's radius is much smaller than the Hill sphere radius. In this phase, the gas accretion rate is limited by the availability of gas in the disk, and there is a gas accretion shock on the surface of the planet. We assume that the accretion shock is supercritical, so that the kinetic energy of infalling material is effectively radiated away and cold accretion occurs \citep{marleyfortney2007,mordasini2013}. This leads to low luminosities and radii at young ages, but this is not  important for the planet's properties at an age of several Gyrs in which we are interested in this work. A certain influence exists since the envelope evaporation rate (Section\cs\ref{sec:methods:evaporation}) depends on the planet's radius. But for the Jovian and Saturnian mass planets studied in this work, the impact of cold vs. hot accretion is in any case not very large, in contrast to more massive giant planets.

\subsubsection{Envelope-planetesimal interaction}\label{appendix:chain_form_planetesimal_enve_interact}
The impact model \citep{mordasinialibert2006} determines the radial mass deposition profile by numerically integrating the trajectory of a planetesimal of initial mass $M_{\rm pl}$ during its flight through the protoplanetary envelope under the actions of gravity, gas drag, thermal ablation, and aerodynamical disruption.
The  \rch{first} governing equation \rch{is} the equation of motion for the planetesimal's position $\mathbf{r}$ in the planetocentric reference frame ($C_{D}$ is the drag coefficient \rch{taken from \citet{Henderson1976}}, $R_{\rm pl}$ the planetesimal radius, \rch{$m$ the protoplanet's mass inside of $r$, and $\rho$ the local gas density in the envelope})
\beq
M_{\rm pl}\ddot{\mathbf{r}}=-\frac{Gm M_{\rm pl}}{r^2}\cdot\frac{\mathbf{r}}{r} - \frac{1}{2}C_{D}\rho\ \dot{r}^2\frac{\dot{\mathbf{r}}}{\dot{r}} \pi R_{\rm pl}^{2}.
\eeq
\rch{The initial velocity is taken from \citet{pollackhubickyj1996} and a central impact geometry is considered. The specific value of the initial velocity is not important for the outcome of the infalling as long as it is smaller than the core's (in the attache phase) respectively planet's (in the detached phase) escape velocity. This is the case in the runaway and oligarchic planetesimal accretion regime occurring during the protoplanet's growth in the nebula  \citep{idamakino1993}. The second governing equation models} the thermal ablation \rch{(mass loss rate)} which is  powered in the most important regime by shock wave radiation due to the planetesimal's hypersonic flight \citep[e.g.][]{zahnle1992}
\beq
\frac{d M_{\rm pl}}{dt}=-C_{H} \sigma T_{\rm shock}^{4} \pi R_{\rm pl}^{2}/Q_{\rm abl}
\eeq
where $C_{H}$ is a heat transfer coefficient \citep{svetsovnemtchinov1995} and $Q_{\rm abl}$ the heat of ablation \citep{opik1958}. \rch{The post-shock temperature $T_{\rm shock}$ is found by solving numerically the normal shock wave jump conditions \citep{LANDAULIFSHITZ1987,chevaliersarazin1994} for a non-ideal gas using the EOS SCvH. The third equation describes the aerodynamical disruption}. Big impactors get aerodynamically disrupted when the aerodynamic load exceeds the tensile strength, leading to a lateral spreading of the impactor \rch{and a rapid destruction}. The rate of lateral expansion of the fluidized impactor can be described with the ``pancake'' equation \citep{zahnle1992,chybathomas1993}:
\beq
\frac{d^{2} R_{\rm pl}}{dt^{2}}=\frac{3}{4}\frac{\rho}{\rho_{\rm b}} \frac{\dot{r}^2}{R_{\rm pl}}.
\eeq
where $\rho_{\rm pl}$ is the material density of the planetesimal. \rch{Planetesimals accreted in- and outside of the water iceline are assumed to consist of silicate rocks and water ice, respectively. The  values of the main material parameters \citep{opik1958,podolakpollack1988,chybathomas1993,svetsovnemtchinov1995} are listed in Table \ref{table:matpropsplanetesimals}.} Further descriptions of the impact model can be found in \citet{fortneymordasini2013,mordasinimolliere2014} \rch{while a discussion of the effect of the impact geometry and the planetesimal's material properties on the mass deposition profile can be found in \citet{mordasini2014}}.
\rch{
\begin{table}
\centering
\begin{tabular}{ll|cc}
Quantity & Unit & Rocky  & Icy   \\ 
\hline \hline
Material density & g/cm$^{3}$ & 3.2  & 1.0 \\
Tensile strength & dyn/cm$^{2}$ & 3.5$\times10^{8}$      &       4.0$\times10^{6}$  \\
Heat of ablation &  erg/g & 8.1$\times10^{10}$    &         2.5$\times10^{10}$  \\
\hline
\end{tabular}
\caption{Material parameters for the planetesimal infall model.}
\label{table:matpropsplanetesimals}
\end{table}
}

\subsubsection{Disk migration}
\label{appendix:chain_form_migration}
At low masses, planets within a gaseous disk undergo type I disk migration \citep{tanakatakeuchi2002a}.
The migration rate $d a/d t$ of a planet at a semi-major axis $a$ under the action of a total torque $\Gamma_{\rm tot}$ we use in our model is 
\beq
\frac{d a}{dt}=2 a \frac{\Gamma_{\rm tot}}{J}
\eeq
where $J=M \sqrt{G M_{\star} a}$ is the angular momentum of a planet of mass $M$. The total torque $\Gamma_{\rm tot}$ can be expressed as \citep{paardekooperbaruteau2010}
\beq
 \Gamma_{\rm tot}=\frac{1}{\gamma}(C_{0}+C_{1}p_{\rm \Sigma}+C_{2}p_{\rm T}) \left(\frac{q}{h}\right)^{2}\Sigma a^{4}\Omega^{2}.
\eeq
In this equation, $q=M/M_{\star}$ is the planet-to-star mass ration, $\gamma$ the adiabatic index of the gas,  $\Sigma$ the gas surface density at the planet's location, $h$ the disk's aspect ratio, and $p_{\rm \Sigma}$ and $p_{\rm T}$  the local power-law slopes of the gas surface density and temperature profile in the disk. These quantities are given by the disk model described in Section \ref{sec:methods:disk_model}. The parameters $C_{0,1,2}$ depend on the local thermodynamical regime in the disk which lead to several sub-regimes of type I migration (isothermal, adiabatic, saturated). Note that these parameters were recently significantly revised \citep{baruteaucrida2013}. In contrast to the original work of \citet{tanakatakeuchi2002a} for isothermal disks, the actual direction of migration for more realistic disk thermodynamics can now also be directed outwards. A detailed description of the (non-isothermal) migration model used in this work is given in  \citet{dittkristmordasini2014}.

Once a planet becomes sufficiently massive to open up a gap in the gaseous disk (of order 100\cu\Mearth), it passes into type II migration. We use the transition criterion of \citet{cridamorbidelli2006} to determine a planet's migration regime.
In the type II regime, the accretion rate can be written as \citep{alexanderarmitage2009}
\beq
\frac{da}{dt}=\mathrm{min}(1,2 \Sigma a^2/M) \times v_{r,gas}
\eeq
where $v_{r,gas}$ is the local radial velocity of the gas. 

\subsection{Evolution}
\label{appendix:chain_evo}
\subsubsection{Envelope evaporation}
\label{appendix:chain_evo_evap}
In our evolutionary model \citep[see][]{jinmordasini2014},  the envelope evaporation rate due to EUV irradiation is modeled at high EUV fluxes with a radiation-recombination limited rate \citep{murray-claychiang2009} as
\beq
\frac{d M_{\rm rr-lim}}{d t}=4 \pi \rho_{\rm s} c_{\rm s} r_{\rm s}^{2}
\eeq
where $\rho_{\rm s}$ and  $c_{\rm s}$ are the density and sound speed at the sonic point which is located at a radius $r_{\rm s}$. At lower EUV fluxes $F_{\rm UV}$, the evaporation rate is energy-limited \citep{watsondonahue1981}:
\beq
\frac{d M_{\rm e-lim}}{d t}=\frac{\epsilon_{\rm UV} \pi F_{\rm UV} R_{\rm UV}^{3}}{G M K_{\rm tide}}.
\eeq 
where $M$ is the planetary mass, $\epsilon_{\rm UV}$ the efficiency factor, $R_{\rm UV}$ the radius where EUV radiation is absorbed, while the $K_{\rm tide}$ factor \citep{erkaevkulikov2007a} takes into account that gas only needs to reach the Hill sphere for escape. The mass loss rate in the X-ray driven regime can be estimated with an analogous equation as in the energy limited UV regime \citep{owenjackson2012a}.

\rch{\section{Planetary enrichment in heavy elements}\label{appendix:relativenerichmentzpzs}
In this appendix we describe the numerical data used to prepare Fig. \ref{fig:hot_jupiters_histo} which shows the enrichment of planets relative to the host star $e_{\rm Z,rel}$, and the parity mass $M_{1}$ where this quantity becomes unity. For gas and ice giant planets, the  enrichment of both the planetary interior and atmosphere relative to the host star can (with the current data) be approximated as a powerlaw of the form \citep{mordasiniklahr2014}
\beq\label{eq:zpzstar}
e_{\rm Z,rel}=\beta\left(\frac{M_{p}}{\mj}\right)^{\alpha}.
\eeq
We have determined the parameters $\alpha$ and $\beta$ for different theoretical and observational data sets by least square fits or by using the published values.  Table \ref{table:zpzstar} lists the values of $\alpha$ and $\beta$ found for the following seven data sets: (1) the observed atmospheric enrichment relative to the sun in carbon based on the CH$_{4}$ abundance in the four giant planets of the Solar System as given in \citet{guillotgautier2014}. (2) the observed mean atmospheric enrichment relative to the sun taking into account all measured heavy elements as quoted in  \citet{guillotgautier2014}. (3) the fit to the atmospheric enrichment relative to the star as a function of mass taking into account the Solar System giants and WASP-43b (Fig. 4 of \citealt{kreidbergbean2014}). Because of condensation and chemical disequilibrium, several of these values may only be lower limits to the bulk abundance \citep{guillotgautier2014}. (4) the $Z_{\rm Pl}/Z_\odot$ found from the  heavy element masses $M_{\rm Z}$ in the interiors of the Solar System giant planets estimated from internal structure models for Jupiter and Saturn by \citet{saumonguillot2004} for different EOS and for Uranus and Neptune by \citet{helledanderson2011,NettelmannHelled2013}. (5) the $Z_{\rm Pl}/Z_*$ based on the $M_{\rm Z}$ derived from interior structure models of weakly irradiated transiting extrasolar giant planets \citep{millerfortney2011}. (6) the $\alpha$ and $\beta$ given by \citet{thorngrenfortney2015} also obtained from interior structure models of  weakly irradiated transiting exoplanets \rt{with an equilibrium temperature of less than $\sim$1000 K}. (7) the  $\alpha$ and $\beta$ found for the bulk enrichment of the synthetic planets around 1 $\msun$ stars in the nominal population of \citet{mordasiniklahr2014}.}

\rch{The parameters $\alpha$ and  $\beta$ and their 1 $\sigma$ errors can then be used to identify (by extrapolation) a parity mass $M_{1}$ where $e_{\rm Z,rel}$=1, given as $M_{1}/\mj=\beta^{-1/\alpha}$ \citep{mordasiniklahr2014}. These masses are also listed in Table \ref{table:zpzstar}. Planets below this mass have a composition increasingly dominated by the accretion of solids, while planets with a mass higher than $M_{1}$  have a composition that may be dominated by the composition of the accreted gas.}

\begin{table*}
\centering
\begin{tabular}{lll|ccccc}
Data set & Quantity & Reference  & $\beta$ & $\alpha$ & $M_{1}$ & $M_{\rm 1,min}$  & $M_{\rm 1,max}$\\ 
\hline \hline
1 & Solar system atmospheres CH$_{4}$ & GG14 & 4.27$\pm$0.25 & -0.75$\pm$0.12 & 7.0 & 5.0 & 10.9\\
2 & Solar  system atmospheres mean & GG14 & 2.32$\pm$0.78 & -1.20$\pm$0.13 & 2.0 & 1.4 & 2.9\\
3 & Solar system atmospheres \& WASP-43b & KB14 & 2.75 & -1.10 & 2.5 &  & \\
4 & Solar system interiors & SG04, HA11 & 6.17$\pm$1.83 & -0.75$\pm$0.10 & 11.4 & 5.7 & 24.9\\
5 & Exoplanet interiors & MF11 & 6.30$\pm$1.00 & -0.71$\pm$0.10 & 13.4 & 7.8 & 26.0\\
6 & Exoplanet interiors & TF15 & 8.10$\pm$1.30 & -0.51$\pm$0.11 & 60.4 & 22.0 & 271.0\\
7 & Population synthesis interiors & M14 & 7.2 & -0.68 & 18.2 & & \\
\hline
\end{tabular}
\caption{Parameters of Eq. \ref{eq:zpzstar} and parity mass $M_{1}$ in units of Jovian masses where $e_{\rm Z,rel}$=1.}
\label{table:zpzstar}
\end{table*}

\bibliographystyle{apj}
\bibliography{references2.bib}{}

\end{document}